\documentclass[onecolumn]{aastex631}

\shorttitle{The $^{13}$CO structures in $^{12}$CO molecular clouds.}
\shortauthors{Yuan et al.}

\graphicspath{{./}{figures/}}
\begin{document}

\title{Molecular Gas Structures traced by $^{13}$CO Emission in the 18,190 $^{12}$CO Molecular Clouds from the MWISP Survey}

\author[0000-0003-0804-9055]{Lixia Yuan}
\email{lxyuan@pmo.ac.cn}
\affiliation{Purple Mountain Observatory and Key Laboratory of Radio Astronomy, Chinese Academy of Sciences, \\
10 Yuanhua Road, Qixia District, Nanjing 210033, PR China}

\correspondingauthor{Ji Yang}
\email{jiyang@pmo.ac.cn}
\affiliation{Purple Mountain Observatory and Key Laboratory of Radio Astronomy, Chinese Academy of Sciences, \\
10 Yuanhua Road, Qixia District, Nanjing 210033, PR China}

\author[0000-0001-7768-7320]{Ji Yang}
\affiliation{Purple Mountain Observatory and Key Laboratory of Radio Astronomy, Chinese Academy of Sciences, \\
10 Yuanhua Road, Qixia District, Nanjing 210033, PR China}

\author[0000-0002-7489-0179]{Fujun Du}
\affiliation{Purple Mountain Observatory and Key Laboratory of Radio Astronomy, Chinese Academy of Sciences, \\
10 Yuanhua Road, Qixia District, Nanjing 210033, PR China}

\author[0000-0002-0197-470X]{Yang Su}
\affiliation{Purple Mountain Observatory and Key Laboratory of Radio Astronomy, Chinese Academy of Sciences, \\
10 Yuanhua Road, Qixia District, Nanjing 210033, PR China}

\author[0000-0001-8315-4248]{Xunchuan Liu} 
\affiliation{Shanghai Astronomical Observatory, Chinese Academy of Sciences, PR China} 

\author[0000-0003-2549-7247]{Shaobo Zhang}
\affiliation{Purple Mountain Observatory and Key Laboratory of Radio Astronomy, Chinese Academy of Sciences, \\
10 Yuanhua Road, Qixia District, Nanjing 210033, PR China}

\author[0000-0002-3904-1622]{Yan Sun}
\affiliation{Purple Mountain Observatory and Key Laboratory of Radio Astronomy, Chinese Academy of Sciences, \\
10 Yuanhua Road, Qixia District, Nanjing 210033, PR China}

\author[0000-0003-2418-3350]{Xin Zhou}
\affiliation{Purple Mountain Observatory and Key Laboratory of Radio Astronomy, Chinese Academy of Sciences, \\
10 Yuanhua Road, Qixia District, Nanjing 210033, PR China}

\author[0000-0003-4586-7751]{Qing-Zeng Yan}
\affiliation{Purple Mountain Observatory and Key Laboratory of Radio Astronomy, Chinese Academy of Sciences, \\
10 Yuanhua Road, Qixia District, Nanjing 210033, PR China}

\author[0000-0002-8051-5228]{Yuehui Ma}
\affiliation{Purple Mountain Observatory and Key Laboratory of Radio Astronomy, Chinese Academy of Sciences, \\
10 Yuanhua Road, Qixia District, Nanjing 210033, PR China}

%% Mark off the abstract in the ``abstract'' environment. 
\begin{abstract}
After the morphological classification of the 18,190 $^{12}$CO molecular clouds, 
we further investigate the properties of their internal molecular gas structures traced by the $^{13}$CO($J=$ 1$-$0) line emissions. 
Using three different methods to extract the $^{13}$CO gas structures within each $^{12}$CO cloud,  
we find that $\sim$ 15$\%$ of $^{12}$CO clouds (2851) have $^{13}$CO gas structures and 
these $^{12}$CO clouds contribute about 93$\%$ of the total integrated flux of $^{12}$CO emission. In each of 2851 $^{12}$CO clouds with $^{13}$CO gas structures, 
the $^{13}$CO emission area generally does not exceed 70$\%$ of the $^{12}$CO emission area, 
and the $^{13}$CO integrated flux does not exceed 20$\%$ of the $^{12}$CO integrated flux.
We reveal a strong correlation between the velocity-integrated intensities of $^{12}$CO lines 
and those of $^{13}$CO lines in both $^{12}$CO and $^{13}$CO emission regions. 
This indicates the H$_{2}$ column densities of molecular clouds are crucial for the $^{13}$CO lines emission. 
After linking the $^{13}$CO structure detection rates of the 18,190 $^{12}$CO molecular clouds to their morphologies, 
i.e. nonfilaments and filaments, we find that the $^{13}$CO gas structures are primarily detected in 
the $^{12}$CO clouds with filamentary morphologies. Moreover, these filaments 
tend to harbor more than one $^{13}$CO structure. That demonstrates filaments not only have larger spatial scales, 
but also have more molecular gas structures traced by $^{13}$CO lines, i.e. the local gas density enhancements.
Our results favor the turbulent compression scenario for filament formation, in which dynamical compression 
of turbulent flows induces the local density enhancements. 
The nonfilaments tend to be in the low-pressure and quiescent turbulent environments of the diffuse interstellar medium. 
\end{abstract}


\section{Introduction} \label{sec:intro}
%Molecular clouds are the site of star formation. The empirical star formation laws on the galaxy-scale,  
%Kennicutt-Schmidt relation \cite{Schmidt1959, Kennicutt1989}, connect the star formation with molecular gas. Furthermore, 
%\cite{GaoSolomon2004, Wu2005} reveal a linear correlation between the surface density of 
%dense gas with the star formation rate. This emphasizes that understanding the gather of dense gas in 
%molecular clouds is key to understand the global conditions of star formation. While the mechanism to 
%hold the molecualr gas together in molecular clouds is still an open question. The mechanism of molecular clouds 
%formation, incuding the Top-down dominated by the gravitation instability, Bottom-up through alggomaration, 
%even for the molecular clouds are live-long by the gravationally bound or transient structures compressed 
%by the turbulent flows is still unclear. Understanding the properties of molecular gas is crucial for 
%answering these questions.

Molecular clouds (MCs) are the fundamental forms of the molecular interstellar medium (ISM),  
which represent its coldest ($\sim$ 10 K), densest components (n $>$ 30 cm$^{-3}$) of the ISM. 
However, how molecular clouds form and what mechanisms determine their 
physical properties are still under debate. 
Previous studies have proposed several mechanisms, 
including large-scale gravitational instabilities \citep{Lin1964, Goldreich1965}, 
agglomeration of smaller clouds \citep{Oort1954, Field1965, Dobbs2014}, turbulent flows 
\citep{Vazquez1995, Passot1995, Ballesteros1999, Semadeni2006, Heitsch2006, Koyama2002, Beuther2020}. 
Understanding the mechanism by which molecular 
clouds form and evolve are crucial for comprehending star formation and galaxy evolution.

Molecular clouds usually present complex and hierarchical structures. 
Since its discovery by \cite{Wilson1970}, CO line emission has been widely used as a tracer of molecular gas.
The boundaries of MCs are usually defined by either the low-$J$ rotational CO emission 
or extinction above some threshold \citep{Heyer2015}. The unbiased Galactic plane CO survey, 
the Milky Way Imaging Scroll Painting (MWISP), is performed using the 13.7m millimeter-wavelength telescope of Purple Mountain Observatory (PMO) and 
observes $^{12}$CO, $^{13}$CO, and C$^{18}$O $(J = 1-0)$ spectra, simultaneously \citep{Su2019}. 
The first phase of the MWISP CO project covering the Galactic longitude from $l=9^{\circ}.75$ to $230^{\circ}.25$ and 
the Galactic latitude from $b =$ -5$^{\circ}$.25 to 5$^{\circ}$.25, has been completed. 
The second phase of MWISP has begun and intend to extend the Galactic latitude from 
$b =$ -10$^{\circ}$.25 to 10$^{\circ}$.25. This high-quality CO survey provides us 
with opportunities to promote the analysis of the molecular clouds properties to a large sample spanning 
wide spatial scales, different evolutionary stages and various environments.

After observations with sufficient sensitivity and high spatial resolution carried out using the Herschel telescope, 
filaments became known to play an important role in the star formation of MCs \citep{Andre2010, Molinari2010, Andre2014, Andre2016, Yuan2019, Yuan2020, peretto2022}. 
Our researchs in \cite{Yuan2021} (Paper I) use the 18,190 MCs identified by the 
$^{12}$CO lines data from MWISP survey and classfied them as filaments and nonfilaments. 
We found that the filaments make up about 10$\%$ of the total number of molecular clouds, 
while contributing about 90$\%$ of the total integrated flux of $^{12}$CO line emission. 
Despite the systematic difference between the filaments and nonfilaments in their spatial areas, 
their averaged H$_{2}$ column densities do not vary significantly. 
\cite{Neralwar2022a} have classified the SEDIGISM clouds into four morphologies and 
found that most of molecular clouds present elongated structures. 
In addition, the ringlike clouds show the peculiar properties, which are speculated to be related to 
the physical mechanisms that regulate their formation and evolution \citep{Neralwar2022b}.
Following our paper I, several questions can be asked, for instance, is there any possible evolution sequence between filaments and nonfilaments? 
What are the physics behind the molecular clouds presenting filaments or nonfilaments? 
Quantifying the amount, distribution, and kinematics of the diffuse and dense gas among them 
may provide new clues to answering these questions.

Compared with the $^{12}$CO(1-0) line emission having a critical density of $\sim$ 10$^{2}$ cm$^{-3}$, 
the less abundant isotope $^{13}$CO(1-0) lines can trace the denser gas with a density of $\sim$ 10$^{3}$ cm$^{-3}$.
The large-scale, unbiased, and highly sensitive data on CO and its isotopic lines from the MWISP survey 
provides us with opportunities to systematically investigate the spatial distribution and 
properties of the diffuse and dense molecular gas in a large sample of Molecular clouds.

In this paper, we use the $^{13}$CO(1-0) line emission to trace relatively dense gas components 
within 18,190 $^{12}$CO molecular clouds and reveal the relationship between the 
$^{13}$CO gas fractions and morphologies in molecular clouds. In section 2, we describe the data set, including the 
$^{13}$CO line emission data and $^{12}$CO molecular cloud catalog; Section 3 introduces three different methods 
used to extract $^{13}$CO molecular gas structures in $^{12}$CO molecular clouds and compares their results. 
In Section 4, we present the distribution of the physical parameters of the extracted $^{13}$CO gas structures within the $^{12}$CO MCs and 
systematically investigate the correlation between $^{12}$CO(1-0) and $^{13}$CO(1-0) line emission in the $^{12}$CO clouds having 
$^{13}$CO structures, in addition, we also link the the $^{13}$CO gas structures and the morphologies of $^{12}$CO molecular clouds to reveal the possible relation between them. 
Section 5 discusses how our observational results provide the clues for us to understand the molecular clouds' formation and evolution. 
We conclude with a summary in Section 6.   

\section{Data} \label{sec:data}
\begin{figure*}[htp]
    \plotone{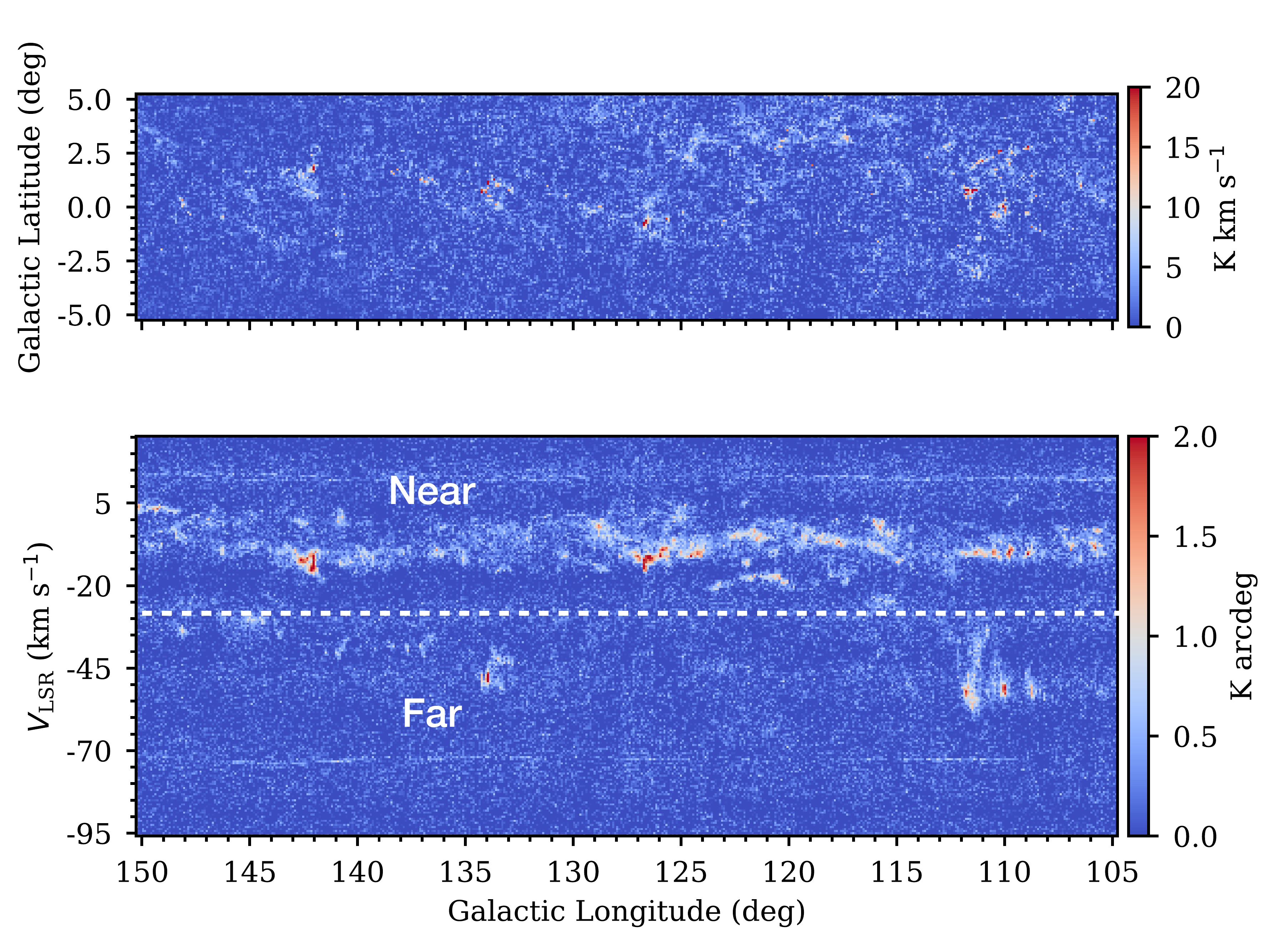}
    \caption{\textbf{Top Panel}: the velocity-integrated intensity map of $^{13}$CO(1-0) emission in the second 
    Galactic quadrant with 104.75$^{\circ}$ $< l <$ 150.25$^{\circ}$ and $|b| <$ 5.25$^{\circ}$. 
    This map is derived by integrating the $^{13}$CO emission over the velocity range between -95 km s$^{-1}$ and 25 km s$^{-1}$. 
    The distribution of noise RMS for the spectrum in each pixel is presented in Figure \ref{fig:fmaprms}, the mean 
    noise RMS is about 0.25 K. 
    The sensitivity for this velocity-integrated map can be calculated as $\sigma \times \sqrt{N} \times dv$ = 1.2 K km/s, 
    where $\sigma$ is the mean noise RMS (0.25 K), $dv$ represents the velocity resolution, its value is 0.17 km/s, 
    $N = $ 760 is the number of the velocity channels. 
    \textbf{Bottom Panel}: the latitude-integrated intensity map of $^{13}$CO(1-0) emission. 
    This map is derived by integrating the $^{13}$CO emission over the latitude range from $-$5.25$^{\circ}$ to 5.25$^{\circ}$.  
    The white dashed line at $V_{\rm LSR} =$ -30 km s$^{-1}$ divides the molecular clouds into two groups, i.e., the Near and Far groups, 
    as described in Section 4.1. The velocity-integrated intensity maps and the latitude-integrated maps 
    for the $^{13}$CO emission in the near and far velocity ranges, are shown in Figure \ref{fig:fmap_near} and \ref{fig:fmap_far}, respectively. \label{fig:f13co}}
\end{figure*}

\subsection{$^{13}$CO $J =$ 1 -- 0 data from MWISP survey}
The $^{13}$CO data is from the MWISP survey, which is an ongoing northern Galactic plane CO survey. 
This survey is conducted by the 13.7m telescope at Delingha, China. 
The detailed introductions for the telescope, the multibeam receiver system, 
observation mode, and data reduction procedures are described in \cite{Su2019}. 
The half-power beamwidth (HPBW) for the telescope at 115 GHz is about 50$^{\prime \prime}$.
The velocity separation of $^{13}$CO lines is about 0.17 km s$^{-1}$. 
The main beam efficiency ($\eta_{\rm MB}$) varied between 40$\%$ and 50$\%$.  
%Under the assumption of the beam filling factor ($f_{\rm b}$) is 1.0, 
%the antenna temperature for the CO data ($T_{\rm A}^{\star}$), has been converted to 
%the main beam brightness temperature ($T_{\rm MB}$), 
%following $T_{\rm MB} =$ $T_{\rm A}^{\star}$/($f_{\rm b} \times \eta_{\rm MB}$).

In this work, we focus on the $^{13}$CO emission in the Second Galactic Quadrant with 
$104^{\circ}.75 < l < 150^{\circ}.25$, $|b| < 5^{\circ}.25$, and $-$95 km s$^{-1}$  $<$ V$_{\rm LSR}$ $<$ 25 km s$^{-1}$. 
Figure \ref{fig:f13co} presents the large-scale $^{13}$CO gas distribution 
as a velocity-integrated intensity map and a latitude-integrated intensity map. 

\subsection{Catalog and morphology classification}
We define a molecular cloud as a contiguous structure in the position-position-velocity (PPV) data cube with 
$^{12}$CO(1-0) line intensities above a certain threshold. As described in \citet{Yan2021}, 
a total of available 18,190 molecular clouds have been identified from the $^{12}$CO data cube in the range of $104^{\circ}.75 < l < 150^{\circ}.25$, $|b| < 5^{\circ}.25$, 
and $-$95 km s$^{-1}$  $<$ V$_{\rm LSR}$ $<$ 25 km s$^{-1}$, using the Density-based Spatial Clustering 
of Applications with Noise (DBSCAN) algorithm \citep{Ester1996, Yan2020}. 

In the paper I, we have compeleted the morphological classification for these $^{12}$CO molecular clouds, 
which are mainly classified into filaments and nonfilaments \citep{Yuan2021}. 
In this work, we aim to analyze the properties of high column density gas traced by the $^{13}$CO line emission 
in these molecular cloud samples. 

\section{Extracting $^{13}$CO $J =$ 1 -- 0 Emission Structures within $^{12}$CO Molelcular Clouds}
In this work, the $^{13}$CO emission structures are defined as molecular structures within the $^{12}$CO molecular 
clouds whose spectral voxels have the $^{13}$CO line intensities above a certain threshold. 
We utilize three different methods, i.e. clipping, DBSCAN \citep{Ester1996,Yan2020}, 
and moment mask \citep{Dame2011} to extract the $^{13}$CO gas structures. 
%Under the consideration of the spatial resolution of the 13.7-m telescope, 
%we further define that their angular sizes are larger than 1 arcmin.
%A common technique is the clipping, which directly extracts the signals above statistical significance level, typically 3 times the RMS noise. 
%Whereas, the clipping can not avoid the positive noise spikes, whose values are above 
%the clipping level, unless enough high clipping levels are defined. 
%Whereas, the high threshold used can lose the faint emission. 
%Essentially, the extracting methods of moment masking \citep{Dame2011} and DBSCAN \citep{Ester1996,Yan2020}, 
%which are based on both the intensity levels and the coherence of signals,
%are performed to determine the extended $^{13}$CO emission in MCs.  
%The moment masking is implemented on a smoothed data cube, 
%which can reduce the effects of the background noises on the extraction of signals \citep{Dame2011}. 
%The DBSCAN algorithm has been used for the identification of $^{12}$CO molecular clouds \citep{Ester1996,Yan2020}. 
%We will compare the performances of three methods and select the optimal one for the $^{13}$CO emission detection.  
\begin{figure*}[htp]
    \plotone{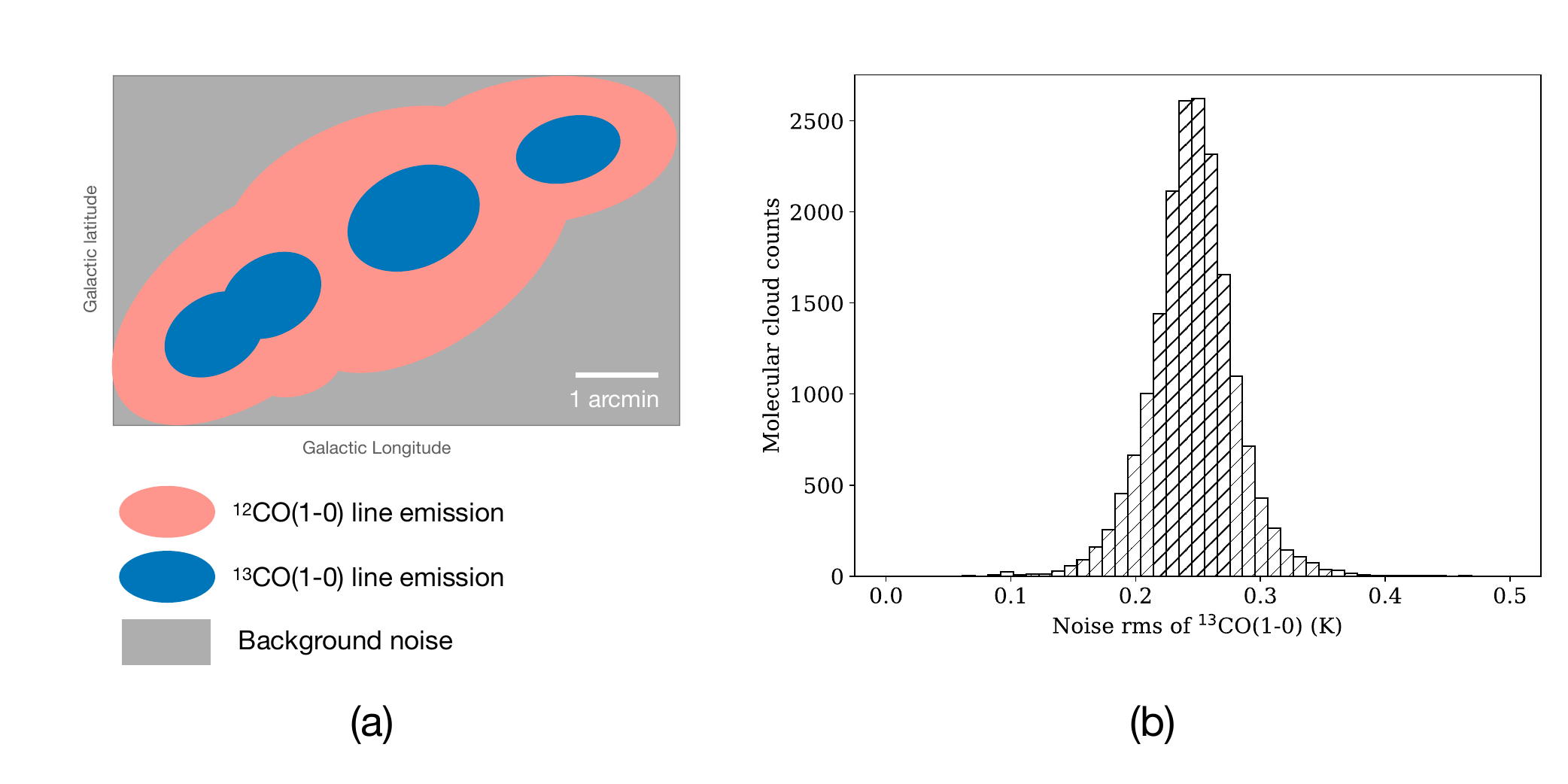}
    \caption{Panel (a): A cartoon demonstrating the distribution of $^{12}$CO molecular cloud, $^{13}$CO line emission, 
    and the background region for estimating the RMS noise. Panel (b): The distribution of the RMS noise of the $^{13}$CO line for 
    each molecular cloud. \label{fig:frms}}
\end{figure*}

\subsection{Background noise}
Before extracting the $^{13}$CO line emission within the $^{12}$CO molecular clouds, 
it is necessary to determine their RMS noises ($\sigma$). 
The $^{13}$CO spectral line data are chopped into the separate data cubes with sizes equivalent to the extent of 
$^{12}$CO emission in the PPV space for $^{12}$CO molecular clouds. 
Figure \ref{fig:frms} presents a sketch map illustrating the distribution of $^{12}$CO 
molecular cloud, the $^{13}$CO emission structures, and the background noise in a separate data cube. 
The voxels within the background region in each $^{13}$CO data cube are utilized to estimate the 
RMS noise of $^{13}$CO lines for each molecular cloud. The distribution of the resultant $^{13}$CO rms noises for 
all of 18,190 molecular clouds are presented in Figure \ref{fig:frms}. 
The typical value is about 0.25 K.  
The corresponding RMS noise for each molecular cloud is used for the $^{13}$CO emission extraction. 
Furthermore, the values of voxels in the background region of each $^{13}$CO data cube are set to zeros, 
so that the $^{13}$CO structures are extracted within the $^{12}$CO emission boundaries. 
We extract the $^{13}$CO emission structures in these chopped 18,190 $^{13}$CO data cubes, which are correspond to the 18,190 $^{12}$CO 
molecular clouds. 

\subsection{Three methods}
\subsubsection{Clipping} 
Clipping is a common technique, which directly extracts the structures containing the $^{13}$CO spectral channels 
above the statistical significance level. 
Spectral channels in each $^{13}$CO data cube having intensities above the defined clipping levels are 
extracted as the significant $^{13}$CO emission.  
Whereas the clipping can not avoid the positive noise spikes with values above the clipping level, 
unless the clipping level is enough high. 
The high threshold used may lead to loss of the faint emission with intensities below the cutoff levels.
 
%We take a molecular cloud G139.725+1.702-38.33 (hereafter G139.73) as a sample and use the clipping 
%technique to extract its $^{13}$CO line emission. 
%Figure \ref{fig:fflux_rms} presents the extracted $^{13}$CO emission fluxes in the G139.73 by the clipping technique 
%at the levels from 2 to 10 times the RMS noise. 
%Figure \ref{fig:fclip} shows their resultant velocity-integrated intensity and latitude-integrated maps 
%under the clipping levels of 3 and 4 times the RMS noise.

\subsubsection{Moment Mask}
The main point of a moment mask is that the $^{13}$CO emission is extracted on the smoothed $^{13}$CO spectral line data, 
to reduce the effects of noise spikes. The extracted structures are further extended to the adjacent voxels, 
whose ranges are determined by the smoothed spatial and velocity resolutions. 
The MWISP $^{13}$CO data has a spatial resolution of $\sim$ 50 arcsec and a velocity 
resolution of 0.17 km s$^{-1}$, we smooth the data with two times the beam size (FWHM$_{\rm S}$ $\sim$ 100$^{\arcsec}$) in position space and 
with four times the velocity channels in velocity space (FWHM$_{\rm V}$ $\sim$ 0.7 km s$^{-1}$).

In the smoothed $^{13}$CO data, we calculate the noise RMS ($\sigma_{\rm sm}$) and extract the $^{13}$CO emissions 
with intensities higher than the defined $\sigma_{\rm sm}$ levels. 
After that, the extracted $^{13}$CO structures are extended to the structures containing the voxels, 
which are adjacent to the extracted voxels. The adjacent voxels are defined as the   
ns pixels in spatially and nv pixels in velocity. 
Among that, ns = 0.5$\times$FWHM$_{\rm S}$/ds, nv = 0.5$\times$FWHM$_{\rm V}$/dv, ds and dv are the spatial and velocity resolutions for the raw data, respectively \citep{Dame2011}. 
According to the resolutions of the smoothed data, we obtain ns$=$1 and nv$=$2. 
Thus, a total of 3$\times$3$\times$5 voxels are determined to be adjacent to an extracted $^{13}$CO voxel.  
Based on the boundaries of enlarged structures, we obtain the determined $^{13}$CO structures from the raw $^{13}$CO data cube. 
%Figure \ref{fig:fflux_rms} shows the $^{13}$CO emission fluxes of G139.73 obtained using the moment mask, 
%under the cutoff levels from 2 to 10 times the smoothed rms noises.
%In Figure \ref{fig:fmom8} and \ref{fig:fmom9}, we present the extracted $^{13}$CO structures in G139.73 at the cutoff levels of 
%8 and 9 times the smoothed noise rms, respectively.   

\subsubsection{DBSCAN}
DBSCAN algorithm, which was designed to discover clusters in arbitrary shape \citep{Ester1996}, has been developed to 
extract a set of contiguous voxels in the PPV space with $^{12}$CO emission above a certain threshold as a molecular cloud \citep{Yan2020}. 
This method is based on both the intensity levels and the connectivity of signals.
%The consecutive structures are confined by two parameters in the DBSCAN, i.e. $\epsilon$ and MinPts. 
%Each point within a consecutive structure is called a core point, 
%the number of points contained in its neighborhood within $\epsilon$ has to exceed a threshold. 
%The parameters of MinPts determine the threshold and the $\epsilon$ represents the radius of the neighborhood. 
%A border point in the consecutive structure is defined as a point inside the $\epsilon$-neighborhood of a core point, 
%but not necessarily contain the MinPts neighbors, as shown in Figure 2 of \citep{ester1996}. 
%In the PPV space of CO data, \cite{Yan2020} has examined all the choices of parameters 
%and $^{12}$CO line intensities cutoffs to identify molecular clouds. 
%The parameters of cutoff = 2$\sigma$, minPts = 4, $\epsilon$ = 1 are used in the DBSCAN algorithm 
%to detect the $^{12}$CO molecular clouds. 
%The post-selection criteria also are utilized to remove the noise contamination \citep{Yan2020}. 
We utilize the identical parameters to extract the $^{13}$CO emission as that used for the $^{12}$CO molecular clouds \citep{Yan2020}, 
except for the post-selection criteria of the peak values. 
For the $^{12}$CO emission, its peak intensity in a $^{12}$CO molecular cloud needs to be larger than 
the intensity of its boundary threshold adding 3$\sigma$. 
Owing to the relatively lower value for the $^{13}$CO line intensity, 
its peak intensity in a $^{13}$CO structure is larger than 
the intensity of its boundary threshold adding the 2$\sigma$, where $\sigma$ is the 
background noise. The parameters used for the DBSCAN extraction are described in detail in Appendix B. 
The chopped $^{13}$CO data cube without any smoothing procedures is used for identifying the $^{13}$CO structures by the DBSCAN algorithm.
%Figure \ref{fig:fflux_rms} shows the $^{13}$CO emission fluxes of G139.73 obtained using the DBSCAN, 
%under the cutoff levels from 2 to 10 times the raw rms noises.
%Figure \ref{fig:fdbscan2} and \ref{fig:fdbscan3} present the extracted $^{13}$CO structures in G139.73 at the cutoff levels of 
%2 and 3 times the raw noise rms, respectively.

\subsection{Comparison among different methods}
\subsubsection{Test on a Case: G139.73}
\begin{figure}[ht]
    \plotone{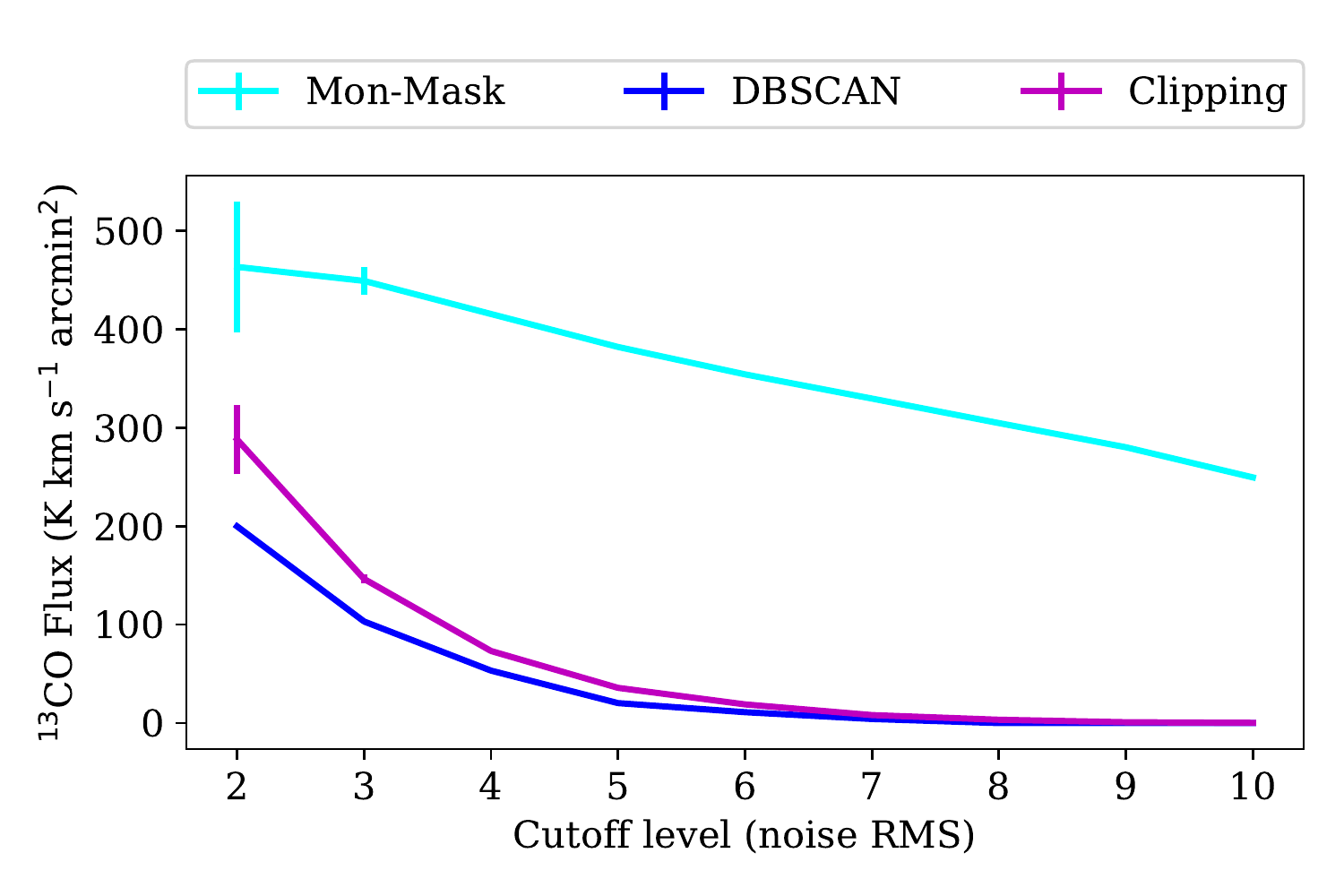}
    \caption{The distribution of $^{13}$CO emission fluxes within the $^{12}$CO cloud G139.73 extracted by three techniques at the cutoff levels from 2$\sigma$ to 10$\sigma$. 
    The $\sigma$ for the clipping and DBSCAN is 0.27 K, which is calculated using the raw $^{13}$CO line data. 
    Whereas the $\sigma$ for the moment mask is 0.05 K, estimated from the smoothed $^{13}$CO line data. \label{fig:fflux_rms}}
\end{figure}

We take a $^{12}$CO molecular cloud G139.725-0.507-038.33 (hereafter G139.73) as an sample to 
compare the performances of the three different methods. In Figure \ref{fig:fexample_raw12co} and \ref{fig:fexample_raw13co}, 
we present the velocity-integrated intensity maps and the latitude-integrated maps of $^{12}$CO and $^{13}$CO emission for the G139.73, 
which are integrated by the chopped data cubes without any clipping. 
Figure \ref{fig:fflux_rms} shows the $^{13}$CO emission fluxes of the G139.73 extracted by three techniques at 
the cutoff levels from 2$\sigma$ to 10$\sigma$. 
We should note the noise $\sigma$ used by the moment mask is estimated using the smoothed data ($\sigma_{\rm sm}$ = 0.05 K), 
but that for the DBSCAN and clipping, their background noise $\sigma$ is calculated using the raw data without any smoothing procedures ($\sigma$ = 0.27 K). 
The distribution of integrated fluxes extracted by the DBSCAN and clipping algorithms have a similar trend and 
the values steep up from 4$\sigma$ to 2$\sigma$. 
For the moment mask with the noise $\sigma_{\rm sm}$ of 0.05 K, 
its extracted fluxes are higher than that from two other methods at the same cutoff levels. 
To quantify the contribution of the background noises to the $^{13}$CO emission fluxes using the three methods, 
we use a $^{13}$CO data cube without the significant $^{13}$CO emission to represent the pure background noises. 
Further, three methods are performed for the extraction of the $^{13}$CO line emission in this noise cube 
at the cutoff levels from 2$\sigma$ to 10$\sigma$. 
This noise data have the same sizes and spatial positions as that of the G139.73 data cube but in the different 
radial velocity range of (-90.3 -78.4) km s$^{-1}$. 
The extracted noise fluxes are shown as the error bars in Figure \ref{fig:fflux_rms}. 
We find that the background noises contribute about 15$\%$ of the integrated fluxes by 
the moment mask and clipping at the cutoff level of 2$\sigma$. 
Whereas, the DBSCAN almost completely avoids the background noise. 
We should note that these effects of the background noises on the integrated fluxes 
are based on a case of G139.73, whose $^{13}$CO line emissions have relatively high intensities. 
For the molecular clouds with faint emission and small spatial scales, 
the effects of noises can be magnified. 
%That may be caused by the overmuch extention or low cutoff levels. 

In Figure \ref{fig:fedgeline_clip}, we present the averaged $^{12}$CO and $^{13}$CO spectral lines 
for pixels along the boundaries, which are determined by the clipping at the cutoff level from 2$\sigma$ to 5$\sigma$,
as well as the corresponding mean $^{13}$CO spectral lines of the extracted $^{13}$CO structures. 
From the cutoff level of 3$\sigma$, the averaged $^{13}$CO spectrum along the determined boundary begins 
to have a significant signal. 
In Figure \ref{fig:fclip3} and \ref{fig:fclip4}, we show the velocity-integrated map and latitude-integrated map of 
the extracted $^{13}$CO structures at cutoff level of 3$\sigma$ and 4$\sigma$. 
We find that there are a lot of positive spikes extracted by the clipping at the cutoff level of 3$\sigma$. 
The same spectral lines for the $^{13}$CO structures but extracted using the DBSCAN are presented in Figure \ref{fig:fedgeline_dbscan}. 
For the DBSCAN, the mean $^{13}$CO spectrum along its boundary determined at 2$\sigma$ begins to have a 
significant ratio of signal to noise (S/N). We also show the integrated maps of the extracted $^{13}$CO structures 
by the DBSCAN at the threshold of 2$\sigma$ and 3$\sigma$ in Figure \ref{fig:fdbscan2} and \ref{fig:fdbscan3}. 
Figure \ref{fig:fedgeline_sm} presents the same $^{13}$CO spectral lines for $^{13}$CO structures but identified using the moment mask 
at the cutoff level from 8$\sigma_{\rm sm}$ to 11$\sigma_{\rm sm}$. 
The averaged-boundary $^{13}$CO spectrum begins to show the effective signal at 9$\sigma_{\rm sm}$.
The maps for the extracted $^{13}$CO structures by the moment mask at 8$\sigma_{\rm sm}$ and 9$\sigma_{\rm sm}$ are 
shown in Figure \ref{fig:fmom8} and \ref{fig:fmom9}. We find that there are tiny $^{13}$CO structures 
extracted by the DBSCAN, not presented in the structures from the moment mask.
We should note that the S/N for the averaged-boundary spectrum is related to the number of the spectrum along the boundaries. 
For the molecular clouds with smaller spatial scales, the averaged-boundary spectrum for $^{13}$CO structures 
extracted using the same method at the same threshold may not have a significant S/N. The cutoff levels of 
4$\sigma$ for the clipping, 2$\sigma$ for the DBSCAN, and 9$\sigma_{\rm sm}$ for the moment mask are adopted to 
extract the $^{13}$CO structures.

\begin{deluxetable}{lcccc}
    \tablenum{1}
    \tablecaption{The number detection rates, area ratios, and the total flux ratios between the $^{13}$CO and $^{12}$CO line emission from three methods. \label{tab:t1}}
    \tablewidth{0pt}
    \tablehead{\colhead{Methods} & \colhead{Number detection rates} & \colhead{Area ratios} & \colhead{Flux ratios}}
    %\decimalcolnumbers
    \startdata
    Clipping & 13.5$\%$ & 10.8$\%$ & 4.1$\%$ \\
    DBSCAN & 15.7$\%$ & 20.3$\%$ & 6.3$\%$ \\
    Monment Mask & 15$\%$ & 20.7$\%$ & 7.3$\%$ \\
    \enddata
    \tablecomments{The number detection rate is the number of $^{12}$CO clouds having $^{13}$CO structures divided by the total 
    number of 18,190. The area ratio is the ratio between the total angular areas of the extracted $^{13}$CO structures and 
    the total $^{12}$CO angular areas of 18,190 $^{12}$CO clouds. The flux ratio is the total integrated fluxes of the extracted $^{13}$CO line emission  divided by that 
    of $^{12}$CO line emission.}
\end{deluxetable}
\subsubsection{Number detection rate}
We determine to extract the $^{13}$CO structures within a large sample of $^{12}$CO molecular clouds, 
which have the angular areas spanning from 1 to 10$^{4}$ arcmin$^{2}$ and the integrated fluxes ranging from $\sim$ 1 
to 10$^{5}$ K km s$^{-1}$ arcmin$^{2}$, using the clipping at the cutoff of 4$\sigma$, the DBSCAN at the cutoff of 2$\sigma$, 
and the moment mask by a threshold of 9$\sigma_{\rm sm}$, to compare the extracted results from the 
three different methods.

For the clipping, there are 4,390 molecular clouds detected $^{13}$CO line emission. 
The extracted structure with values above the threshold, whose spatial size is 
one pixel (0.25 arcmin) or its velocity span is just one channel (0.17 km s$^{-1}$),  
is determined as the noise spike. After removing these noise spikes, 2,462 molecular clouds 
are regarded as having the significant $^{13}$CO line emission. 
However, the DBSCAN and Moment mask algorithms do not extract individual noise spike. 
For the DBSCAN algorithm, 2,851 molecular clouds are identified to have the $^{13}$CO emission. 
The moment mask extracts the $^{13}$CO line emission in the 2,735 molecular clouds. 
The number detection rates in the total 18,190 MCs by three methods are listed in Table \ref{tab:t1}. 

\subsubsection{Area ratios between the $^{13}$CO and $^{12}$CO line emission}
The total angular area for the 18,190 $^{12}$CO molecular clouds is about 228.2 deg$^{2}$. 
The total angular area for the extracted $^{13}$CO structures within the 18,190 $^{12}$CO 
clouds by the clipping is 24.7 deg$^{2}$. 
The value is 46.2 deg$^{2}$ extracted by the DBSCAN and 47.2 deg$^{2}$ from the moment mask.
The ratios between the total $^{13}$CO angular areas of the extracted $^{13}$CO structures and 
the total $^{12}$CO angular areas of the 18,190 $^{12}$CO clouds are listed in Table \ref{tab:t1}.

\subsubsection{Integrated flux ratios between the $^{13}$CO and $^{12}$CO line emission}
The total $^{12}$CO(1-0) emission fluxes for the 18,190 $^{12}$CO molecular clouds is 3.7$\times$10$^{6}$ K km s$^{-1}$ arcmin$^{2}$. 
The total extracted $^{13}$CO(1-0) emission flux in this catalog is 1.5$\times$10$^{5}$ K km s$^{-1}$ arcmin$^{2}$ 
by the clipping. The value is 2.3$\times$10$^{5}$ K km s$^{-1}$ arcmin$^{2}$ by the DBSCAN algorithm 
and 2.7$\times$10$^{5}$ K km s$^{-1}$ arcmin$^{2}$ by the moment mask method. 
These total integrated flux ratios between the $^{13}$CO and $^{12}$CO line emission 
by three methods are listed in Table \ref{tab:t1}.

We find that the number detection rates of the DBSCAN and moment mask are consistent with $\sim$ 15$\%$, 
the value of the clipping is about 2$\%$ lower than that of the other two techniques. 
In addition, the clipping extracts plenty of positive noise spikes. 
For the total angular areas of the extracted $^{13}$CO structures, the values from the clipping is about 50$\%$ of that from the DBSCAN and moment mask. 
For the extracted integrated fluxes of $^{13}$CO line emission, 
the values from the clipping are about 60$\%$ of that from the DBSCAN or Moment mask. 
That indicates the clipping method extracts amounts of noise spikes with 
intensities larger than the threshold of 4$\sigma$, meanwhile, it also loses substantial faint $^{13}$CO emission. 
For either the total angular areas or the total integrated fluxes of the extracted $^{13}$CO structures, 
the values from the moment mask are a bit higher than that from the DBSCAN, while the number detection rate of the moment mask is a bit lower 
than that from the DBSCAN. Overall, these values from the DBSCAN and moment mask are close.   

\begin{figure*}[ht]
    \plotone{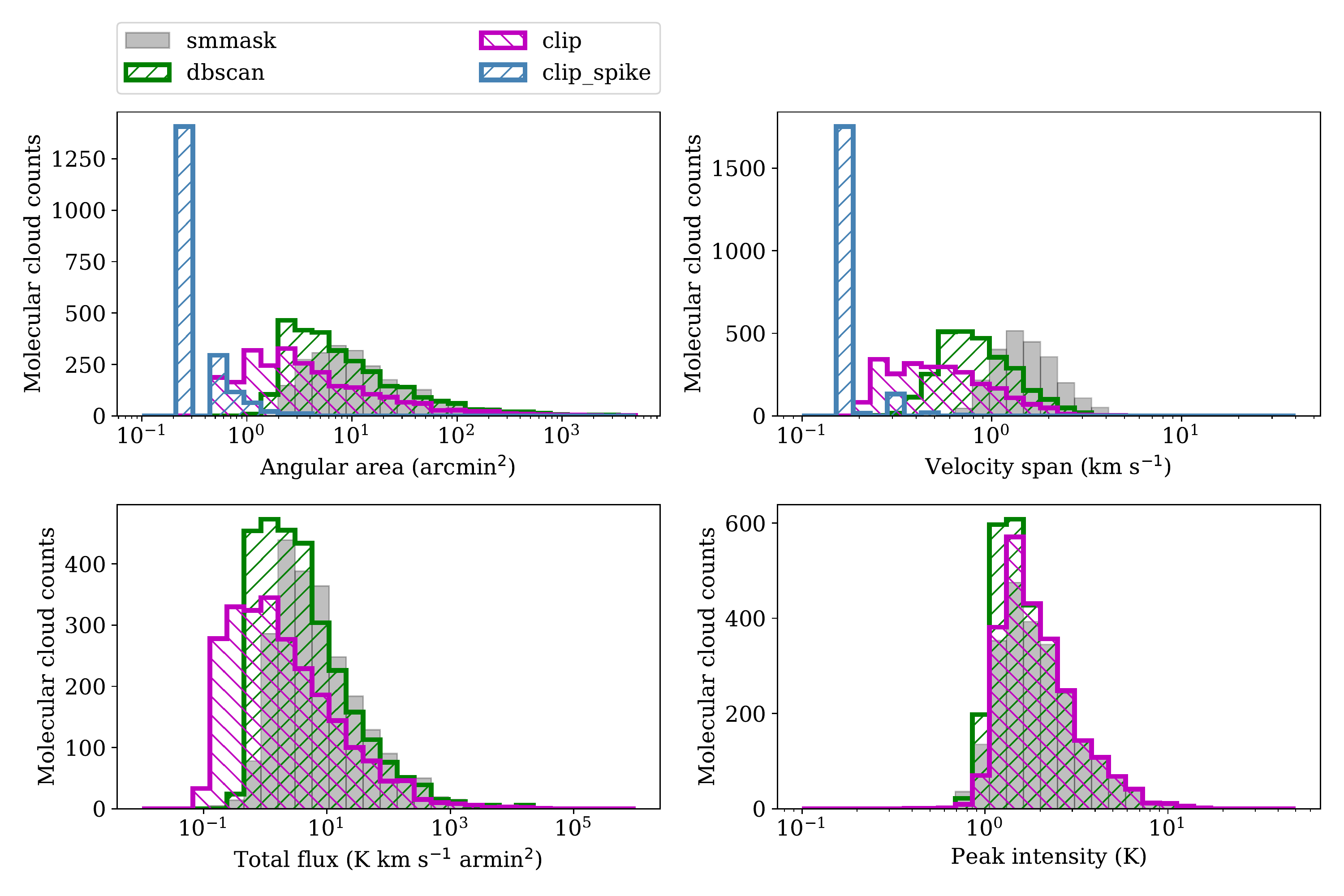}
    \caption{The number distributions of angular sizes, mean velocity spans, the integrated fluxes of $^{13}$CO line emission, and the peak 
    intensities of $^{13}$CO(1-0) line emission for the $^{13}$CO structures extracted by three techniques. For the clipping, 
    the extracted $^{13}$CO emission structures with a pixel in spatial scale or a velocity channel in velocity, 
    are determined as the noise spikes and shown as blue histograms. \label{fig:f3method}}
\end{figure*}

\subsubsection{Differences of extracted basic parameters}
The $^{13}$CO emissions are extracted within the $^{12}$CO clouds by three methods. 
All the $^{13}$CO emissions within the boundary of a $^{12}$CO cloud belong to the same cloud. 
We take all the $^{13}$CO line emissions within a $^{12}$CO cloud as a whole, referred to as $^{13}$CO molecular gas structures. 
The equivalent angular area of $^{13}$CO molecular structures ($\mathcal{A}_{\rm ^{13}CO}$) for a $^{12}$CO cloud 
is the sum of the pixel areas of the extracted $^{13}$CO emission regions projected in the l-b panel. 
The mean velocity span of $^{13}$CO molecular structures ($V_{\rm aver, ^{13}CO}$) within a $^{12}$CO cloud is 
the averaged extracted velocity span of each pixel in the $^{13}$CO emission regions, 
weighting by its corresponding velocity-integrated intensity, 
i.e. $V_{\rm aver, ^{13}CO}=\Sigma(V_{\rm span}(l,b) \times W_{\rm ^{13}CO}(l,b))/\Sigma W_{\rm ^{13}CO}(l,b)$.
The total $^{13}$CO integrated flux ($F_{^{13}CO}$) is the integrated flux of all the $^{13}$CO structures in an individual molecular cloud, 
$F_{\rm ^{13}CO}$ = $\int$T$_{\rm mb}(l,b,v)dldbdv$ = 0.166$\times$0.25$\Sigma$T$_{\rm mb}(l,b,v)$ K km s$^{-1}$ arcmin$^{2}$, 
where $T_{\rm mb}(l, b, v)$ is the $^{13}$CO line intensity at the coordinate of $(l, b, v)$ in PPV space, 
$dv =$ 0.166 km s$^{-1}$ is the velocity resolution. 
The peak value ($T_{\rm{peak}, ^{13}CO}$) is the maximal value of the $^{13}$CO line intensities for the extracted $^{13}$CO structures. 

The number distributions of angular sizes ($\mathcal{A}_{\rm ^{13}CO}$), mean velocity spans ($V_{\rm{aver}, ^{13}CO}$), 
integrated fluxes ($F_{^{13}CO}$), and peak intensities ($T_{\rm{peak}, ^{13}CO}$) of $^{13}$CO structures,  
which are identified by three techniques, are presented in Figure \ref{fig:f3method}.
We find the angular sizes of $^{13}$CO structures extracted by the clipping are systematically smaller than 
that from the DBSCAN and moment mask. 
We compare the angular sizes of $^{13}$CO structures from the DBSCAN and moment mask, 
their distributions in the range of the angular sizes larger than 6 arcmin$^{2}$ are similar. 
There are more structures from the DBSCAN in the range of 1 -- 6 arcmin$^{2}$. 
To check the reliability of these extra structures, which are identified by the DBSCAN, 
but not by the moment mask, we present their $^{13}$CO line intensity maps 
integrated along with three different directions ($l, b, v$) and 
their averaged $^{13}$CO line spectra. As shown in Figure \ref{fig:fdbscan_diff1}, we find 
that these structures usually are located around the regions contoured at the levels of half of 
the peak velocity-integrated intensity of $^{12}$CO emission.  
In addition, their $^{13}$CO spectral profiles usually present the Gaussian-like profiles. 
Thus, we determine these $^{13}$CO structures are valid.  
The number distributions of $T_{\rm{peak}, ^{13}CO}$ of $^{13}$CO line emission from the three techniques 
are similar in the range of $T_{\rm{peak}, ^{13}CO} >$ 2.0 K.  
After the smooth procedure, a portion of tiny structures with $T_{\rm{peak}, ^{13}CO} <$ 2.0 K may be missed by the moment mask. 
For the distribution of the mean velocity span of $^{13}$CO structures, 
we find the values from the moment mask tend to be larger than that from the DBSCAN, and the values from the DBSCAN 
are prone to be larger than that from the clipping. 
That may be due to that the extracted voxels in the smoothed data by the moment mask are further extended to the adjacent voxels, 
while the intensities of voxels extracted by the DBSCAN are larger than 2$\sigma$, and the $^{13}$CO structures 
from the clipping only contain the velocity channels with intensities larger than 4$\sigma$. 
Thus the velocity span of $^{13}$CO emission in each spatial pixel is derived from the velocity channels 
with intensities larger than 2$\sigma$ for the DBSCAN and 4$\sigma$ for the clipping. 
The number distributions of $F_{\rm ^{13}CO}$ have a similar trend with that of $\mathcal{A}_{\rm ^{13}CO}$. 
That means the differences of the $F_{\rm ^{13}CO}$ distributions from three 
techniques are mainly attributed to their $\mathcal{A}_{\rm ^{13}CO}$ distributions.

\subsection{Summary of methods}
Above all, the clipping extracts plenty of noise spikes with values larger than 4$\sigma$ and 
meanwhile loses the faint significant emission having intensities less than 4$\sigma$. 
That leads to the extracted parameters being systematically smaller than that from the other two methods. 
In addition, the moment mask leaves out a part of faint and tiny $^{13}$CO structures, 
owing to the smooth procedure. The moment mask is more suitable for the structures with relative 
large angular sizes and high emission intensities.
The DBSCAN algorithm can not only avoid the noise spikes 
but also preserve the faint and tiny $^{13}$CO structures not identified by the moment mask. 
Each voxel in the PPV space of the structures extracted by the DBSCAN is larger than 2$\sigma$. 
We take the resultant $^{13}$CO structures from the DBSCAN, which is consistent 
with the extraction algorithm used for the $^{12}$CO clouds identification, for the follow-up analysis.

\begin{longrotatetable}
    \begin{deluxetable*}{lcrrcccccccl}
    \tablenum{2}
    \tabletypesize{\scriptsize}
    \tablecaption{A catalog of $^{13}$CO line emission parameters \label{tab:t2}}
    \tablewidth{0pt}
    \tablehead{\colhead{Name} & \colhead{$l_{\rm cen}$} & \colhead{$b_{\rm cen}$} & \colhead{$V_{\rm LSR}$} & \colhead{$\mathcal{A}_{\rm ^{12}CO}$} & \colhead{$V_{\rm{span}, ^{12}CO}$} &\colhead{$T_{\rm{peak}, ^{12}CO}$} & \colhead{$F_{\rm ^{12}CO}$} & \colhead{$\mathcal{A}_{\rm ^{13}CO}$} & \colhead{$V_{\rm{span}, ^{13}CO}$} & \colhead{$T_{\rm{peak}, ^{13}CO}$} & \colhead{$F_{\rm ^{13}CO}$} \\
     & \colhead{(degree)} & \colhead{(degree)} & \colhead{(km s$^{-1}$)} &
    \colhead{(arcmin$^{2}$)} & \colhead{km s$^{-1}$} & \colhead{(K)} & \colhead{(K km s$^{-1}$ arcmin$^{2}$)} & \colhead{(arcmin$^{2}$)} & \colhead{km s$^{-1}$} & \colhead{(K)} & \colhead{(K km s$^{-1}$ arcmin$^{2}$)}} 
    \decimalcolnumbers
    \startdata
    G104.794+02.286-063.40 & 104.794 & 02.286 & -63.40 & 03.50 & 01.67 & 2.95 & 4.40 & 01.50 & 01.16 & 0.99 & 0.55 \\
    G104.803-02.869-004.14 & 104.803 & -2.869 & -4.14 & 17.75 & 02.67 & 5.30 & 40.34 & 04.00 & 01.00 & 1.26 & 1.15 \\
    G104.810+01.058-009.45 & 104.810 & 01.058 & -9.45 & 12.50 & 03.51 & 4.55 & 20.54 & 04.00 & 01.00 & 1.72 & 2.09 \\
    G104.822-02.335-034.52 & 104.822 & -2.335 & -34.52 & 11.75 & 04.01 & 5.14 & 38.85 & 02.75 & 01.33 & 1.11 & 0.70 \\
    G104.870-00.447-009.32 & 104.870 & -0.447 & -9.32 & 42.25 & 05.68 & 6.70 & 131.49 & 07.25 & 01.33 & 1.44 & 2.80 \\
    G104.871+00.646-001.89 & 104.871 & 00.646 & -1.89 & 46.00 & 02.67 & 5.59 & 86.97 & 02.50 & 01.00 & 1.01 & 0.63 \\
    G104.872+01.008-049.95 & 104.872 & 01.008 & -49.95 & 33.25 & 04.84 & 4.40 & 68.78 & 04.25 & 01.66 & 1.18 & 1.83 \\
    G104.886+00.134-010.58 & 104.886 & 00.134 & -10.58 & 28.25 & 03.51 & 8.71 & 86.43 & 04.00 & 01.16 & 1.47 & 1.50 \\
    G104.923-03.111-046.89 & 104.923 & -3.111 & -46.89 & 04.25 & 02.00 & 6.59 & 9.49 & 02.00 & 00.83 & 1.50 & 0.88 \\
    G104.935+04.396-044.62 & 104.935 & 04.396 & -44.62 & 07.25 & 02.34 & 5.90 & 16.76 & 03.50 & 01.16 & 1.40 & 1.48 \\
    G104.980+00.900-055.53 & 104.980 & 00.900 & -55.53 & 19.50 & 03.51 & 4.43 & 30.89 & 02.50 & 00.83 & 1.10 & 0.62 \\
    G104.983-02.680-042.49 & 104.983 & -2.680 & -42.49 & 60.75 & 06.68 & 5.95 & 185.27 & 03.25 & 01.83 & 1.32 & 1.69 \\
    G104.997-02.669-004.75 & 104.997 & -2.669 & -4.75 & 26.25 & 04.34 & 6.56 & 79.88 & 07.25 & 01.66 & 1.69 & 2.78 \\
    G105.023+00.538-049.82 & 105.023 & 00.538 & -49.82 & 44.25 & 04.51 & 6.06 & 159.96 & 19.25 & 01.83 & 2.61 & 14.01 \\
    %G105.040+00.737-054.12 & 105.040 & 00.737 & -54.12 & 19.00 & 03.84 & 3.85 & 40.21 & 03.25 & 00.69 & 1.44 & 1.27 \\
    %G105.090-04.021-047.84 & 105.090 & -4.021 & -47.84 & 22.75 & 03.34 & 5.26 & 55.56 & 06.75 & 00.75 & 1.49 & 2.95 \\
    %G105.097+00.846-047.20 & 105.097 & 00.846 & -47.20 & 38.00 & 05.18 & 5.97 & 99.68 & 05.75 & 00.74 & 1.27 & 2.35 \\
    %G105.112+02.246-018.93 & 105.112 & 02.246 & -18.93 & 23.75 & 04.01 & 4.09 & 49.75 & 02.00 & 00.49 & 1.00 & 0.44 \\
    %G105.139+00.323-069.24 & 105.139 & 00.323 & -69.24 & 87.25 & 07.01 & 7.36 & 347.49 & 20.00 & 02.57 & 3.79 & 36.73 
    \enddata

    \tablecomments{The central Galactic coordinates ($l_{\rm cen}$, $b_{\rm cen}$) for each $^{12}$CO cloud 
    are the averaged Galactic coordinates in its velocity-integrated $^{12}$CO(1-0) intensity map, 
    weighting by the value of the velocity-integrated $^{12}$CO(1-0) intensity. 
    The central velocity ($V_{\rm LSR}$) for each cloud is the averaged radial velocity in its radial 
    velocity field, weighting by the value of the velocity-integrated $^{12}$CO(1-0) intensity. 
    The $\mathcal{A}_{\rm ^{12}CO}$ and $\mathcal{A}_{\rm ^{13}CO}$ are the angular areas of $^{12}$CO(1-0) and $^{13}$CO(1-0) lines emission, respectively.  
    The $V_{\rm{span}, ^{12}CO}$ represents the velocity span of each cloud cube in the velocity axis of PPV space, 
    which is calculated using the number of velocity channels in the $^{12}$CO cloud cube multiplied by a velocity resolution of 0.167 km s$^{-1}$.
    The $V_{\rm{span}, ^{13}CO}$ is the velocity range between the maximal and minimal velocity for all the extracted $^{13}$CO gas structures within each $^{12}$CO molecular cloud. 
    The $T_{\rm{peak}, ^{12}CO}$ and $T_{\rm{peak}, ^{13}CO}$ represent the peak values of 
    $^{12}$CO(1-0) and $^{13}$CO(1-0) line intensity in each cloud, respectively. 
    The $F_{\rm ^{12}CO}$ is the integrated flux of $^{12}$CO(1-0) line emission for each cloud, 
    $F_{\rm ^{12}CO}$ = $\int$T$_{\rm mb}(l,b,v)dldbdv$ = 0.167$\times$0.25$\Sigma$T$_{\rm mb}(l,b,v)$ K km s$^{-1}$ arcmin$^{2}$, 
    where $T_{\rm mb}(l, b, v)$ is the $^{12}$CO line intensity at the coordinate of $(l, b, v)$ in PPV space, 
    $dv =$ 0.167 km s$^{-1}$ is the velocity resolution, $dldb =$ 0.5 arcmin $\times$ 0.5 arcmin $=$ 0.25 arcmin$^{2}$, 
    the angular size of a pixel is 0.5 arcmin. The $F_{\rm ^{13}CO}$ is calculated using the $^{13}$CO(1-0) line emission within each cloud 
    through the same formula, but T$_{\rm mb}(l, b, v)$ is the $^{13}$CO line intensity.
    This table is available in its entirety from the online journal. 
    A portion is shown here for guidance regarding its form and content.}
\end{deluxetable*}
\end{longrotatetable}

\section{Results}
\subsection{Comparing the physical properties of molecular clouds with and without $^{13}$CO molecular structures \label{sec:41}} 
The whole catalog of 18,190 molecular clouds is identified using the $^{12}$CO(1-0) line data by the DBSCAN algorithm \citep{Yan2021}. 
Among that, the 2,851 $^{12}$CO clouds have $^{13}$CO structures, which are also extracted by the DBSCAN algorithm.  
Since the boundary of a molecular cloud is defined by the 3D surface of $^{12}$CO(1-0) line emission in the PPV space, 
all the $^{13}$CO emissions within this surface belong to the same $^{12}$CO cloud. 
Thus its internal $^{13}$CO emission components are characterized as its substructures, which 
are referred to as $^{13}$CO molecular structures. \textbf{All the extracted $^{12}$CO(1-0) line emission data of 
18,190 molecular clouds and the extracted $^{13}$CO(1-0) line emission data within the 2,851 $^{12}$CO clouds have 
been published in ScienceDB \citep{Yuan2022}.}

We should note that a single molecular cloud may have more than one individual $^{13}$CO molecular structure. 
We take all the separate $^{13}$CO molecular structures in a single $^{12}$CO molecular cloud as a unity. 
The $^{13}$CO emission parameters are derived from all the $^{13}$CO structures in a $^{12}$CO cloud as a whole.
The equivalent angular area of $^{13}$CO molecular structures for a $^{12}$CO cloud is the sum of the pixel areas 
of the extracted $^{13}$CO emission regions projected in the l-b panel. 
The velocity span of $^{13}$CO molecular structures within a $^{12}$CO cloud is 
the range between the maximal and minimal velocity of the extracted $^{13}$CO structures along the velocity axis. 
For a $^{12}$CO cloud with multiple $^{13}$CO structures, the minimal velocity is the minimal value in the 
velocity ranges for all the extracted $^{13}$CO structures and the maximal velocity is the maximal value of that. 
Figure \ref{fig:fedgeline_dbscan} illustrates the velocity span for the extracted $^{13}$CO structures in the $^{12}$CO cloud G139.73. 
The $^{13}$CO integrated fluxes are the integrated fluxes of all the $^{13}$CO structures in an individual molecular cloud. 
The peak value is the maximal value of the $^{13}$CO line intensities within the boundary of a $^{12}$CO MC. 
The parameters of $^{13}$CO molecular structures for the 2,851 $^{12}$CO molecular clouds are listed in Table \ref{tab:t2}.
The rest 15,339 $^{12}$CO molecular clouds do not have the significant $^{13}$CO structures.  
We systematically compare the basic physical parameters of the $^{12}$CO molecular clouds having $^{13}$CO structures ($^{13}$CO-detects) 
to that of molecular clouds without $^{13}$CO structures (Non$^{13}$CO-detects). 
\begin{figure*}[hp]
    \plotone{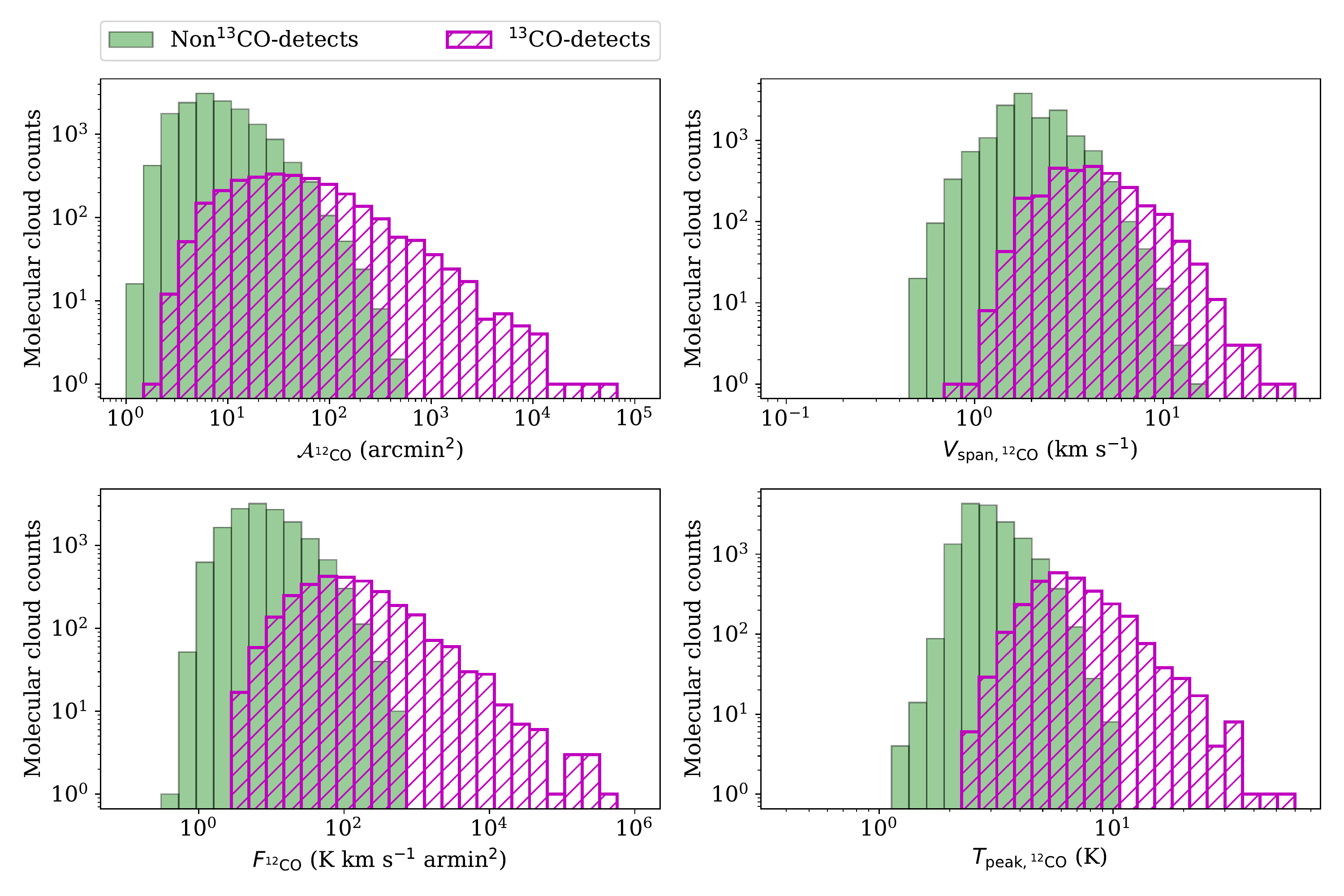}
    \caption{The number distributions of angular areas ($\mathcal{A}_{\rm ^{12}CO}$), velocity spans ($V_{\rm{span, ^{12}CO}}$), the peak intensities ($T_{\rm{peak, ^{12}CO}}$) and 
    the integrated fluxes ($F_{\rm ^{12}CO}$) of $^{12}$CO line emission for $^{12}$CO molecular clouds with and without $^{13}$CO structures. 
    The green histgrams represent $^{12}$CO molecular clouds not having the $^{13}$CO structures (Non$^{13}$CO-detects). 
    The magenta ones are the $^{12}$CO molecular clouds having the $^{13}$CO structures ($^{13}$CO-detects). \label{fig:fno13_with13}}
\end{figure*}
\begin{figure*}
    \plotone{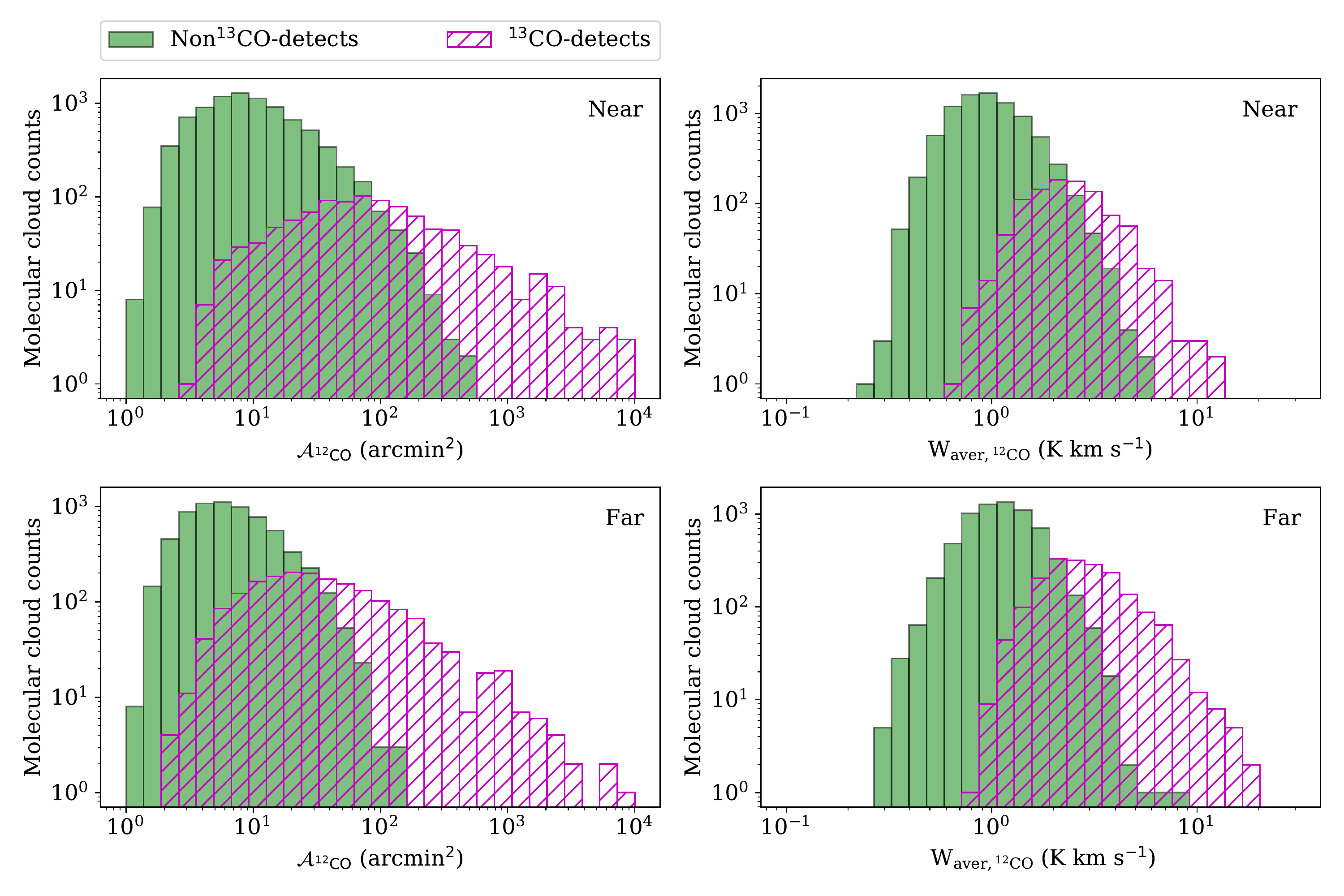}
    \caption{The number distributions of angular areas ($\mathcal{A}_{\rm ^{12}CO}$) and 
    the averaged velocity-integrated intensities ($W_{\rm{aver, ^{12}CO}}$) of $^{12}$CO line emission for $^{12}$CO molecular clouds 
    in near and far groups, respectively. The near $^{12}$CO clouds are 
    in a velocity range of (-30 25) km s$^{-1}$, the far $^{12}$CO clouds are in (-95 -30) km s$^{-1}$. 
    The green histgrams represent $^{12}$CO molecular clouds not having the $^{13}$CO structures (Non$^{13}$CO-detects). 
    The magenta ones are the $^{12}$CO molecular clouds having the $^{13}$CO structures ($^{13}$CO-detects). \label{fig:fcow}}
\end{figure*}

In Figure \ref{fig:fno13_with13}, we present the number distributions of 
angular areas ($\mathcal{A}_{\rm ^{12}CO}$), velocity spans ($V_{\rm{span}, ^{12}CO}$), peak intensities ($I_{\rm{peak}, ^{12}CO}$), and 
integrated fluxes ($F_{\rm ^{12}CO}$) of $^{12}$CO line emission for the $^{13}$CO-detects and Non$^{13}$CO-detects, respectively. 
%For the Non$^{13}$CO-detects, their $\mathcal{A}_{^{12}CO}$ are in a range of 1 -- 463.75 arcmin$^{2}$ and focus on 4.25 -- 13.75 arcmin$^{2}$. 
%The $V_{\rm{span}, ^{12}CO}$ are from 0.5 to 15.4 km s$^{-1}$ and mainly from 1.5 to 2.67 km s$^{-1}$. 
%The range of $T_{\rm{peak}, ^{12}CO}$ is 1.2 -- 10.0 K, and their typical range is 2.5 -- 3.6 K.
%The values of $F_{^{12}CO}$ are from 0.52 to 714.8 K km s$^{-1}$ arcmin$^{2}$ and concentrate in 3.87 -- 16.6 K km s$^{-1}$ arcmin$^{2}$.
The quantiles at 0.05, 0.25, 0.5, 0.75, and 0.95 of these parameters for the $^{13}$CO-detects and Non$^{13}$CO-detects are listed in Table \ref{tab:t12co}, respectively. 
We find that these quantiles of the parameters in the $^{13}$CO-detects are systematically larger 
than that from the Non$^{13}$CO-detects. 
We calculate the total $F_{\rm ^{12}CO}$ of the whole Non$^{13}$CO-detects. 
The value is about 2.5$\times$10$^{5}$ K km s$^{-1}$ arcmin$^{2}$, which makes up about 6.8$\%$ of 
that of the total 18,190 $^{12}$CO molecular clouds (3.7$\times$10$^{6}$ K km s$^{-1}$ arcmin$^{2}$). 
The rest $\sim$ 93$\%$ are from the $^{13}$CO-detects. That indicates the $^{13}$CO-detects are the main 
contributor of $^{12}$CO emission fluxes, although their number only take a percentage of $\sim$ 15$\%$ in 
the total number. The total $\mathcal{A}_{\rm ^{12}CO}$ for the 18,190 $^{12}$CO clouds is about 
228.2 deg$^{2}$. Among that, the sum of the $\mathcal{A}_{\rm ^{12}CO}$ from the $^{13}$CO-detects 
take a percentage of 76.2$\%$ and that from the Non$^{13}$CO-detects take about the rest of 23.8$\%$.

\begin{deluxetable*}{lccccc}
    \tablenum{3}
    \tablecaption{The physical parameters of $^{12}$CO line emission for Non$^{13}$CO-detects and $^{13}$CO-detects\label{tab:t12co}}
    \tablewidth{0pt}
    \tablehead{\colhead{Types}  & \colhead{Quantile} & \colhead{$\mathcal{A}_{\rm ^{12}CO}$} & \colhead{$V_{\rm span, ^{12}CO}$} & \colhead{$T_{\rm{peak}, ^{12}CO}$} & \colhead{$F_{^{12}CO}$} \\
    & &\colhead{(arcmin$^{2}$)} &\colhead{(km s$^{-1}$)} &\colhead{(K)} & \colhead{(K km s$^{-1}$ arcmin$^{2}$)} 
    }
    %\decimalcolnumbers
    \startdata 
    & 0.05 & 2.5 & 1.0  & 2.2 & 1.7 \\
    & 0.25 & 4.25 & 1.5 & 2.5 & 3.9 \\
    Non$^{13}$CO-detects & 0.5  & 7.25 & 2.0 & 2.9 & 7.6 \\
    & 0.75 & 13.75 & 2.67 & 3.5 & 16.6 \\
    & 0.95 & 39.5 & 4.2 & 5.0 & 58.8 \\
    & Mean & 12.8 & 2.2 & 3.1 & 16.5 \\
    \hline
    & 0.05 & 6.25 & 1.84 & 3.8 & 11.8 \\
    & 0.25 & 16.4 & 2.84 & 5.1 & 38.9 \\
    $^{13}$CO-detects & 0.5  & 39 & 4.0 & 6.3 & 101.2 \\
    & 0.75 & 102.5 & 5.68 & 8.3 & 299.4 \\
    & 0.95 & 640.0 & 10.35 & 13.3 & 2422.3 \\
    & Mean & 219.5 & 4.73 & 7.3 & 1211.9 \\
    \enddata
    \tablecomments{The quantiles at 0.05, 0.25, 0.5, 0.75 and 0.95 for each parameter in its sequential data and its mean value.}
\end{deluxetable*}

\begin{deluxetable}{lccc}
    \tablenum{4}
    \tablecaption{The angular areas ($\mathcal{A}_{\rm ^{12}CO}$) and H$_{2}$ column densities (N$_{\rm H_{2}}$) 
    for $^{13}$CO-detects and Non$^{13}$CO-detects, which are in the near and far groups, respectively. \label{tab:tw}}
    \tablewidth{0pt}
    \tablehead{
    \colhead{Types}  & \colhead{Quantile} & \colhead{$\mathcal{A}_{\rm ^{12}CO}$} & \colhead{N$_{\rm H_{2}}$} \\
    & &\colhead{(arcmin$^{2}$)} & \colhead{10$^{20}$ cm$^{-2}$} \\
    & &\colhead{Near, Far} & \colhead{Near, Far}} 
    
    \startdata
    & 0.05 & 2.5, 2.25 & 1.05, 1.2 \\ 
    & 0.25 & 5.0, 3.75 & 1.46, 1.72 \\
    Non$^{13}$CO-detects & 0.5  & 8.5, 6.0 & 1.88, 2.2 \\
    & 0.75 & 16.75, 10.75 & 2.5, 2.88 \\
    & 0.95 & 51.1, 26.5 & 3.9, 4.26 \\
    & Mean & 15.6, 9.1 & 2.1, 2.4 \\
    \hline
    & 0.05 & 8.5, 5.75 & 2.4, 2.8 \\
    & 0.25 & 31.88, 13.25 & 3.5, 4.1 \\
    $^{13}$CO-detects & 0.5 & 73.13, 28.75 & 4.6, 5.4 \\
    & 0.75 & 185.13, 70.0 & 6.2, 7.48 \\
    & 0.95 & 1134.4, 322.5 & 9.6, 13.3 \\ 
    & Mean & 413.8, 116.3 & 5.2, 6.4 \\
    \enddata
    \tablecomments{The quantiles at 0.05, 0.25, 0.5, 0.75 and 0.95 for the $\mathcal{A}_{\rm ^{12}CO}$ and 
    N$_{\rm H_{2}}$ in their sequential data and their mean values. 
    The H$_{2}$ column density (N$_{\rm H_{2}}$) are calculated using the N$_{\rm H_{2}}$ = X$_{\rm CO}$W$_{\rm ^{12}CO}$, 
    where X$_{\rm CO}$=2$\times$10$^{20}$ cm$^{-2}$ (K km s$^{-1}$)$^{-1}$. 
    The Near represents the molecular clouds in the near range ($V_{\rm LSR}$ from $-$30 km s$^{-1}$ to 25 km s$^{-1}$), 
    the Far means the molecular clouds in the far range ($V_{\rm LSR}$ from $-$95 km s$^{-1}$ to $-$30 km s$^{-1}$).}
\end{deluxetable}

%For the $^{13}$CO-detects, the values of their $\mathcal{A}_{\rm ^{12}CO}$ are in a range of 2.0 -- 60745.25 arcmin$^{2}$ and 
%concentrate on the scale of 16.4 -- 102.5 arcmin$^{2}$. 
%Their $V_{\rm{span}, ^{12}CO}$ are distributed in (0.84 43.42) km s$^{-1}$ and centrally located in 2.84 -- 5.68 km s$^{-1}$.
%Their $T_{\rm{peak}, ^{12}CO}$ range from 2.4 to 50.8 K and concentrate in 5.1 -- 8.3 K.
%The values of $F_{^{12}CO}$ are distributed in a range of 3.0 -- 390096 K km s$^{-1}$ arcmin$^{-2}$ and 
%mainly in 38.9 -- 299.4 K km s$^{-1}$ arcmin$^{-2}$.

%Owing to that angular areas of molecular clouds are not their physical sizes, whose values depend on their distances, 
%we divide the molecular clouds into the near and far groups. 
Following the paper I \citep{Yuan2021}, the $^{12}$CO clouds are divided into two groups 
by a $V_{\rm LSR}$ threshold of - 30 km s$^{-1}$, shown as a white-dashed line in Figure \ref{fig:f13co}. 
The $^{12}$CO clouds with central velocities in a range of (-30 25) km s$^{-1}$ are in the near group,  
and the $^{12}$CO clouds with central velocities ranging from -95 km s$^{-1}$ to -30 km s$^{-1}$ are in the far group. 
In the near group, there are 9,544 $^{12}$CO molecular clouds, among which the $^{13}$CO-detects take a percentage of 10.4$\%$. 
In the far group, there are 8,646 $^{12}$CO molecular clouds and 21.5$\%$ of them have $^{13}$CO structures. 

The number detection rate of the $^{13}$CO-detects in the near is lower than that in the far group. 
That may be due to that there are more MCs with faint $^{12}$CO emission, but no $^{13}$CO emission detected in the near group. 
The number distributions of the $\mathcal{A}_{\rm ^{12}CO}$ of $^{13}$CO-detects and Non$^{13}$CO-detects 
in the near and far group are presented in Figure \ref{fig:fcow}, respectively.
%For the near group, the $\mathcal{A}_{\rm ^{12}CO}$ of Non$^{13}$CO-detects concentrate in 5.0 -- 16.75 arcmin$^{2}$, 
%the central values for $^{13}$CO-detects are in 32 -- 185 arcmin$^{2}$. 
%For the far group, the $\mathcal{A}_{\rm ^{12}CO}$ of Non$^{13}$CO-detects are mainly in 3.75 -- 10.75 arcmin$^{2}$, 
%that of $^{13}$CO-detects focus on 13 -- 70 arcmin$^{2}$. 
The quantiles at 0.05, 0.25, 0.5, 0.75, and 0.95 of their $\mathcal{A}_{\rm ^{12}CO}$ values are listed in Table \ref{tab:tw}. 
According to the spiral structure of the Milky Way, the kinematical distances, 
which are estimated using the Bayesian distance calculator in \cite{Reid2016}, 
center on about $\sim$ 0.5 kpc for molecular clouds in the Local arm and $\sim$ 2 kpc for that 
in the Perseus arm. Considering these typical distances, 
the molecular cloud in the local region with an angular size of 1$^{\prime}$ has a physical scale of $\sim$ 0.15 pc, 
the value is $\sim$ 0.6 pc for that in the Perseus arm.

In addition, we estimate the averaged velocity-integrated intensities of $^{12}$CO lines emission, 
W$_{\rm{aver, ^{12}CO}}$ = $\int$T$_{\rm mb}(l, b, v)dvdldb$/$\int$$dldb$, for molecular clouds 
in the near and far gourps, respectively. Their distributions are shown in Figure \ref{fig:fcow}. 
%In the near range, for the Non$^{13}$CO-detects, their W$_{\rm{aver, ^{12}CO}}$ range from 0.26 to 5.2 K km s$^{-1}$ 
%and focuses on 0.73 -- 1.25 K km s$^{-1}$. 
%For $^{13}$CO-detects, the values of  W$_{\rm ^{12}CO}$ are in a range of 0.7 -- 13.1 K km s$^{-1}$ 
%and concentrates on 1.75 -- 3.1 K km s$^{-1}$. 
%In the far range, the W$_{\rm{aver, ^{12}CO}}$ of Non$^{13}$CO-detects is from 0.29 to 8.9 K km s$^{-1}$ and 
%centre on 0.86 -- 1.44 K km s$^{-1}$, that of $^{13}$CO-detects are in a range of 0.79 -- 18.5 K km s$^{-1}$ 
%and mainly in 2.1 -- 3.7 K km s$^{-1}$. 
The H$_{2}$ column density (N$_{\rm H_{2}}$) can be calculated through 
the N$_{\rm H_{2}}$ = X$_{\rm CO}$W$_{\rm ^{12}CO}$, where X$_{\rm CO}$ = 2$\times$10$^{20}$ cm$^{-2}$ (K km s$^{-1}$)$^{-1}$ 
is the CO-to-H$_{2}$ conversion factor \citep{Bolatto2013}. 
The quantile values of N$_{\rm H_{2}}$ for Non$^{13}$CO-detects and $^{13}$CO-detects are also listed in Table \ref{tab:tw}.
We find that the typical values of $\mathcal{A}_{\rm ^{12}CO}$ and N$_{\rm H_{2}}$ of $^{13}$CO-detects are larger than 
those of Non$^{13}$CO-detects, either in the near or far groups. 
Thus the $^{13}$CO emission seems to be related to the properties of $^{12}$CO emissions independent of 
the distance.

Compared with the Non$^{13}$CO-detects, the $^{13}$CO-detects tend to have larger $\mathcal{A}_{\rm ^{12}CO}$, 
higher $T_{\rm{peak, ^{12}CO}}$, and $W_{\rm{aver, ^{12}CO}}$, either in near or far group.  
\begin{figure*}
    \plotone{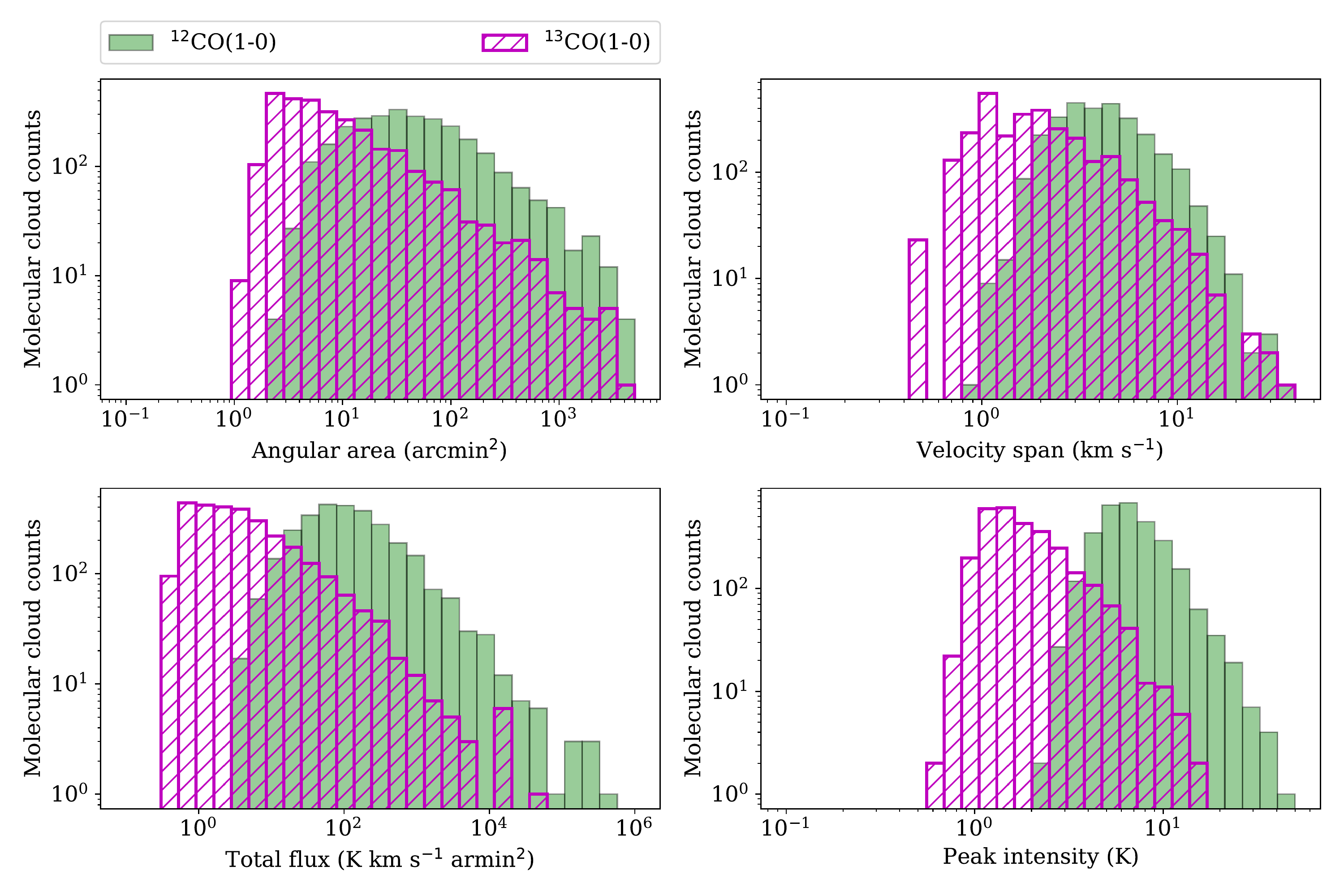}
    \caption{The number distributions of the angular sizes, velocity spans, the total fluxes and the peak 
    intensities of $^{13}$CO(1-0) and $^{12}$CO(1-0) line emission for the 2,851 $^{13}$CO-detects. 
    The green histgrams represent parameters of the $^{12}$CO line emission. 
    The magenta ones are the parameters of the $^{13}$CO line emission. \label{fig:f12co_13co}}
\end{figure*}

\begin{figure*}
    \plotone{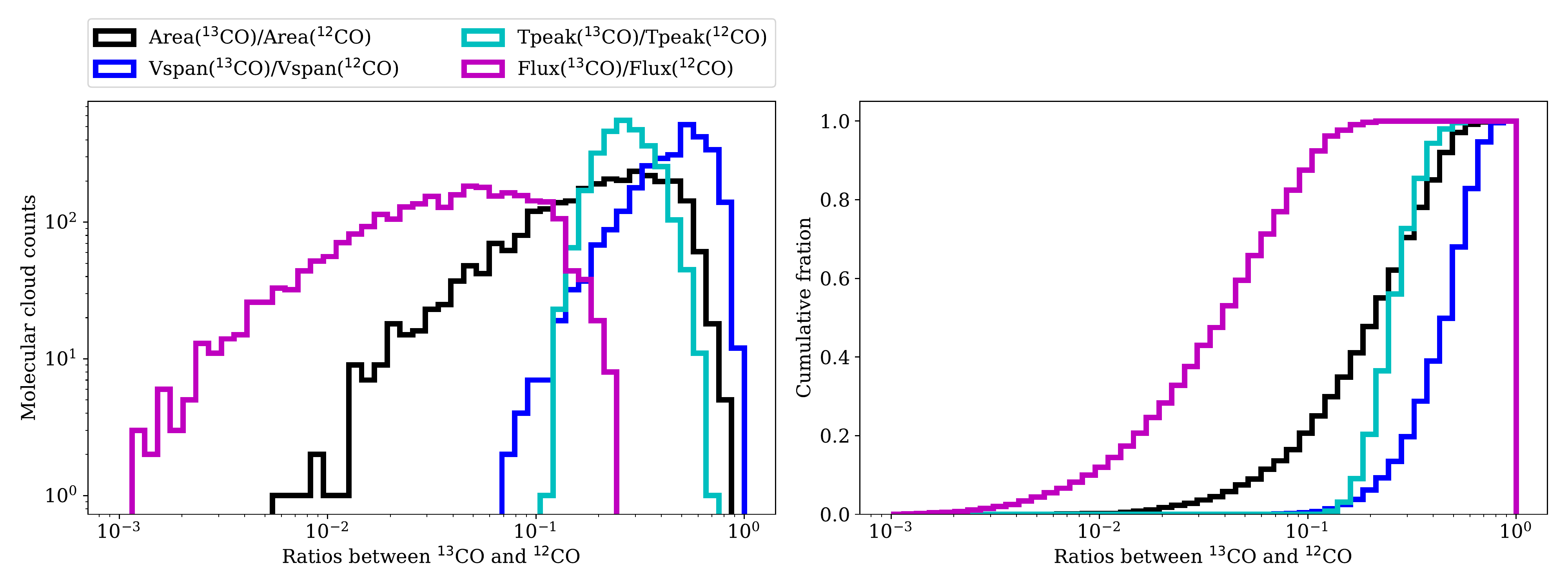}
    \caption{The counts distributions (\textbf{left panel}) and cumulative fractions (\textbf{right panel}) 
    of the ratios between the parameters of $^{13}$CO structures and that of $^{12}$CO line emission for the 2,851 $^{13}$CO-detects, 
    these parameter ratios include their angular sizes (black), velocity spans (blue), the peak intensities (cyan) and the integrated fluxes (magenta). \label{fig:fratio}}
\end{figure*}

\subsection{$^{12}$CO and $^{13}$CO lines emission in $^{12}$CO molecular clouds having $^{13}$CO structures} 
To systematically analyze what properties of MCs determine the $^{13}$CO line emission in the $^{13}$CO-detects,
we examine the correlation between their physical properties of $^{13}$CO line emission with that of the $^{12}$CO line emission.

\subsubsection{Distribution of $^{13}$CO and $^{12}$CO emission parameters} 
Figure \ref{fig:f12co_13co}  presents the number distributions of angular sizes ($\mathcal{A}_{\rm ^{12}CO}$, $\mathcal{A}_{\rm ^{13}CO}$), 
velocity spans ($V_{\rm{span}, ^{12}CO}$, $V_{\rm{span}, ^{13}CO}$), the integrated fluxes ($F_{\rm ^{12}CO}$, $F_{\rm ^{13}CO}$), 
and peak intensities ($T_{\rm{peak}, ^{12}CO}$, $T_{\rm{peak}, ^{13}CO}$) of $^{12}$CO and $^{13}$CO lines emission of the 2,851 $^{13}$CO-detects. 
The quantiles at 0.05, 0.25, 0.5, 0.75, and 0.95 of these parameters for $^{13}$CO line emission are listed in Table \ref{tab:t13co}.
Those for $^{12}$CO line emission are listed in Table \ref{tab:t12co}. 
We find that the values of angular areas, velocity spans, peak intensities, and the integrated fluxes of $^{13}$CO 
structures in the MCs are systematically smaller than that of their $^{12}$CO line emission. 

\subsubsection{Ratios between $^{13}$CO and $^{12}$CO emission parameters}
Since only a portion molecular gas components in a $^{12}$CO cloud have $^{13}$CO emission, 
what are the specific fractions of $^{13}$CO gas in these $^{12}$CO MCs? 
Figure \ref{fig:fratio} presents the distributions of the ratios between the $^{13}$CO emission parameters and 
$^{12}$CO emission parameters in the each $^{12}$CO cloud with $^{13}$CO structures.  
%We find that the ratios of $\mathcal{A}_{\rm ^{13}CO}$ to $\mathcal{A}_{\rm ^{12}CO}$ distribute from 0.006 to 0.8 and have a typical range of 0.12 -- 0.35. 
%The ratios of $V_{\rm{span}, ^{13}CO}$ to $V_{\rm{span}, ^{12}CO}$ are in a range of 0.04 -- 0.59 and concentrate on 0.15 -- 0.29.
%These values for ratios of $T_{\rm{peak}, ^{13}CO}$ and $T_{\rm{peak}, ^{12}CO}$ are in 0.12 -- 0.71 and centrally in 0.22 -- 0.33.
%Finally, the ratios of $F_{\rm ^{13}CO}$ and $F_{\rm ^{12}CO}$ are from 0.0009 to 0.24 and mainly in 0.02 -- 0.07. 
The quantiles at 0.05, 0.25, 0.5, 0.75, and 0.95 of these ratio values are listed in Table \ref{tab:t13co}. 
We find that the median and mean values of $\mathcal{A}_{\rm ^{13}CO}$/$\mathcal{A}_{\rm ^{12}CO}$ and 
$T_{\rm{peak}, ^{13}CO}$/$T_{\rm{peak}, ^{12}CO}$ are close to $\sim$ 0.25. In addition, 
the 95$\%$ of the $^{13}$CO-detects have the $\mathcal{A}_{\rm ^{13}CO}$/$\mathcal{A}_{\rm ^{12}CO}$ with values less than 0.53 and 
the $T_{\rm{peak}, ^{13}CO}$/$T_{\rm{peak}, ^{12}CO}$ with values less than 0.44. 
For the $V_{\rm{span}, ^{13}CO}$/$V_{\rm{span}, ^{12}CO}$ in the $^{13}$CO-detects, their median and mean values are about 0.5 and 
95$\%$ of them are less than 0.76. The median and mean values of $F_{\rm ^{13}CO}$/$F_{\rm ^{12}CO}$, 
are 0.04 and 0.05, respectively. Moreover, the $F_{\rm ^{13}CO}$/$F_{\rm ^{12}CO}$ for the 95$\%$ of $^{13}$CO-detects is not larger than 0.13. 
That implies the fractions of $^{13}$CO gas in the $^{12}$CO molecular clouds are typically less than 13$\%$. 
Considering the $^{12}$CO lines are more optically thick, this value should be much lower. 

\begin{deluxetable*}{lccccc}
    \tablenum{5}
    \tablecaption{The physical parameters $^{13}$CO emission and the ratios between the parameters of the $^{13}$CO and $^{12}$CO emission.\label{tab:t13co}}
    \tablewidth{0pt}
    \tablehead{\colhead{Types}& \colhead{Quantile} & \colhead{Angular area} & \colhead{V$_{\rm span}$} &\colhead{Peak Intensity} & \colhead{Flux} \\
    & &\colhead{(arcmin$^{2}$)} &\colhead{(km s$^{-1}$)} &\colhead{(K)} & \colhead{(K km s$^{-1}$ arcmin$^{2}$)} 
    }
    %\decimalcolnumbers
    \startdata
    & 0.05 & 2.0 & 0.66 & 1.0 & 0.6 \\
    & 0.25 & 3.25 & 1.16 & 1.3 & 1.2 \\
    $^{13}$CO emission & 0.5 & 6.25 & 1.66 & 1.6 & 3.1 \\
    & 0.75 & 16.5 & 2.66 & 2.3 & 10.7 \\
    & 0.95 & 120.0 & 6.31 & 4.7 & 121.9 \\
    & Mean & 58.4 & 2.37 & 2.1 & 82.0 \\
    %& Peak & 2.0 & 1.0 & 1.3 & 0.53 \\
    \hline
    & 0.05 & 0.04 & 0.20 & 0.17 & 0.006 \\
    & 0.25 & 0.12 & 0.36 & 0.22 & 0.02 \\
    $^{13}$CO/$^{12}$CO & 0.5 & 0.22 & 0.50 & 0.27 & 0.04 \\
    & 0.75 & 0.35 & 0.61 & 0.33 & 0.07 \\
    & 0.95 & 0.53 & 0.76 & 0.44 & 0.13 \\
    & Mean & 0.25 & 0.48 & 0.28 & 0.05 \\
    %& Peak & 0.28 & 49.4$\%$ & 24.4$\%$ & 4.5$\%$ \\  
    \enddata
    \tablecomments{The quantiles at 0.05, 0.25, 0.5, 0.75 and 0.95 for each parameter in its sequential data 
    and its mean value.}
\end{deluxetable*}
%From the cumulative fractions distribution in Figure \ref{fig:fratio}, we find that ?

\begin{figure*}
\plotone{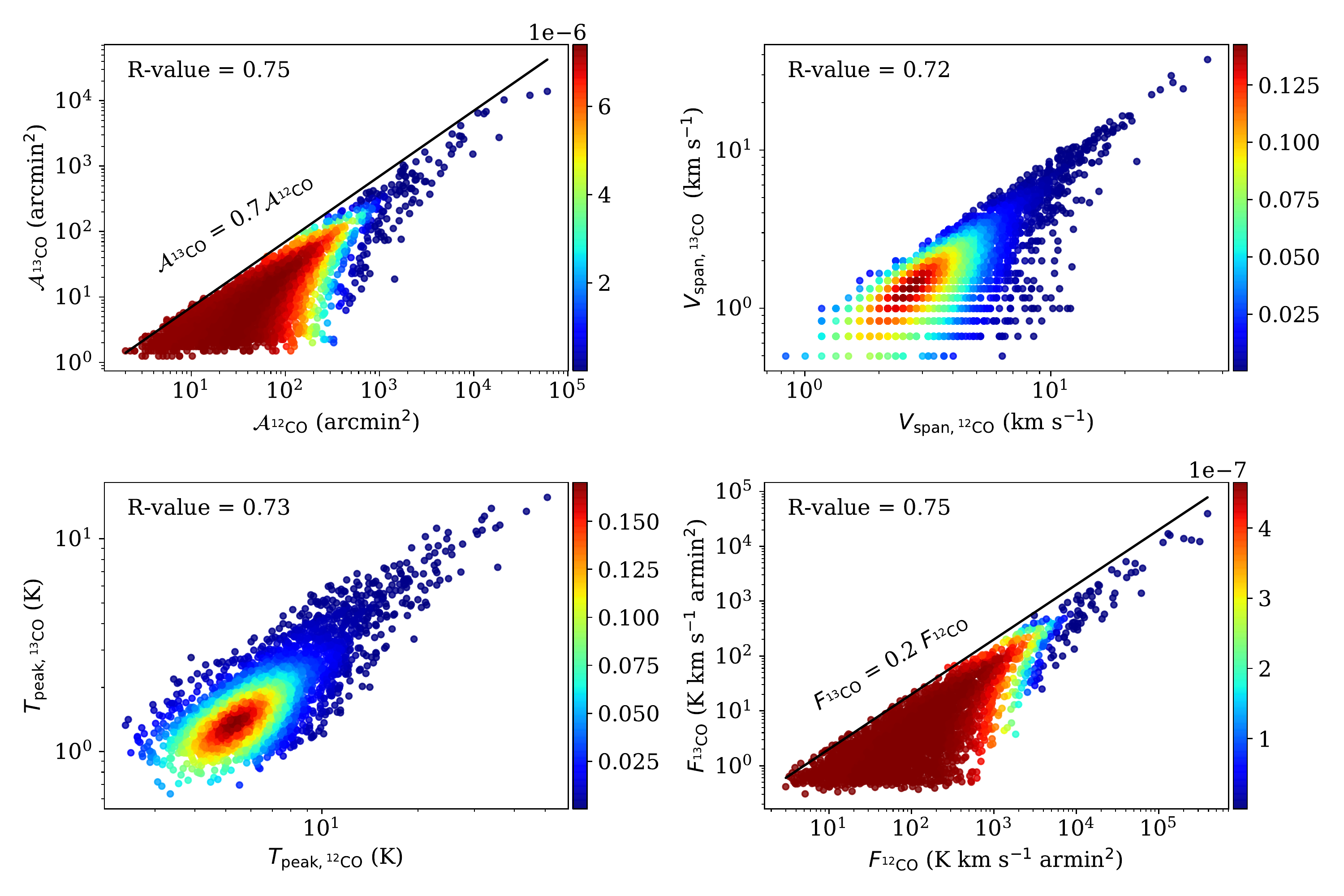}
\caption{The correlations of angular sizes, velocity spans, peak intensities, and integrated fluxes 
of $^{12}$CO(1-0) emission between that of $^{13}$CO(1-0) line emission. 
Each dot represents a $^{12}$CO cloud with $^{13}$CO structures. 
The $^{12}$CO emission parameters of a $^{12}$CO cloud represent the global physical parameters of a $^{12}$CO clouds.
The $^{13}$CO emission parameters are derived from all the $^{13}$CO structures inside a single $^{12}$CO cloud. 
The black lines represent the upper limits ($\mathcal{A}_{\rm ^{13}CO}$ vs $\mathcal{A}_{\rm ^{12}CO}$, $F_{\rm ^{13}CO}$ vs $F_{\rm ^{12}CO}$). 
The colors on the points represent the distribution of the probability density function of the $^{13}$CO-detects counts (2D-PDF), 
which are calculated utilizing the Kernel-density estimation through Gaussian kernels in the PYTHON package 
\href{https://docs.scipy.org/doc/scipy/reference/generated/scipy.stats.gaussian\_kde.html}{scipy.stats.gaussian\_kde}. 
\label{fig:fglobal_corre}}
\end{figure*}

\begin{figure*}
    \plotone{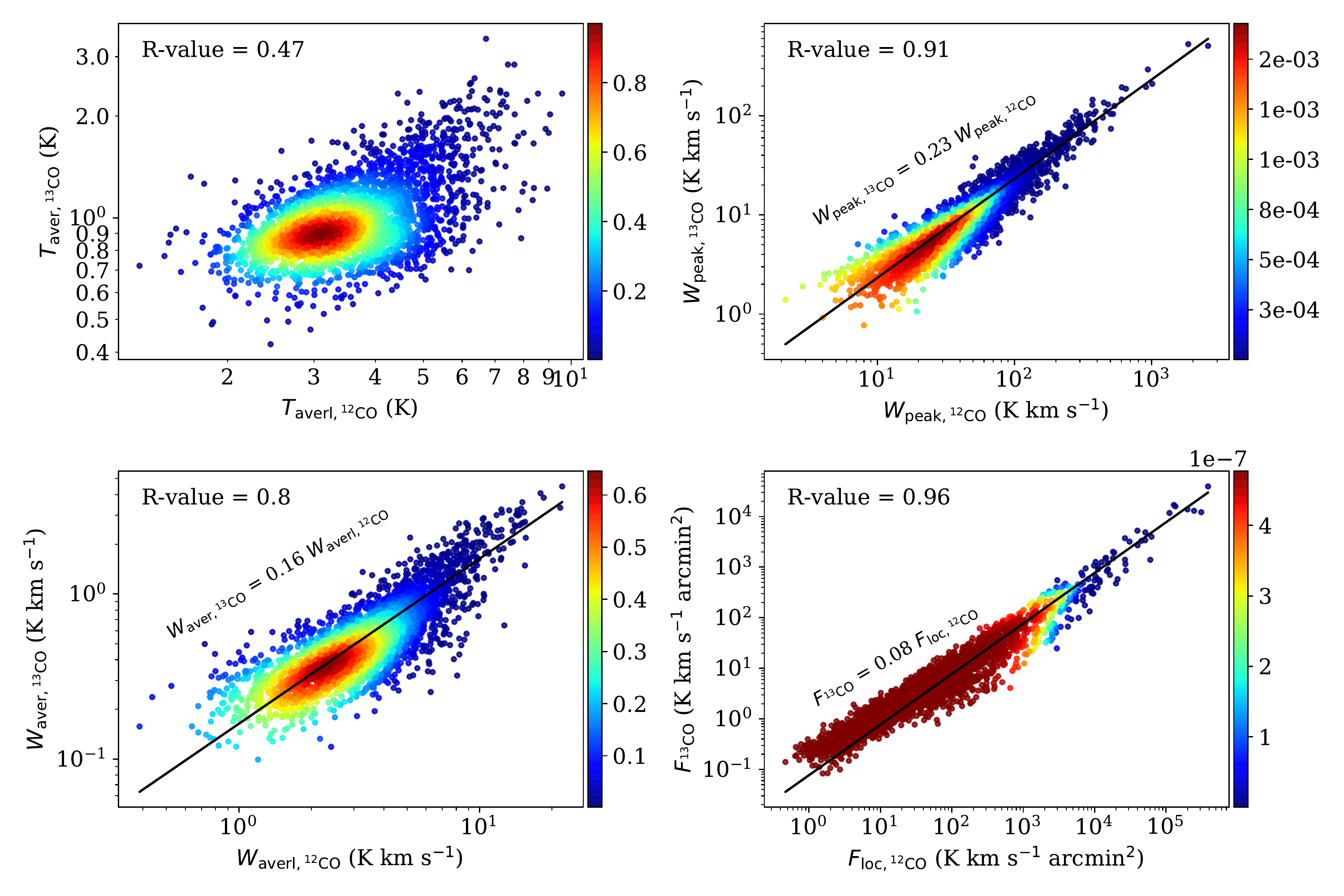}
    \caption{The correlations between the averaged line intensities ($T_{\rm{averl}, ^{12}CO}$ vs $T_{\rm{aver}, ^{13}CO}$), 
    peak ($W_{\rm{peak}, ^{12}CO}$ vs $W_{\rm{peak}, ^{13}CO}$) and averaged ($W_{\rm{averl}, ^{12}CO}$ vs $W_{\rm{aver}, ^{13}CO}$) velocity-integrated intensities, and the 
    integrated fluxes ($F_{\rm loc, ^{12}CO}$ vs $F_{\rm ^{13}CO}$) of $^{12}$CO and $^{13}$CO emission towards the same areas where both the $^{12}$CO and $^{13}$CO emissions are detected in each $^{13}$CO-detects. 
    Each dot represents a single $^{12}$CO cloud with $^{13}$CO structures.
    The parameters of $^{12}$CO line emission are the physical parameters toward the $^{13}$CO-emitting regions within $^{12}$CO clouds. 
    The $^{13}$CO emission parameters are derived from all the $^{13}$CO structures inside a single $^{12}$CO cloud. 
    The black lines show the linear least-squares fits to the data. 
    The colors on the points show the distribution of the probability density function of the $^{13}$CO-detects counts (2D-PDF). \label{fig:flocal_corre}}
\end{figure*}
    
\subsubsection{Correlation between $^{13}$CO and $^{12}$CO emission parameters}

%Although the $^{13}$CO-detects are prone to large angular areas, there are also a part of 
%Non$^{13}$CO-detects having angular areas in the area-range of $^{13}$CO-detects, 
%as well as their $T_{\rm{peak, ^{12}CO}}$ and $W_{\rm{aver, ^{12}CO}}$. 
%That is, the $\mathcal{A}_{\rm ^{12}CO}$, $T_{\rm{peak, ^{12}CO}}$ and $W_{\rm{aver, ^{12}CO}}$
%are not the critical factors. What is the critical parameter for the $^{13}$CO emission? 

Since the parameters of $^{13}$CO line emission are distributed in a certain range,  
how do the global properties of molecular clouds affect their $^{13}$CO line emission? 
Figure \ref{fig:fglobal_corre} presents the correlations between the parameters of $^{13}$CO emission and that of $^{12}$CO emission in each $^{13}$CO-detects. 
The spearman's rank correlation coefficients (R-value) for these relations are estimated and 
the resultant R-values are noted in the Figure \ref{fig:fglobal_corre}. 
%For the correlation between $\mathcal{A}_{\rm ^{12}CO}$ and $\mathcal{A}_{\rm ^{13}CO}$, 
%we estimate their spearman's rank correlation coefficient, and the value is 0.75. 
%There is a roughly positive correlation between them, but it still presents a little dispersed. 
We find the R-value (0.75) for $\mathcal{A}_{\rm ^{12}CO}$ and $\mathcal{A}_{\rm ^{13}CO}$ is consistent with 
that for $F_{\rm ^{13}CO}$ and $F_{\rm ^{12}CO}$, a little higher than the R-value (0.73) for 
$T_{\rm{peak}, ^{13}CO}$ and $T_{\rm{peak}, ^{12}CO}$ and the value of 0.72 for the $V_{\rm{span}, ^{12}CO}$ and $V_{\rm{span}, ^{13}CO}$.
That implies there are roughly positive correlations between them, but they still present a little dispersed. 

There are sharp upper limits for the ratios of $\mathcal{A}_{\rm ^{13}CO}$/$\mathcal{A}_{\rm ^{12}CO}$, 
and $F_{\rm ^{13}CO}$/$F_{\rm ^{12}CO}$ independent of the angular areas and CO integrated fluxes in wide ranges. 
As shown in Figure \ref{fig:fglobal_corre}, we outline their upper limits with slopes of 0.7 and 0.2, respectively. 
That indicates the area of $^{13}$CO emission in a molecular cloud generally does not exceed 
the 70$\%$ of the $^{12}$CO emission area, independent of the $^{12}$CO emission area. 
For the integrated fluxes, the $^{13}$CO emission fluxes are usually less than 20$\%$ of the $^{12}$CO emission fluxes. 

Overall, the global physical parameters of molecular clouds, such as the angular areas and integrated fluxes of $^{12}$CO emission, 
show roughly positive correlations and provide upper limits for that of $^{13}$CO emission. 
Whereas they cannot critically determine the $^{13}$CO emission. 

Since the correlation between the global physical properties of each $^{12}$CO cloud and that of its interior $^{13}$CO structures 
exhibits a bit scattered, 
we further focus on the local areas having both $^{12}$CO and $^{13}$CO line emissions in each $^{12}$CO cloud with $^{13}$CO structures. 
Figure \ref{fig:flocal_corre} presents the correlations between the parameters of $^{12}$CO and $^{13}$CO emission towards the same areas 
where both the $^{12}$CO and $^{13}$CO emissions are detected in each $^{13}$CO-detects. 
The $T_{\rm{averl}, ^{12}CO}$ is calculated by averaging the $^{12}$CO line intensities within the $^{13}$CO emission region. 
The $T_{\rm{aver}, ^{13}CO}$ is the mean value of the $^{13}$CO spectra intensities.  
The $W_{\rm{aver}, ^{13}CO}$ and $W_{\rm{averl}, ^{12}CO}$ are the averaged values of the velocity-integrated 
intensities of $^{13}$CO line and that of $^{12}$CO line in the $^{13}$CO emission region, respectively. 
The $W_{\rm{peak}, ^{13}CO}$ and $W_{\rm{peak}, ^{12}CO}$ are the peak values of the velocity-integrated intensity of $^{13}$CO lines and 
that of $^{12}$CO lines at the same positions, respectively.  
The $F_{\rm ^{13}CO}$ and $F_{\rm loc, ^{12}CO}$ represent the integrated fluxes of $^{13}$CO line emission and 
that of $^{12}$CO emission in the $^{13}$CO emission area, respectively. 
We calculate their spearman's rank correlation coefficients (R-value) and the resultant R-values are noted in Figure \ref{fig:flocal_corre}. 
Based on these relations between $W_{\rm{aver}, ^{13}CO}$ and $W_{\rm{averl}, ^{12}CO}$ (R-value = 0.8), 
$W_{\rm peak, ^{13}CO}$ and $W_{\rm peak, ^{12}CO}$ (R-value = 0.91), and 
$F_{\rm ^{13}CO}$ and $F_{\rm loc, ^{12}CO}$ (R-value = 0.96), their correlations are more and more tightly.  
That indicates the properties of $^{12}$CO line emission in the area where both $^{12}$CO and $^{13}$CO are 
detected, its velocity-integrated intensity ($W_{\rm ^{12}CO}$) in this area, 
is a more direct link for that of $^{13}$CO line emission.   

Furthermore, we also implement the linear least-squares to these linear relations and 
the fitted slopes are noted in Figure \ref{fig:flocal_corre}. These relations 
indicate that the $^{13}$CO fluxes linearly increase as the $^{12}$CO fluxes incrases in 
the region, where the $W_{\rm ^{12}CO}$ is larger than a value of $\sim$ 1 K km s$^{-1}$, 
which is close to the sensitivities of MWISP data. 

\subsubsection{The counts of $^{13}$CO molecular structures in a single $^{12}$CO molecular cloud}
\begin{figure*}
    \plotone{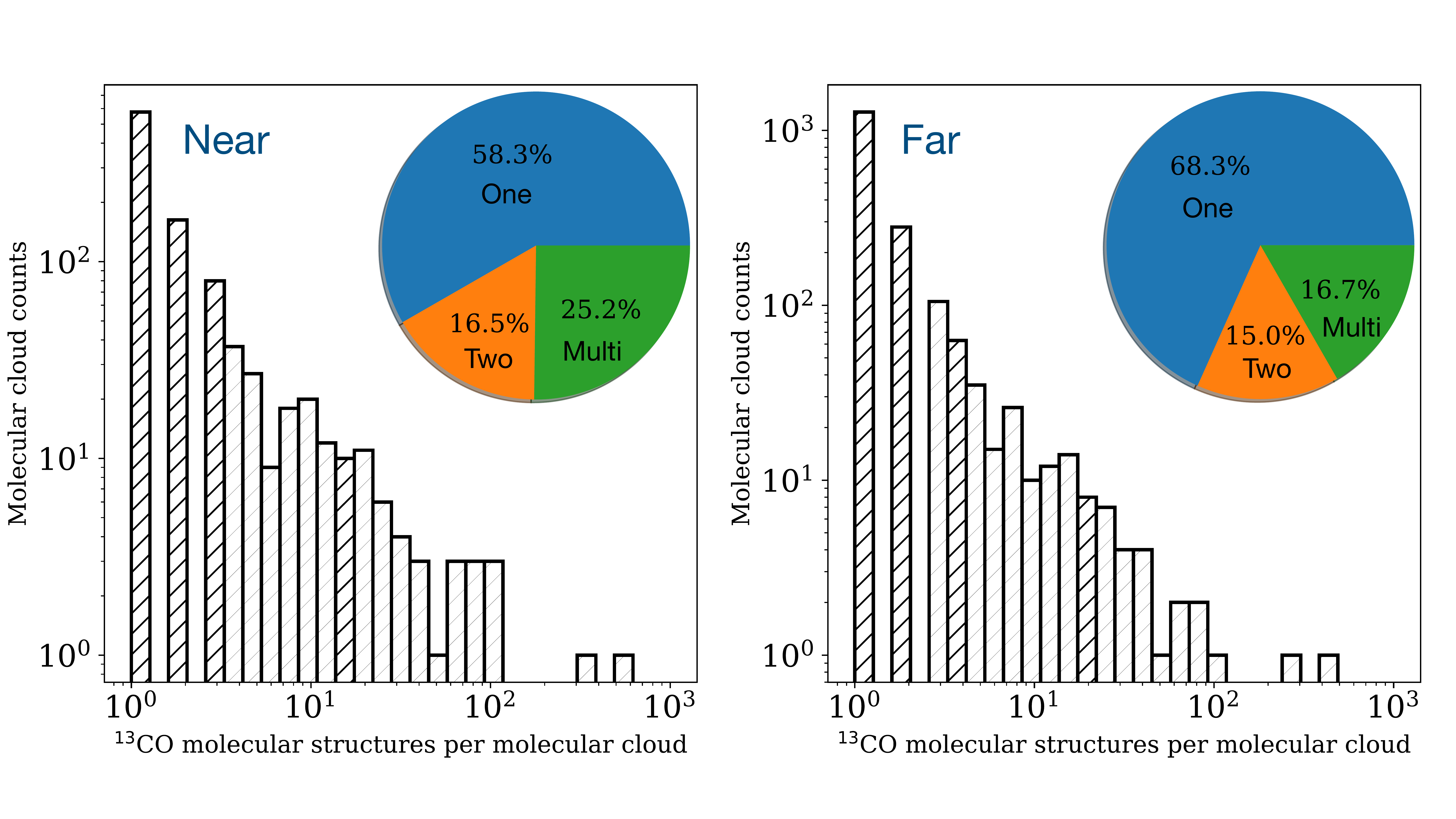}
    \caption{The distribution of the $^{13}$CO structure counts in a single $^{12}$CO cloud for the 
    $^{13}$CO-detects in the near (left panel) and far (right panel) groups, respectively. In the right-up corner of each panel, 
    the pie chart illustrates the percentages of $^{12}$CO clouds in this sample with one $^{13}$CO structure, 
    two $^{13}$CO structures, and multiple (more than two) $^{13}$CO structures, respectively. \label{fig:f13co_coms}}
\end{figure*}

The boundary of a molecular cloud is defined by its $^{12}$CO(1-0) line emission, 
its internal $^{13}$CO molecular structures can present several individual structures, 
as shown in Figure \ref{fig:fdbscan2}. The counts of $^{13}$CO molecular structures in a single 
$^{12}$CO cloud can provide essential clues to the development of dense gas content and 
the internal sub-structures of molecular clouds. 
We statistic the number of separate $^{13}$CO molecular structures in each $^{13}$CO-detects. 
Owing to that the molecular cloud distance may affect the spatial physical resolution of 
$^{13}$CO molecular structures, we divide the 2,851 $^{13}$CO-detects into the near and far groups, 
as introduced in Section \ref{sec:41}. 
Figure \ref{fig:f13co_coms} presents the distributions of $^{13}$CO molecular structure counts in a single 
$^{12}$CO cloud for the $^{13}$CO-detects in near and far groups, respectively. We find that the $^{13}$CO-detects 
with one $^{13}$CO structure are dominant and take a percentage of about 60$\%$.  
Then the molecular clouds having two $^{13}$CO molecular structures occupied about 15$\%$ of the $^{13}$CO-detects. 
The rest $\sim$ 20$\%$ of $^{13}$CO-detects have more than two $^{13}$CO structures, 
the $^{13}$CO structure counts in a single $^{12}$CO cloud can be up to $\sim$ 600. 
It should be noted that the number fraction of $^{12}$CO clouds with one $^{13}$CO velocity structure 
varies about 10$\%$ for that in the near and far groups, as well as the fraction of $^{12}$CO clouds with multiple $^{13}$CO structures.

\begin{deluxetable}{lccc}
    \tablenum{6}
    \tablecaption{The number detection rates and the total flux ratios 
    in the total molecular clouds, nonfilaments, and filaments. \label{tab:tdetec}}
    \tablewidth{0pt}
    \tablehead{\colhead{Types} & \colhead{All} & \colhead{Nonfilament} & \colhead{filament} }
       %\decimalcolnumbers
    \startdata
    Number detection rate & 15.7$\%$ & 14.0$\%$ & 56.5$\%$ \\
    Flux ratio & 6.3$\%$ & 2.9$\%$ & 6.7$\%$ \\
    \enddata
    \tablecomments{The number detection rate is the number of extracted $^{13}$CO-detects divided by the number of 
    molecular clouds in the total samples, nonfilaments and filaments, respectively. 
    The flux ratio is the total integrated fluxes of $^{13}$CO line emission divided by that 
    of $^{12}$CO line emission for the total molecular clouds, nonfilaments and filaments, respectively.}
\end{deluxetable}
\subsection{Linking the internal $^{13}$CO gas structures to the morphologies of $^{12}$CO clouds}
\subsubsection{The $^{13}$CO structures detection rates and morphologies} 
%The 13co detection rate of Clipping is 24.1$\%$ in the whole molecular clouds, and 23.2$\%$ in the non-filaments (11680) and 67.7$\%$ in the filaments (2062).
In paper I \citep{Yuan2021}, we took the morphological classification for the total 18,190 molecular clouds, which were 
classified as unresolved, non-filaments (11,680), and filaments (2,062).
Among the 2,851 $^{13}$CO-detects, 1,641 $^{12}$CO molecular clouds belong to nonfilaments and 1,166 clouds are classified as filaments. 
We find that the number detection rate of $^{13}$CO structures is 15.7$\%$(2851/18190) in the whole $^{12}$CO molecular clouds, 
14$\%$ (1641/11680) in the $^{12}$CO clouds classified as nonfilaments, and 56.5$\%$ (1166/2062) in the $^{12}$CO clouds 
classified as filaments. 
For the ratio of the total integrated fluxes of $^{13}$CO line emission to that of $^{12}$CO line emission, 
the value is 6.3$\%$ for the total molecular clouds, 2.9$\%$ for the nonfilaments, 
and 6.7$\%$ for the filaments. Those values are listed in Table \ref{tab:tdetec}. 
Thus, compared with nonfilaments, the filaments tend to have higher density gas structures, which are traced by $^{13}$CO lines.

\begin{figure*}
    \plotone{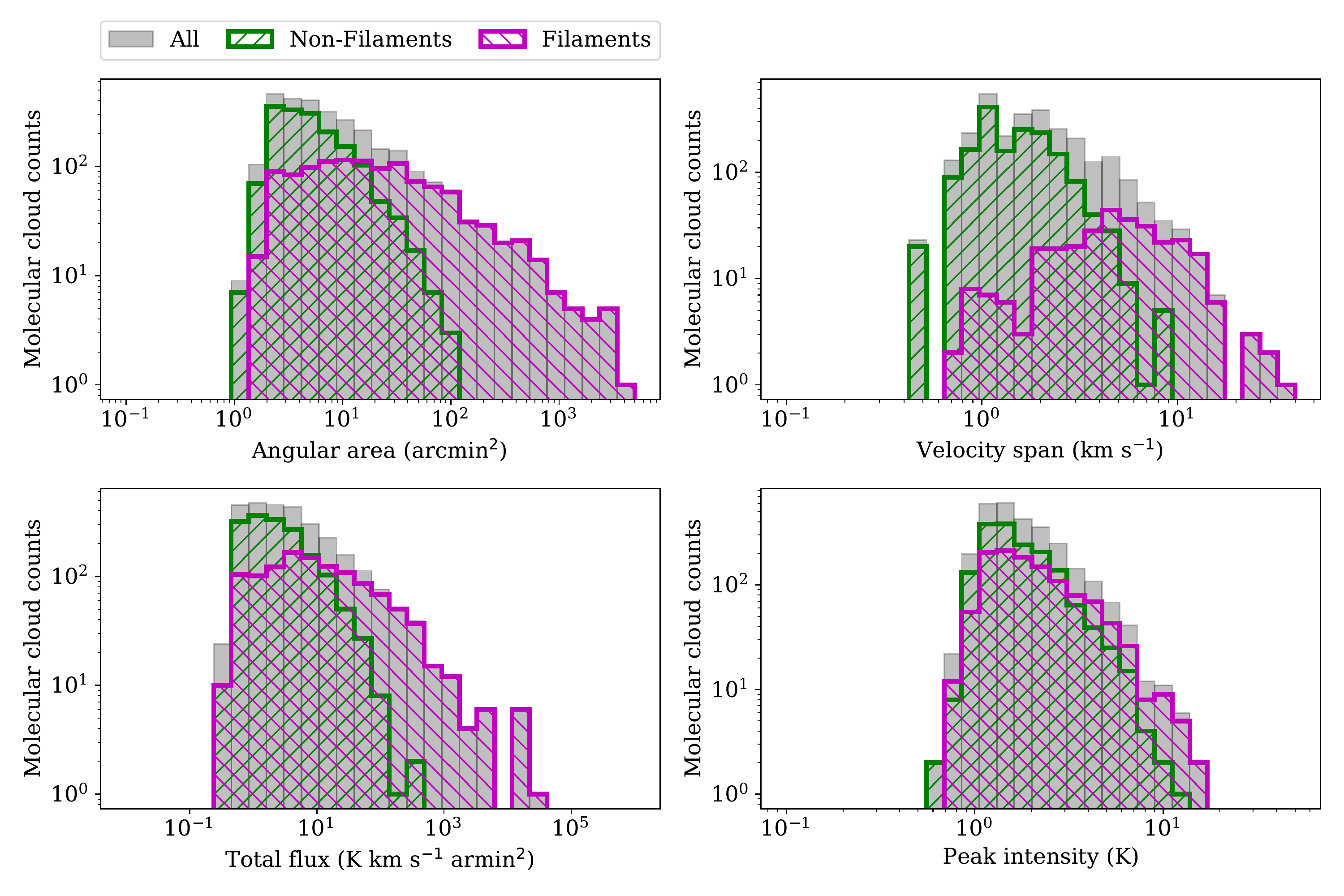}
    \caption{The number distributions of angular sizes, velocity spans, the total fluxes and the peak 
    intensities of $^{13}$CO(1-0) line emission for the $^{13}$CO-detects classified as filaments and nonfilaments. \label{fig:fmorp_13co}}
\end{figure*}

\begin{figure*}
    \plotone{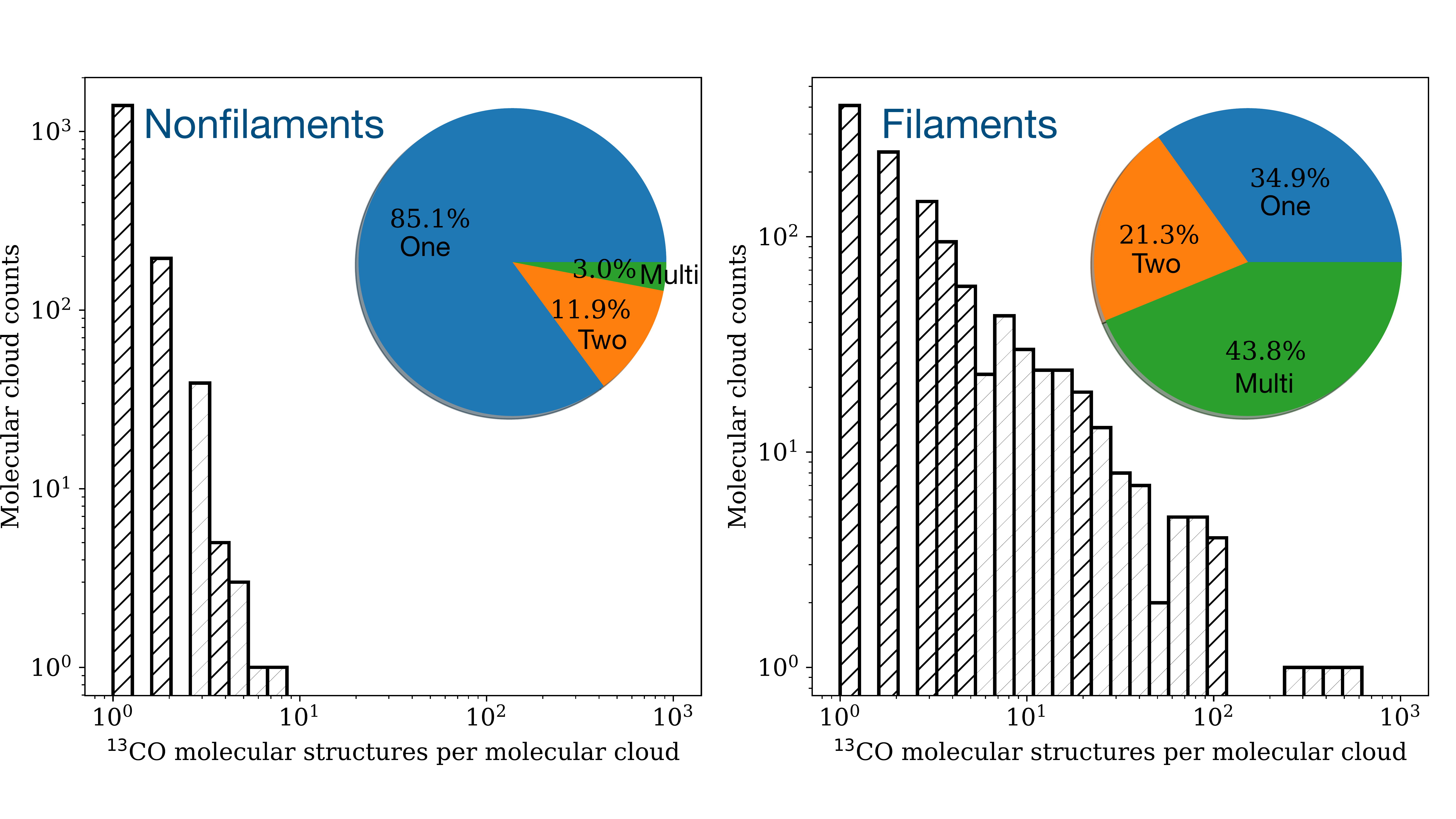}
    \caption{The distribution of the $^{13}$CO molecular structure counts in a single $^{12}$CO cloud for the 
    $^{13}$CO-detects classified as nonfilaments (left panel) and filaments (right panel) groups, respectively. 
    In the right-up corner of each panel, the pie chart illustrates the percentages of MCs in this sample with one $^{13}$CO molecular structure, 
    two $^{13}$CO molecular structures, multiple (larger than two) $^{13}$CO molecular structures, respectively.
    \label{fig:f13co_coms_morph}}
\end{figure*}

\subsubsection{The $^{13}$CO structures parameters and morphologies}
Furthermore, we focus on the properties of the $^{13}$CO molecular gas structures in the filaments 
and nonfilaments of the 2,851 $^{13}$CO-detects. 
Figure \ref{fig:fmorp_13co} compares the $^{13}$CO structures parameters of these filaments and nonfilaments. 
We find that the angular areas ($\mathcal{A}_{\rm ^{13}CO}$), velocity spans ($V_{\rm{span}, ^{13}CO}$) and integrated fluxes ($F_{\rm ^{13}CO}$) 
of $^{13}$CO structures in these filaments tend to be larger than that in these nonfilaments. 
While the number distributions of their peak intensities are similar. We find that filaments not only tend to have $^{13}$CO gas structures, but also 
their internal $^{13}$CO structures have larger angular sizes, velocity span, and integrated fluxes. 
That indicates the $^{12}$CO cloud classified as filaments gather more high-density gas structures in the local areas where 
both $^{12}$CO and $^{13}$CO emissions are detected. 

We also link the $^{13}$CO molecular structure counts in a single $^{12}$CO cloud 
to its morphology. Figure \ref{fig:f13co_coms_morph} illustrates the distribution of $^{13}$CO 
molecular structure counts within one $^{12}$CO cloud for the $^{13}$CO-detects classified as filaments and nonfilaments. 
We find that about 85$\%$ of these nonfilaments have one $^{13}$CO molecular structure and only 
3$\%$ of nonfilaments have more than two $^{13}$CO molecular structures. 
While for these filaments, those with one $^{13}$CO molecular structure only occupy about 35$\%$, 
and those having more than two $^{13}$CO molecular structures take about 44$\%$. Moreover, only filaments 
could harbor more than ten $^{13}$CO structures. That indicates the filament tend to have more separate $^{13}$CO molecular structures in its interior. 
That implies the development of dense gas content in filaments is separate and inhomogenous.

\section{Discussion}
\subsection{Comparison with previous works}

In the Milky Way, \cite{Torii2019} used the CO $J=1-0$ data from the FUGIN project and found that 
the ratio of the integrated intensities of $^{13}$CO and $^{12}$CO lines (W($^{13}$CO)/W($^{12}$CO)) 
along the Galactical longitude l = 10$^{\circ}$ -- 50$^{\circ}$ were distributed in a range of 1$\%$ -- 10$\%$. 
While the W($^{13}$CO)/W($^{12}$CO) in the star-forming cloud Orion B can achieve about 15$\%$ \citep{Gratier2021}. 
\cite{Roman-Duval2016} investigated the Galactic distribution of molecular gas 
components traced by $^{12}$CO and $^{13}$CO lines along the Galactic radius. 
The values of W($^{13}$CO)/W($^{12}$CO) have an approximately constant value of 20$\%$ out to 
the galactic radius of 6.5 kpc, decrease to $\sim$ 8$\%$ -- 10$\%$ in the solar neighborhood 
and about 5$\%$ -- 10$\%$ out to the radius of 14 kpc.   
%The average Galactic integrated intensity corresponds to the total luminosity in a Galactocentric 
%radius bin, divided by the surface area covered by the survey in that bin, projected onto the Galactic disk. 
%Igal is therefore the integrated intensity as seen from above the plane, averaged over radial bins of width 0.1 kpc.)

In the nearby galaxies, \cite{Cormier2018} carried out the CO observations for nine nearby spiral galaxies 
using IRAM 30-m telescope, which has a spatial resolution of $\sim$ 1.5 kpc, 
the resultant W($^{13}$CO)/W($^{12}$CO) ratio has a median value of $\sim$ 9$\%$ and varies by a factor of 2. 
For the five nearby star-forming galaxies,  
\cite{Gallagher2018} combined the IRAM $^{12}$CO(1-0) maps and ALMA observations of $^{13}$CO(1-0) lines and 
presented the distribution of the W($^{13}$CO)/W($^{12}$CO) along the radius, which 
have a mean value of 8.8$\%$ with a radius less than 1 kpc and 3.9$\%$ in a radius larger than 1 kpc. 
Furthermore, \cite{Mendez-Hernandez2020} presented the ALMA observations towards 27 low-redshift (0.02 $< z <$ 0.2) 
star-forming galaxies, their averaged value of W($^{13}$CO)/W($^{12}$CO) is 5.6$\pm$1$\%$ and varies by 
a factor of 2. 

\begin{figure*}
\plotone{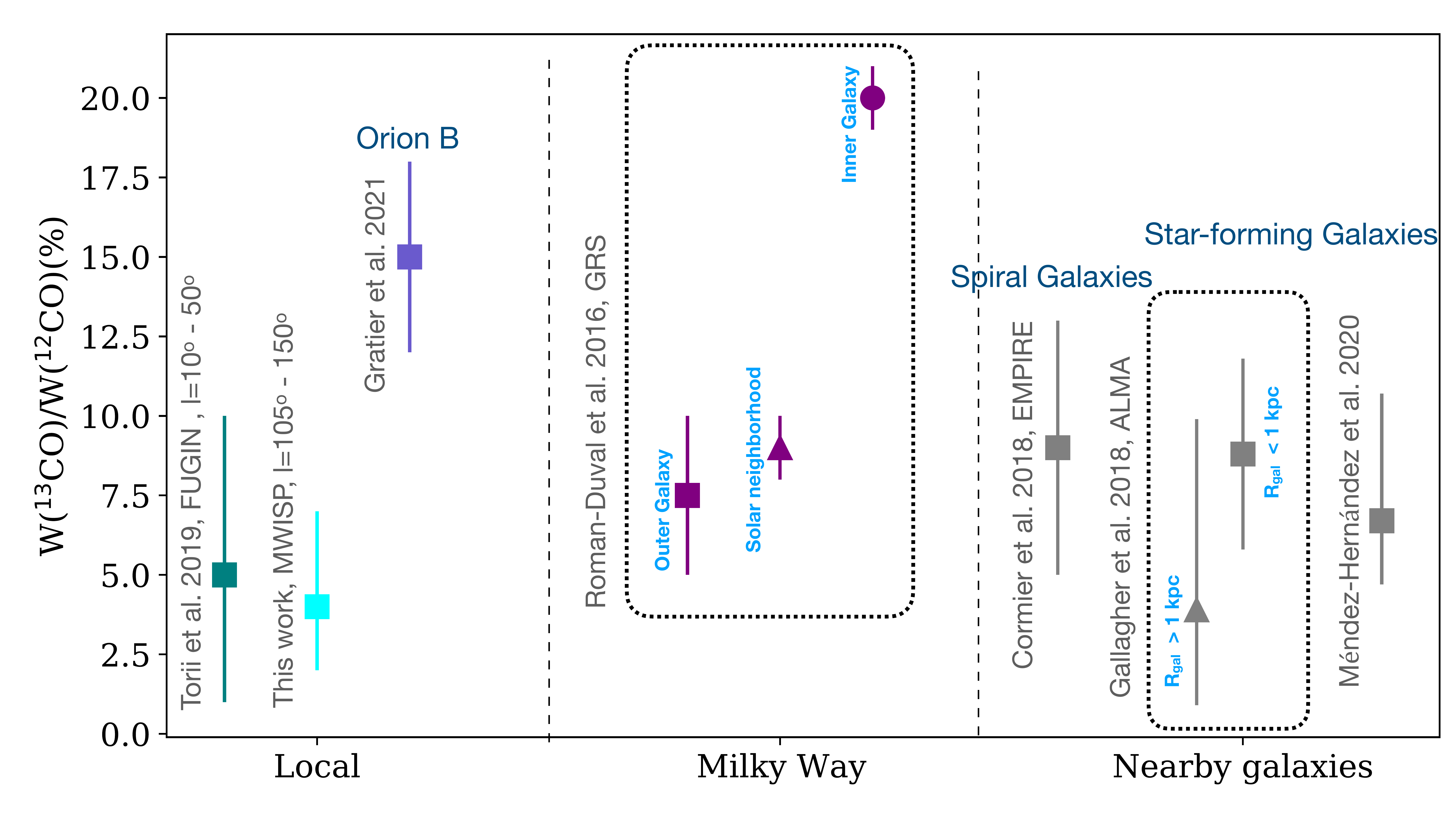}
\caption{Comparisons among the integrated intensity ratios between $^{13}$CO and $^{12}$CO lines (W($^{13}$CO)/W($^{12}$CO)). 
The colored points show the average or median values of W($^{13}$CO)/W($^{12}$CO). 
The error bars on the individual points reflect the distributions for the ratios. 
From left to right, we present the literature results of W($^{13}$CO)/W($^{12}$CO) in the local regions of Milky Way 
and in the nearby galaxies, respectively. 
The W($^{13}$CO)/W($^{13}$CO), which is presented in Figure 12 of \cite{Torii2019}, 
is calculated through the total velocity-integrated intensities of $^{13}$CO emission 
in a Galactic longitude of 1 degree divided by that of $^{12}$CO emission in the same bin. 
The presented values of this work is the ratio of the total integrated intensities of $^{13}$CO emission to that of $^{12}$CO 
emission in each molecular cloud. The W($^{13}$CO)/W($^{13}$CO) for the Orion B is the mean integrated intensity of $^{13}$CO emission divided 
by that of $^{12}$CO emission, whose values are listed in Table 2 of \cite{Gratier2021}. 
The values of W($^{12}$CO) and W($^{12}$CO) in \cite{Roman-Duval2016, Cormier2018, Gallagher2018,Mendez-Hernandez2020} 
are corresponds to their total luminosities in a Galactocentric radius bin, divided by the surface areas covered by the surveys in that bin, 
further projected onto the Galactic disk. \label{fig:fdense_gas}}
\end{figure*}

Figure \ref{fig:fdense_gas} compares the above literature results with our results.
Overall, the W($^{13}$CO)/W($^{12}$CO) depends on not only the molecular cloud conditions but also their positions in the galaxy.  
Moreover,  the W($^{13}$CO)/W($^{12}$CO) in the local molecular clouds with active star formation rates is higher.   

%Figure \ref{fig:fpre} presents the distribution of the W($^{13}$CO)/W($^{12}$CO) in the Milky Way and galaxies. 

\subsection{Implications of molecular clouds formation and evolution}
The questions of how molecular clouds form and what mechanisms determine their physical properties still remain open.   
Several mechanisms have been invoked to explain the gathering mass of molecular clouds. The agglomeration 
of smaller clouds \citep{Oort1954, Field1965, Kwan1983, Tomisaka1984}, the turbulence flows in the diffuse ISM 
\citep{Vazquez1995, Passot1995, Ballesteros1999}, and the large-scale gravitational instability of the Galactic disk 
\citep{Lin1964, Roberts1969, Tasker2009}. 

From our observations of 18,190 molecular clouds using $^{12}$CO and $^{13}$CO lines. 
In terms of their morphologies, i.e., nonfilaments and filaments, filaments tend to have larger spatial scales. 
Whereas their averaged H$_{2}$ column densities do not vary significantly \citep{Yuan2021}. 
Furthermore, we find that $^{13}$CO gas emission determined by the its H$_{2}$ column density 
is primarily detected in the filaments. That indicates the filament gathers more mass on a global scale 
and meanwhile has local density enhancements where both $^{12}$CO and $^{13}$CO emissions are detected. 
In addition, the filament also tends to have more than one individual structure traced by $^{13}$CO lines in its interior. 
That implies the development of dense gas content in filaments is separate and inhomogenous.
The formation of filament often arises from the shock compression in the ISM \citep{Arzoumanian2018, Abe2021, Arzoumanian2022}. 
The shock compression may be caused by supersonic turbulence in the molecular clouds \citep{Padoan1999, Pudritz2013, Matsumoto2015}, 
cloud collisions \citep{Inoue2013, Inoue2018, Tokuda2019}, feedback from massive stars, and galactic spiral shock. 
While the supercritical filaments may be driven by the gravitational contraction/accretion \citep{Arzoumanian2013,Gong2021,Yuan2020} and 
further fragment into smaller components owning to turbulence and gravitational instabilities \citep{Hacar2011, Henshaw2016, Kainulainen2017, Lu2018, Lin2019,Yuan2019}. 

We try to investigate the relation between the filaments and nonfilaments. 
If the filaments fragment into nonfilaments due to the gravitational instability, 
the high-density gas fraction in nonfilaments should be comparable to that of filaments. 
That is unlikely owing to our observational results of nonfilaments with less dense gas.
Our observed properties of filaments and nonfilaments favor 
that molecular clouds be explained as the density fluctuations induced by the 
turbulent compression in the diffuse ISM and broken up by the combination of dynamical and thermal instabilities, 
like the physical processes of shear, rotation, cooling, and magnetic fields \citep{Parades1999, Koyama2002, Heitsch2006, Semadeni2006,
Beuther2020}. Filaments tend to be under shock compressions and nonfilaments tend to be in low-pressure 
environments. Meanwhile they present the different spatial scales and internal structures. 
In addition, we are not able to rule out the hypothesis that nonfilament collisions to form filaments to some degree. 
%The molecular clouds appear to be transient objects, rather than long-lived, 
%equilibrium structures. 
%That represents a brief molecular phase formed by the large-scale supersonic compressions in the diffuse warm 
%neutral medium. Their internal substructures of clumps traced by $^{13}$CO are also transient density 
%enhancements induced by the turbulent flows. The filaments under the large turbulent compression, and nonfilaments in low-pressure 
%environments. Thus they present the different spatial scales and internal structures. 

\section{Summary and conclusions}
We identify the $^{13}$CO gas structures in the 18,190 $^{12}$CO molecular clouds and systematically compare the 
physical properties of $^{12}$CO clouds having $^{13}$CO gas structures ($^{13}$CO-detects) and 
those of $^{12}$CO clouds without $^{13}$CO gas structures (Non$^{13}$CO-detects). 
Furthermore, we systematically analyze the $^{13}$CO and $^{12}$CO emission parameters in the 2851 $^{13}$CO-detects,  
%fractions and correlations of the $^{12}$CO and $^{13}$CO emission parameters in the $^{13}$CO-detects. 
%The parameters include their angular areas ($\mathcal{A}_{^{12}CO}$, $\mathcal{A}_{^{13}CO}$), velocity spans ($V_{\rm{span}, ^{12}CO}$, $V_{\rm{span}, ^{13}CO}$), 
%peak intensities ($T_{\rm{peak}, ^{12}CO}$, $T_{\rm{peak}, ^{13}CO}$), and integrated fluxes 
%($F_{^{12}CO}$, $F_{^{13}CO}$). 
and link the internal $^{13}$CO gas structures of each $^{12}$CO cloud with its morphology, i.e., filament or nonfilament. 
The main conclusions are as follows: 

1. In the whole sample of 18,190 $^{12}$CO molecular clouds, 
$\sim$ 15.7$\%$ $^{12}$CO clouds (2851) have the $^{13}$CO molecular gas structures. 
The total integrated fluxes of $^{12}$CO line emission for the $^{13}$CO-detects are about 93$\%$ of 
that for the whole sample of molecular clouds.   

2. In the 2851 $^{13}$CO-detects, 
we find the $^{13}$CO structures' area in a $^{12}$CO cloud generally does not exceed 70$\%$ of the $^{12}$CO emission area, 
independently of the $^{12}$CO emission area, and its interior integrated fluxes of $^{13}$CO emission are usually less than 
20$\%$ of those of its $^{12}$CO emission. 

3. In the 2851 $^{13}$CO-detects, 
we find a strong correlation between the velocity-integrated intensities of $^{12}$CO lines and those of $^{13}$CO lines 
emission in the same areas where both the $^{12}$CO and $^{13}$CO emissions are detected. 

4. In the 2851 $^{13}$CO-detects, 
we find that there are $\sim$ 60$\%$ of $^{12}$CO clouds have one individual $^{13}$CO structure, 
about 15$\%$ of $^{12}$CO clouds have two separate $^{13}$CO structures, and the rest of them have more than two 
separate $^{13}$CO structures.

5. We link the $^{13}$CO gas fractions in the $^{13}$CO-detects with their morphologies, i.e., 
filaments or nonfilaments, and find that the $^{13}$CO line emissions are primarily detected in 
the $^{12}$CO clouds classified as filaments. In addition, a filament tends to have more than one individual $^{13}$CO structure in its interior. 

\begin{acknowledgments}
We gratefully thank the anonymous referee for the constructive comments that helped improve the quality of this paper.
This research made use of the data from the Milky Way Imaging Scroll Painting (MWISP) project, 
which is a multi-line survey in $^{12}$CO/$^{13}$CO/C$^{18}$O along the northern galactic plane with PMO-13.7m telescope. 
We are grateful to all of the members of the MWISP working group, particulaly to the staff members at the PMO-13.7m telescope, 
for their long-term support. This work was supported by the National Natural Science Foundation of China through grant 12041305. 
MWISP was sponsored by the National Key R\&D Program of China with grant 2017YFA0402701 
and the CAS Key Research Program of Frontier Sciences with grant QYZDJ-SSW-SLH047. 
\end{acknowledgments}

\software{Astropy \citep{astropy2013, astropy2018}, Matplotlib \citep{Hunter2007}}

\clearpage
\appendix
%\restartappendixnumbering
\renewcommand{\thefigure}{\Alph{section}\arabic{figure}}
\renewcommand{\theHfigure}{\Alph{section}\arabic{figure}}
\setcounter{figure}{0}
\section{$^{13}$CO($J=$ 1-0) lines data} 
\begin{figure*}
    \plotone{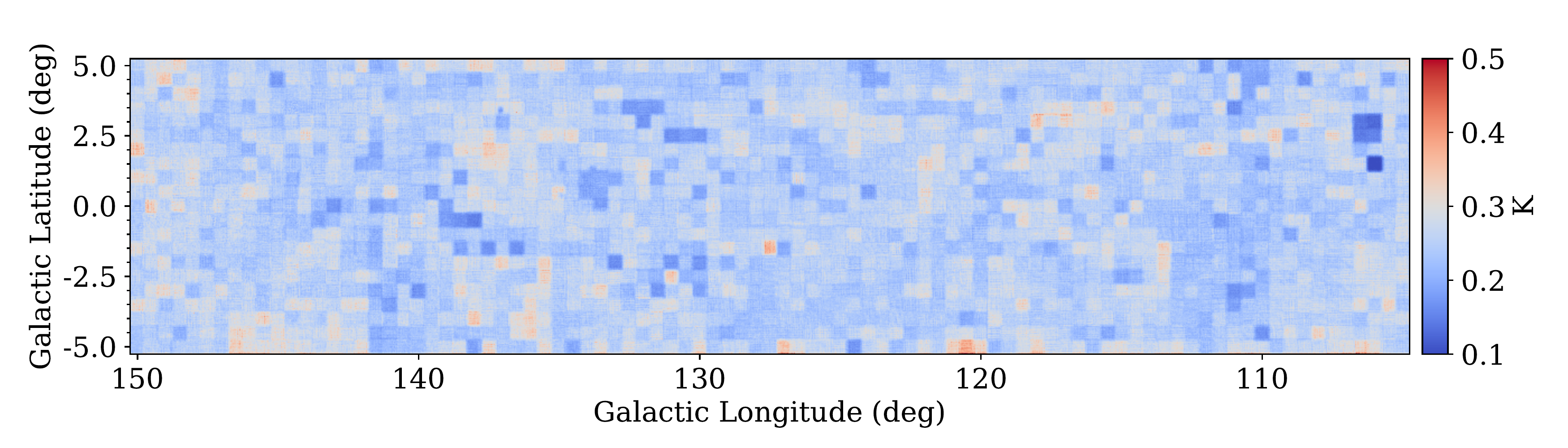}
    \caption{The distribution of the background noise RMS in the second Galactic quadrant with 
    104.75$^{\circ}$ $< l <$ 150.25$^{\circ}$ and $|b| <$ 5.25$^{\circ}$. \label{fig:fmaprms}}
\end{figure*}

\begin{figure*}
    \plotone{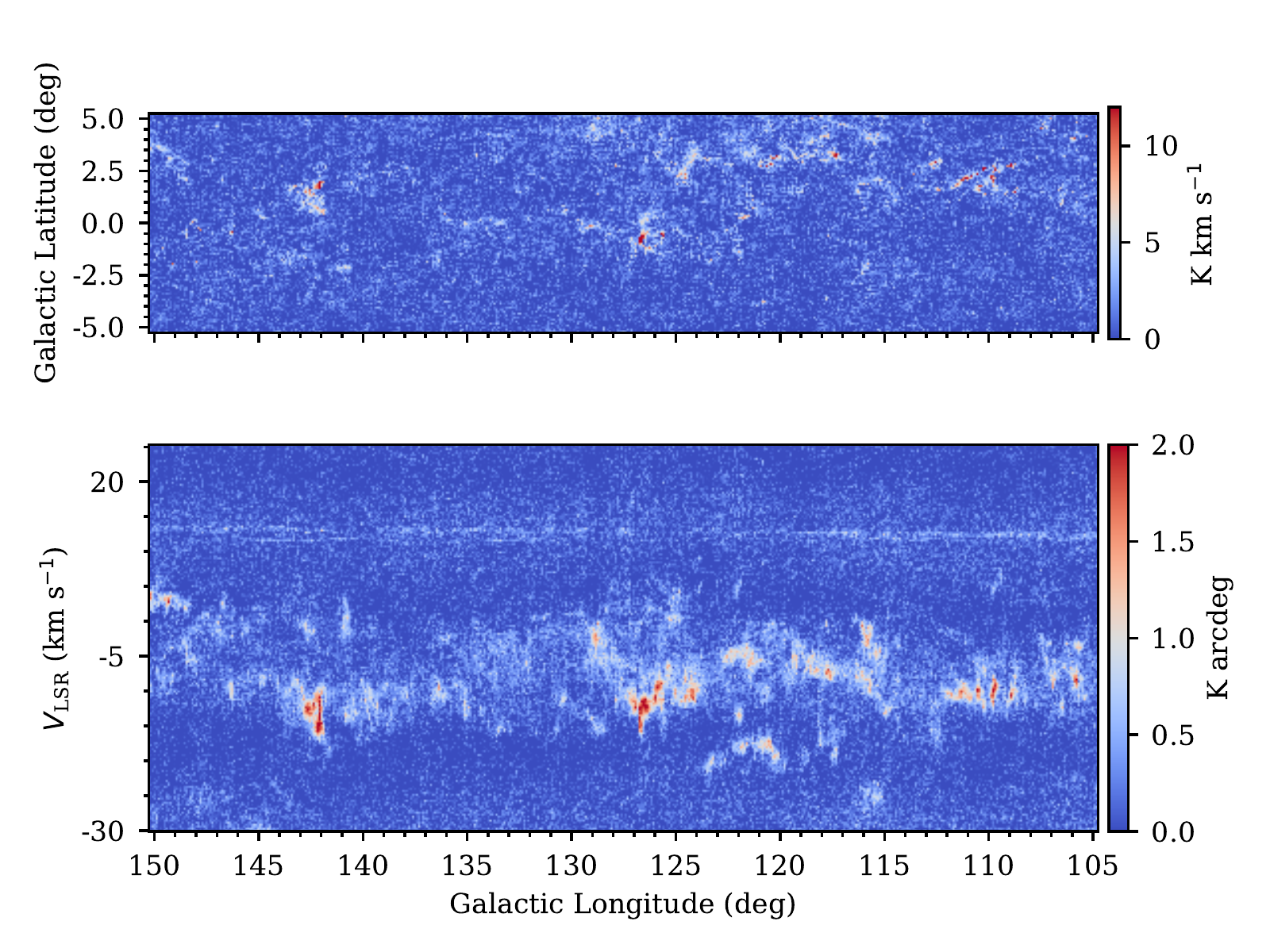}
    \caption{Top Panel: the velocity-integrated intensity map of $^{13}$CO(1-0) emission in the near group. 
    This map is derived by integrating the $^{13}$CO emission over the velocity range between $-$30 km s$^{-1}$ and 25 km s$^{-1}$.  
    The sensitivity for this velocity-integrated is about 0.77 K km s$^{-1}$. 
    Bottom Panel: the latitude-integrated intensity map of $^{13}$CO(1-0) emission in the near group. 
    This map is derived by integrating the $^{13}$CO emission over the latitude range from $-$5.25$^{\circ}$ to 5.25$^{\circ}$. \label{fig:fmap_near}}
\end{figure*}

\begin{figure*}
    \plotone{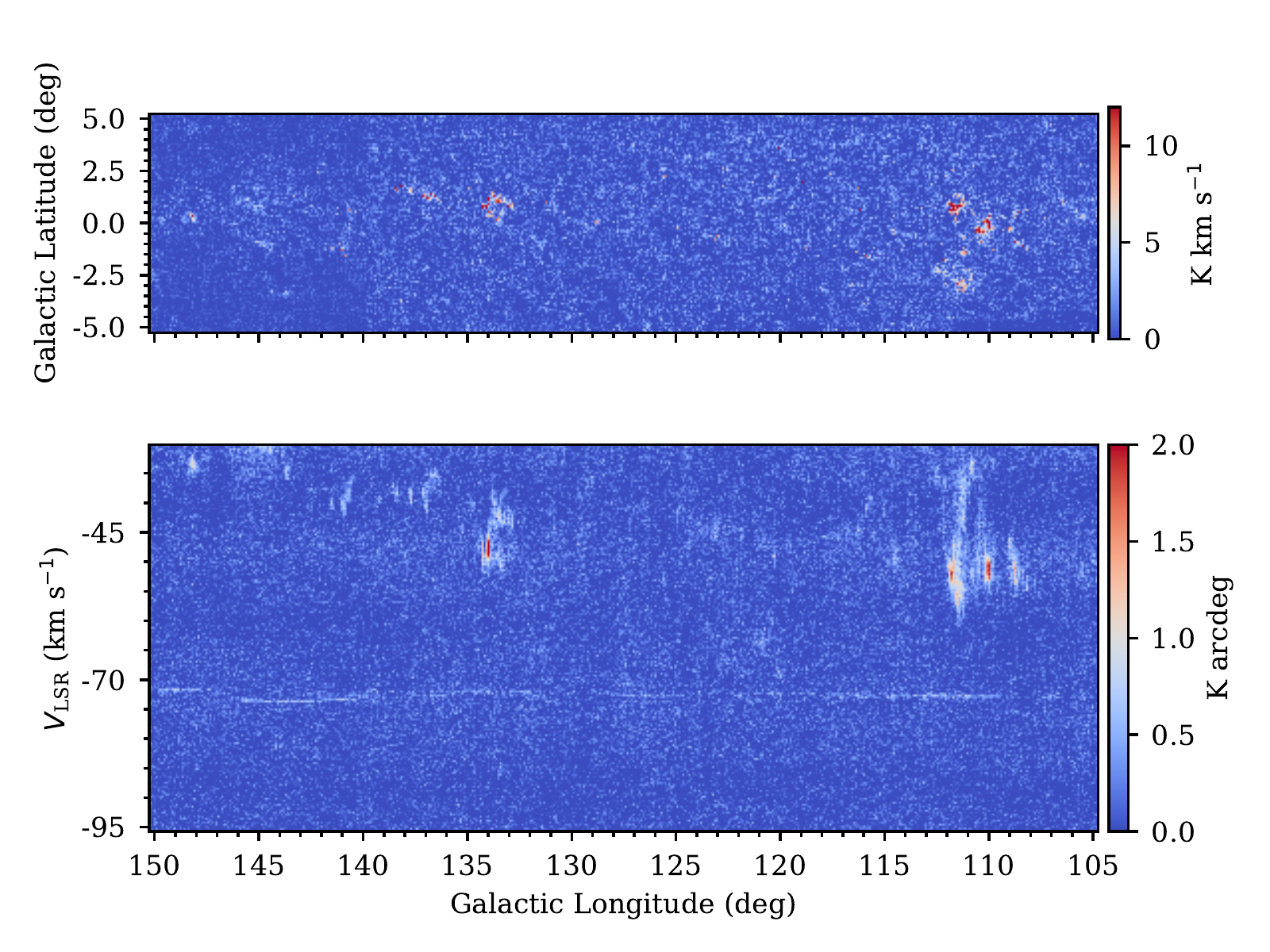}
    \caption{Top Panel: the velocity-integrated intensity map of $^{13}$CO(1-0) emission in the far group. 
    This map is derived by integrating the $^{13}$CO emission over the velocity range between $-$95 km s$^{-1}$ and $-$30 km s$^{-1}$.  
    The sensitivity for this velocity-integrated is about 0.83 K km s$^{-1}$. 
    Bottom Panel: the latitude-integrated intensity map of $^{13}$CO(1-0) emission in the far group. 
    This map is derived by integrating the $^{13}$CO emission over the latitude range from $-$5.25$^{\circ}$ to 5.25$^{\circ}$. \label{fig:fmap_far}}
\end{figure*}

\section{The parameters of DBSCAN algorithm}

The DBSCAN algorithm extracts the consecutive structures in the PPV space of CO lines data, 
based on a line intensity threshold and two parameters, i.e. $\epsilon$ and MinPts. 
Two parameters of $\epsilon$ and MinPts define the connectivity of structures in the PPV space. 
Each point within the extracted consecutive structure is called a core point. 
For a core point, its adjacent points contained in its neighborhood has to exceed a threshold. 
The parameter of MinPts determines the threshold of the number of adjacent points and 
the $\epsilon$ represents the radius of the neighborhood. 
A border point in the consecutive structure is defined as a point inside the $\epsilon$-neighborhood of a core point, 
but not necessarily contain the MinPts neighbors, as shown in Figure 2 of \citep{Ester1996}. 
In the PPV space of CO data, \cite{Yan2020} has examined all the choices of parameters 
and $^{12}$CO line intensities cutoffs to identify molecular clouds. 
The parameters of cutoff = 2$\sigma$, minPts = 4, $\epsilon$ = 1 are used in the DBSCAN algorithm 
to extract the $^{12}$CO molecular clouds in the $^{12}$CO data cube, as well as the 
$^{13}$CO molecular structures within the $^{12}$CO molecular cloud.
The post-selection criteria are examined and utilized to avoid the noise contamination \citep{Yan2020}. 
These criteria for a extracted structure include: (1) the minimum voxel number is 16; (2) the peak intensity 
is larger than the value of cutoff + 3$\sigma$ for $^{12}$CO or cutoff + 2$\sigma$ for $^{13}$CO; 
(3) the angular area is large than one beam size (2$\times$2 pixels); (4) the number of velocity 
channels are larger than 3.

\section{The $^{13}$CO structures of the molecular cloud G139.73 extracted by three methods} 
\begin{figure*}
    \plotone{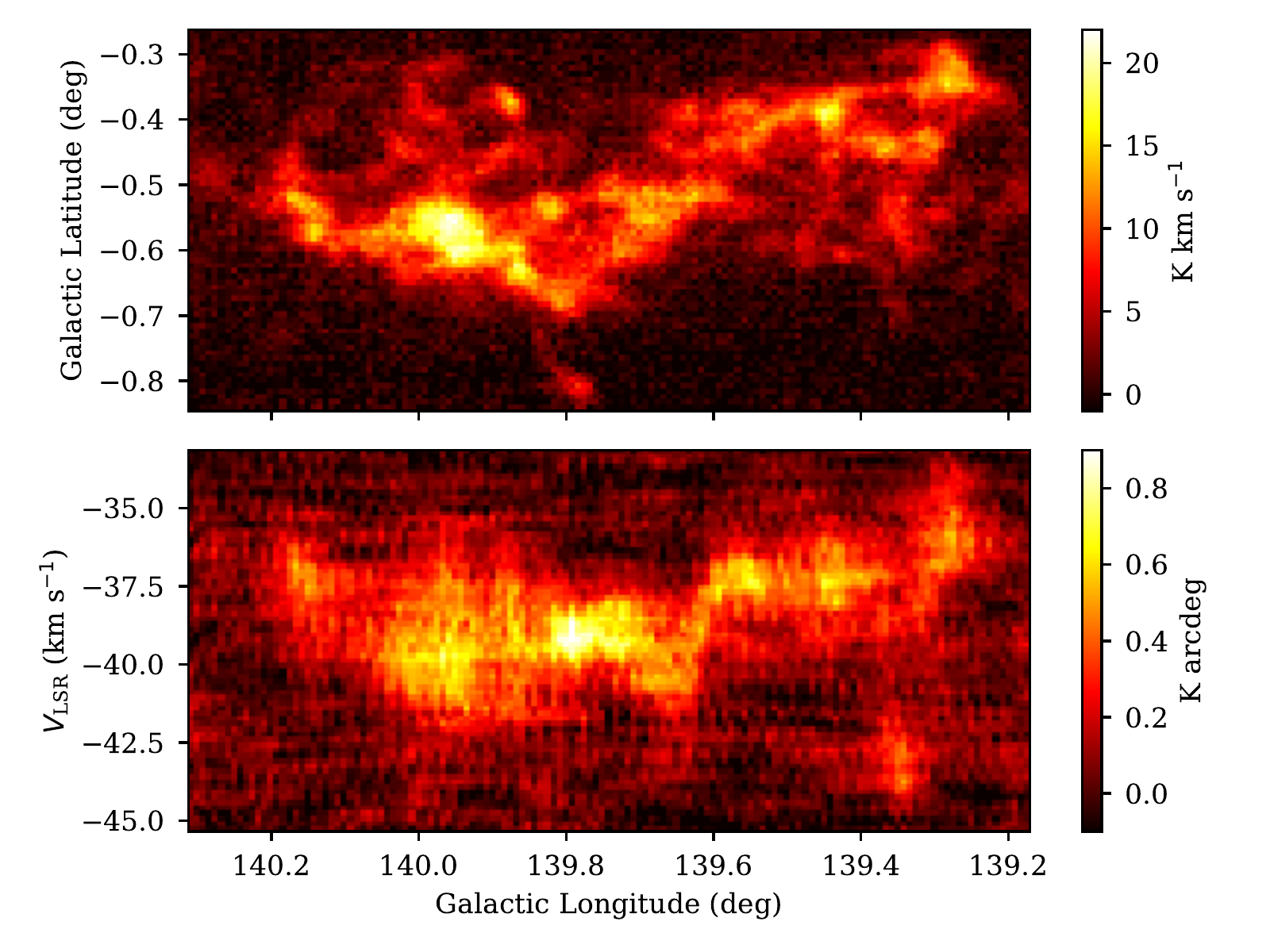}
    \caption{The velocity-integrated intensity map (\textbf{upper}) and latitude-integrated intensity map (\textbf{lower}) 
    of $^{12}$CO emission for the molecular cloud G139.73, which are derivied by the raw chopped $^{12}$CO data cube without any clipping. 
    \label{fig:fexample_raw12co}}
\end{figure*}

\begin{figure*}
    \plotone{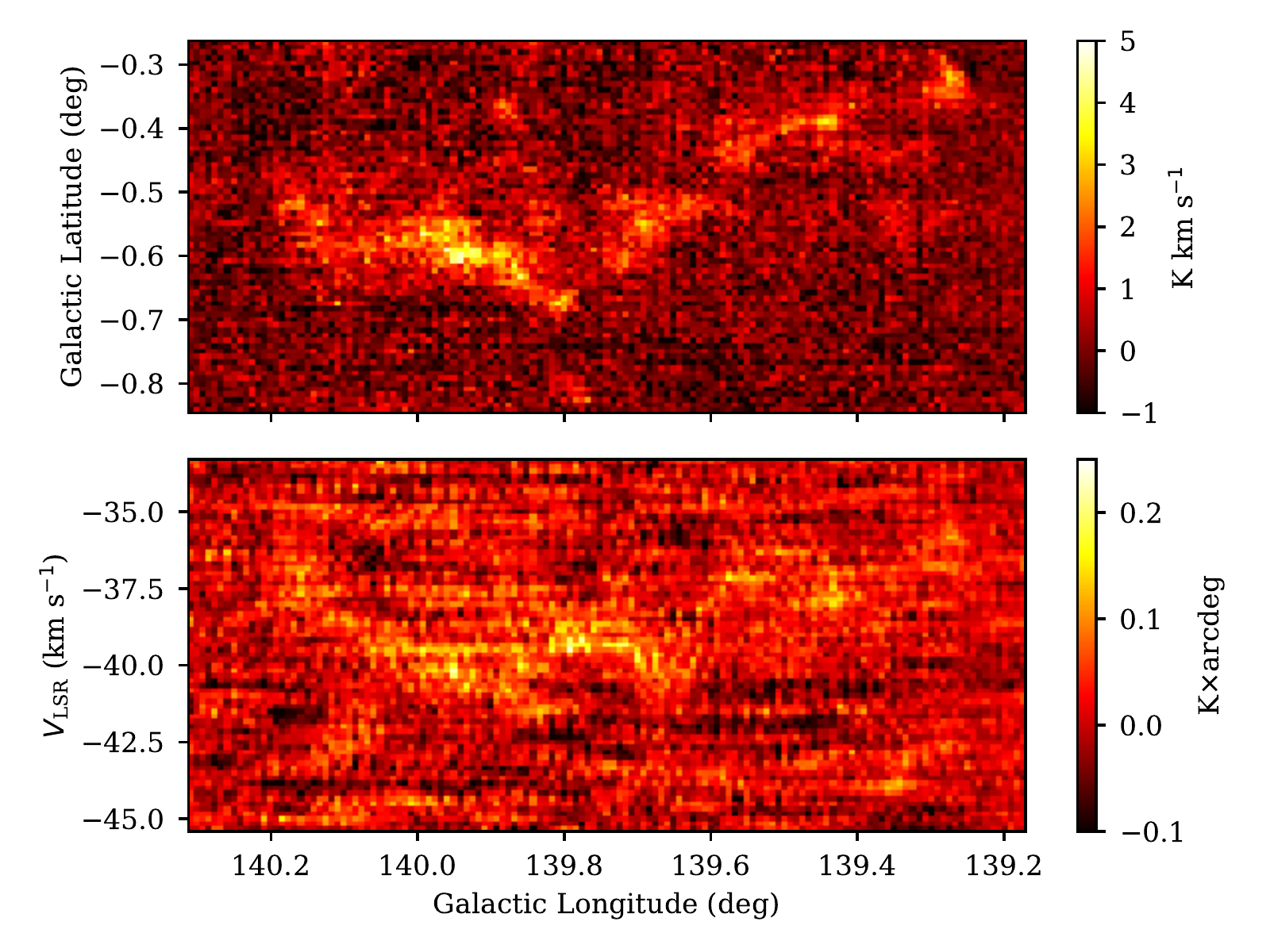}
    \caption{The velocity-integrated intensity map (\textbf{upper}) and latitude-integrated intensity map (\textbf{lower}) 
    of $^{13}$CO emission for the molecular cloud G139.73, which are derivied by the raw chopped $^{13}$CO data cube without any clipping. 
    \label{fig:fexample_raw13co}}
\end{figure*}

\begin{figure*}
    \plotone{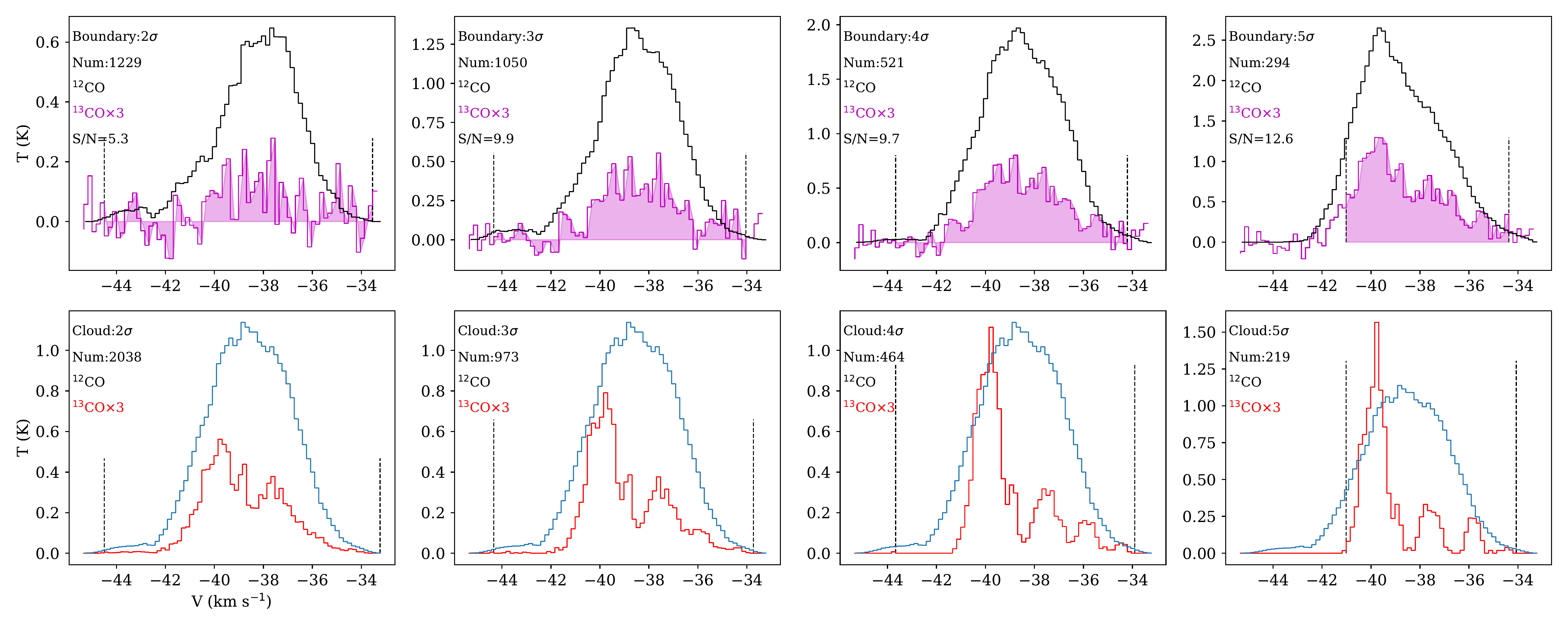}
    \caption{\textbf{Upper panels}: the mean spectrum of $^{12}$CO (black) and $^{13}$CO (magenta) spectral lines 
    along the boundaries of $^{13}$CO structures extracted in the molecular cloud G139.73 by the clipping at the cutoff level of 
    2$\sigma$, 3$\sigma$, 4$\sigma$, and 5$\sigma$, respectively. The $^{13}$CO spectral lines are from the raw chopped $^{13}$CO data cube without 
    any clipping, as shown in Figure \ref{fig:fexample_raw13co}. The $^{12}$CO spectral lines are from the extracted $^{12}$CO cloud of G139.73.
    The noted Num in each panel is the number of the spectrum along the corresponding boundary. 
    The S/N is the ratio of the peak intensity to the noise RMS for the averaged-boundary $^{13}$CO spectrum. 
    \textbf{Lower panels}: the mean spectrum of the extracted $^{12}$CO cloud (blue) in \cite{Yan2021} and the mean spectrum of the 
    $^{13}$CO structures (red) extracted by the clipping at the cutoff level of 2$\sigma$, 3$\sigma$, 4$\sigma$, and 5$\sigma$. 
    The vertical dashed lines illustrate the velocity span for the extracted $^{13}$CO structures within G139.73 by the 
    clipping at the corresponding cutoff levels of 2$\sigma$, 3$\sigma$, 4$\sigma$, and 5$\sigma$, respectively. 
    All the $^{13}$CO spectra are multiplied by a factor of 3. The $\sigma$ value is estimated from the raw data and equal to 0.27 K\label{fig:fedgeline_clip}}
\end{figure*}

\begin{figure*}
\plotone{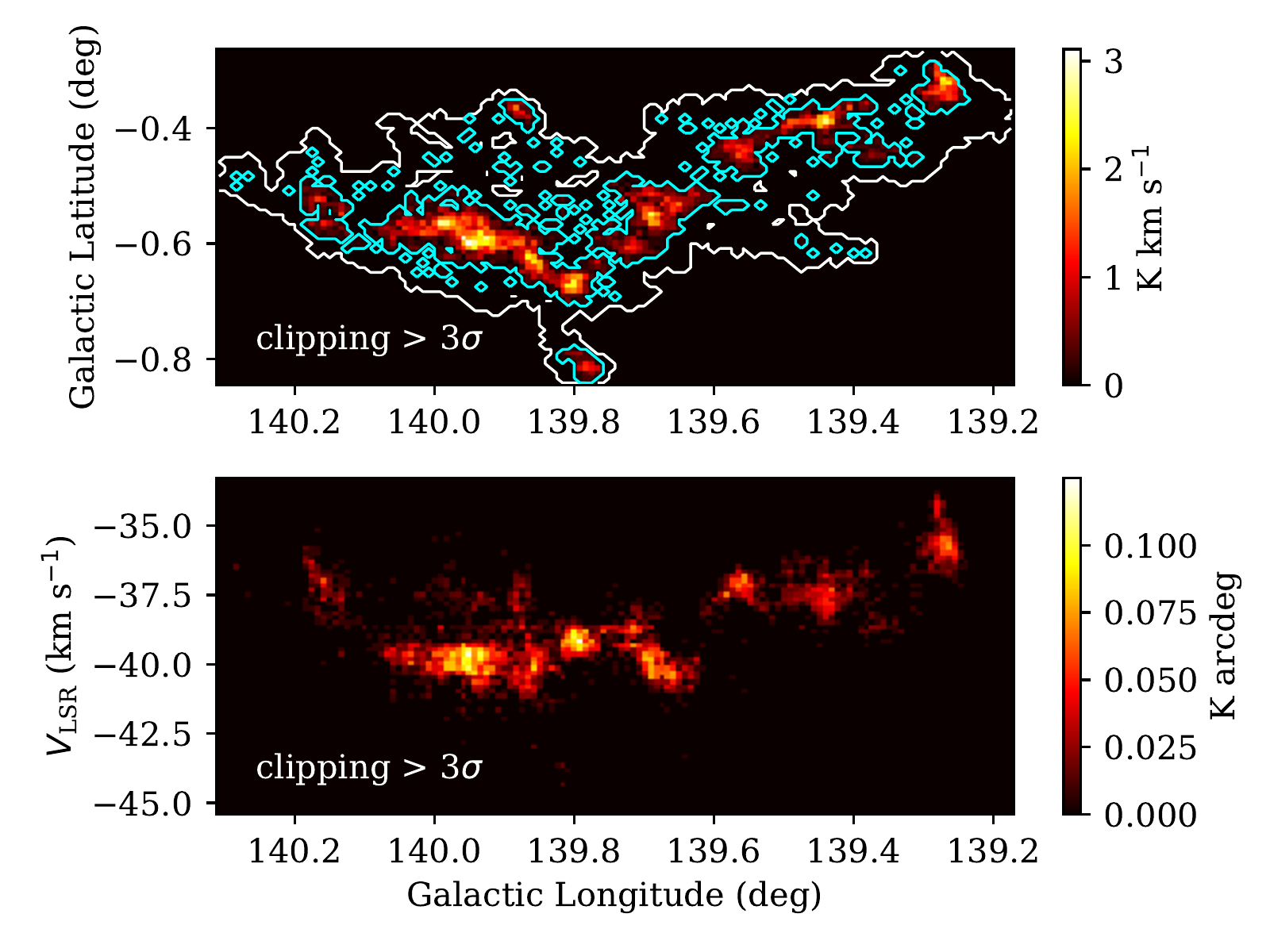}
\caption{The velocity-integrated intensity map (\textbf{upper}) and latitude-integrated intensity map (\textbf{lower}) 
of $^{13}$CO emission extracted within the $^{12}$CO cloud G139.73 using the clipping at the cutoff level of 3$\sigma$ (0.81 K). 
The white contours are the boundaries of $^{12}$CO molecular cloud. The cyan contours are the determined boundaries of $^{13}$CO line emission. \label{fig:fclip3}}
\end{figure*}

\begin{figure*}
\plotone{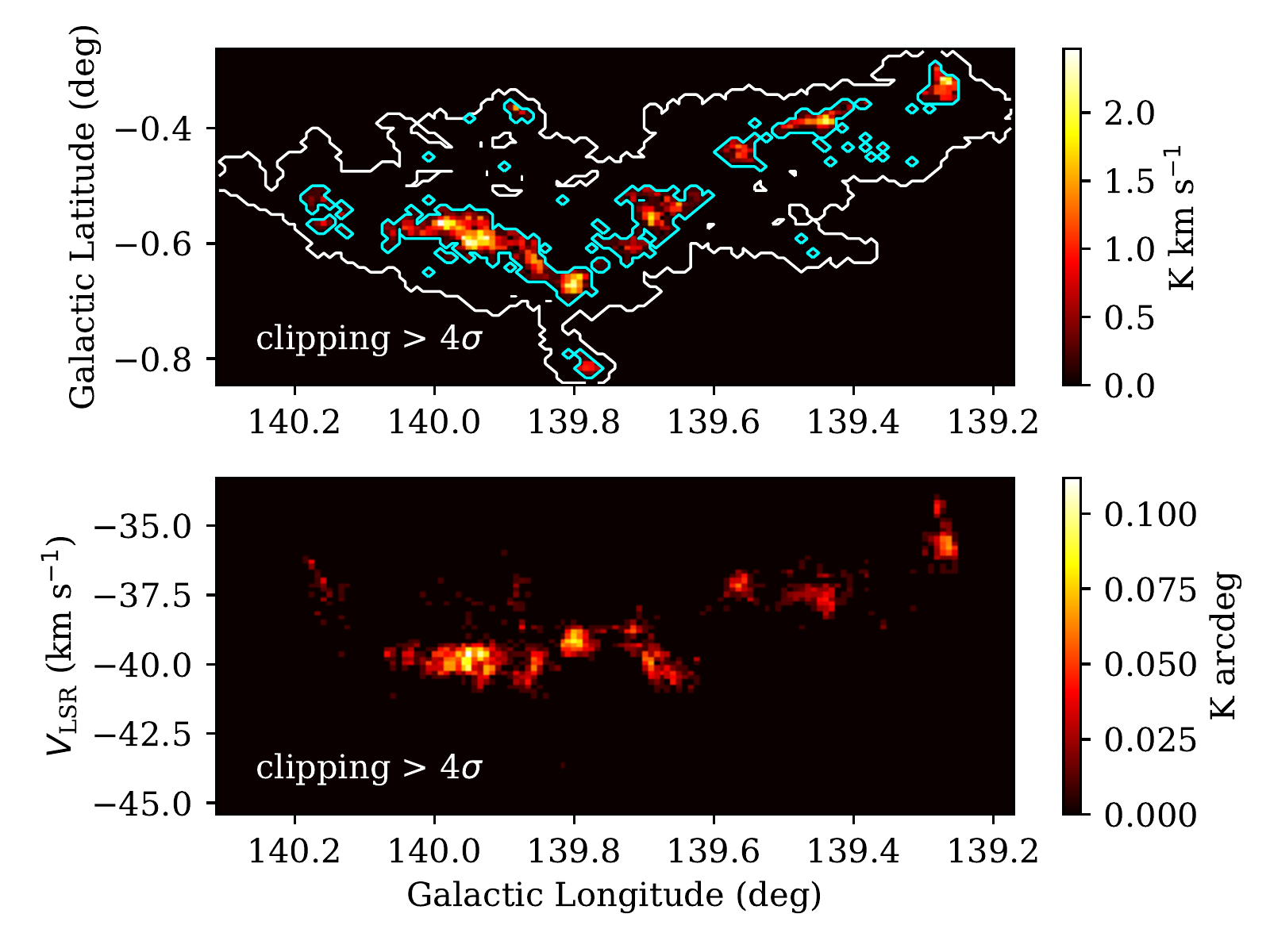}
\caption{Same as Figure \ref{fig:fclip3}, but using the clipping at the cutoff level of 4$\sigma$ (1.08 K). \label{fig:fclip4}}
\end{figure*}

\begin{figure*}
    \plotone{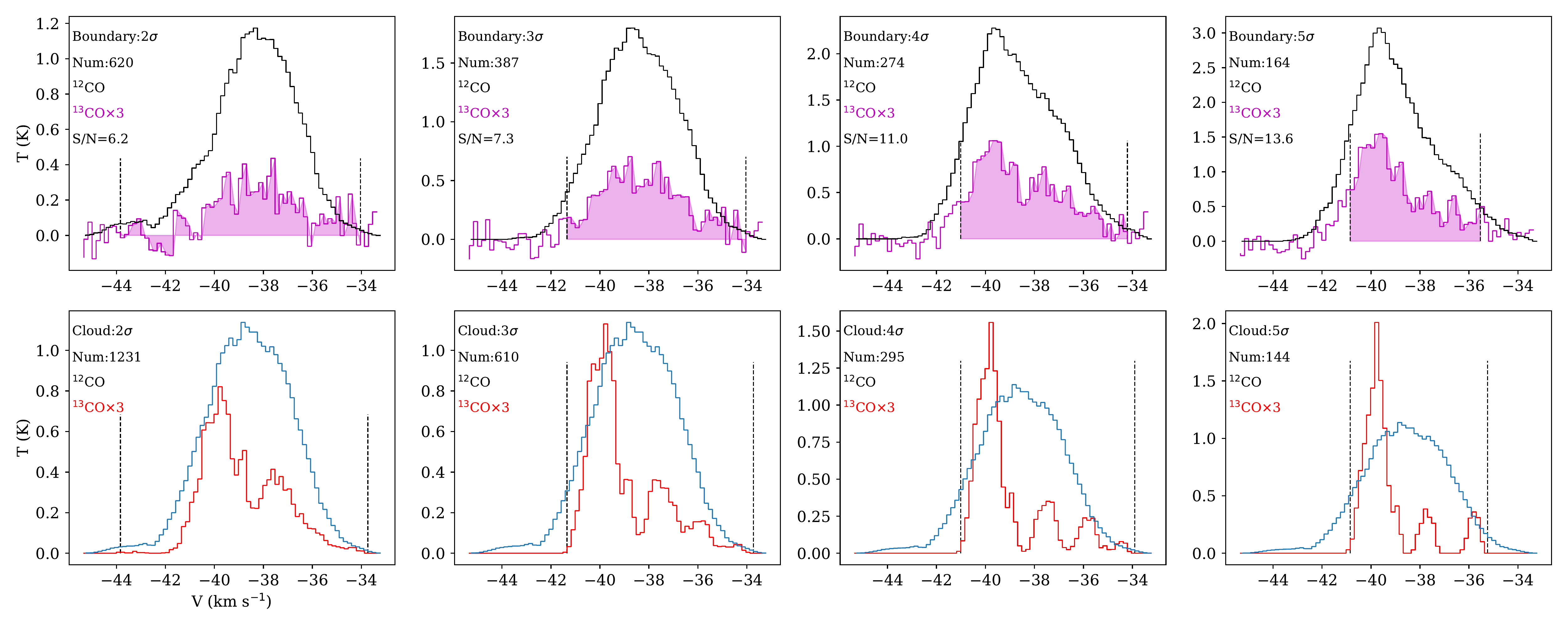}
    \caption{Same as Figure \ref{fig:fedgeline_clip}, but for the $^{13}$CO structures extracted by the 
    DBSCAN algorithm at the cutoff level of 2$\sigma$, 3$\sigma$, 4$\sigma$, and 5$\sigma$. \label{fig:fedgeline_dbscan}}
\end{figure*}

\begin{figure*}
\plotone{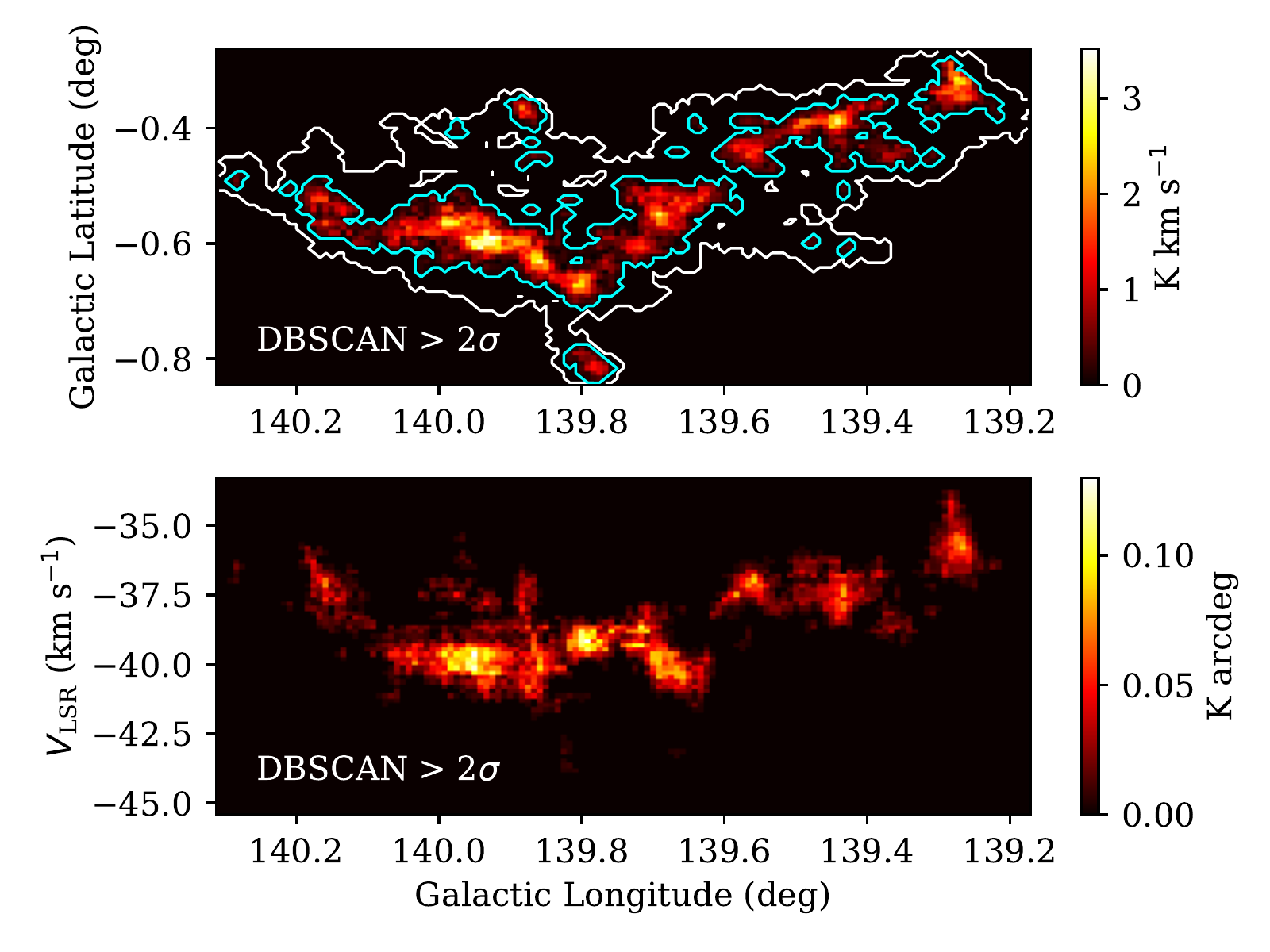}
\caption{Same as Figure \ref{fig:fclip3}, but using the DBSCAN at the cutoff level of 2$\sigma$ (0.54 K). \label{fig:fdbscan2}}
\end{figure*}

\begin{figure*}
\plotone{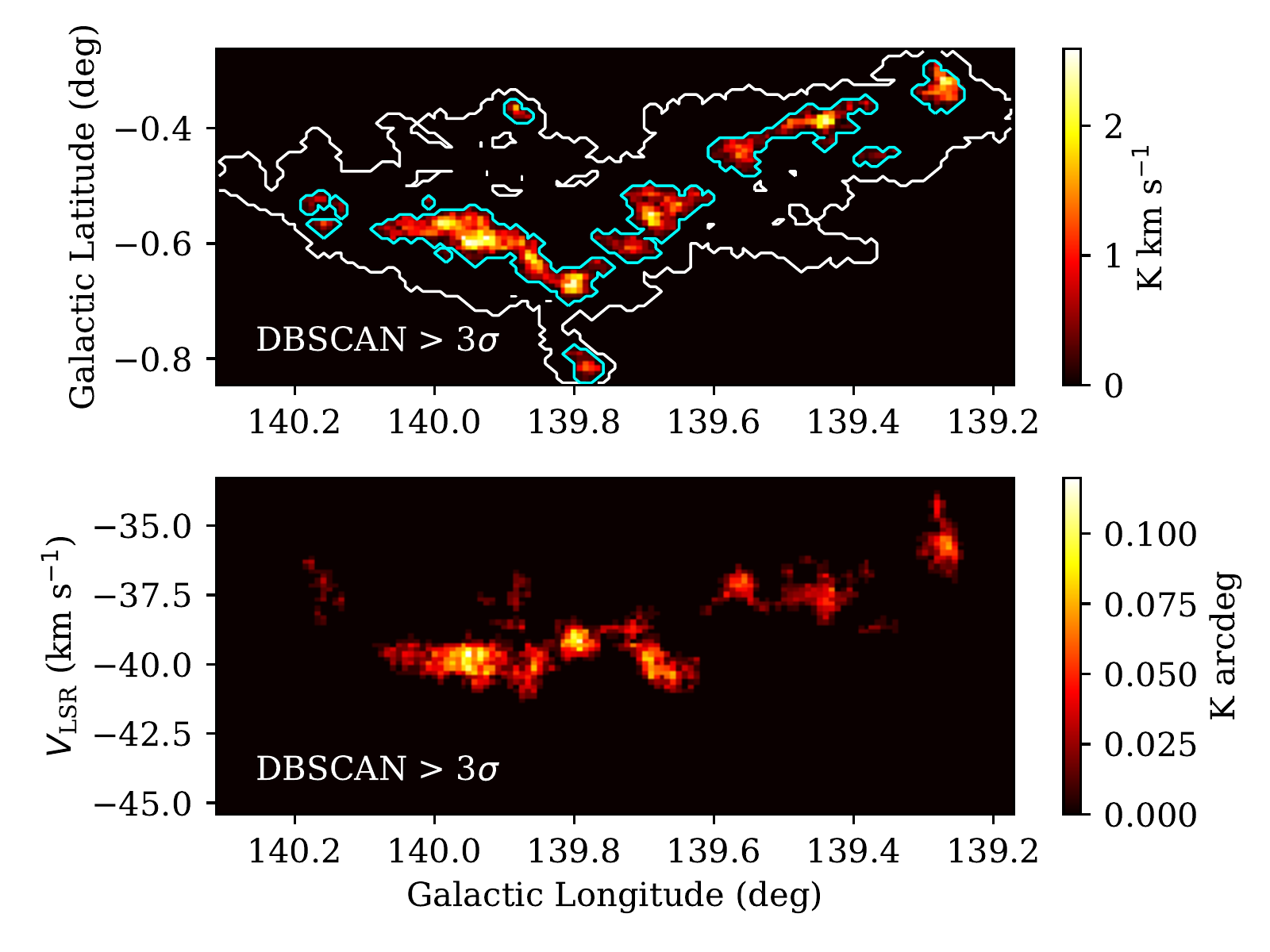}
\caption{Same as Figure \ref{fig:fclip3}, but using the DBSCAN at the cutoff level of 3$\sigma$ (0.81 K). \label{fig:fdbscan3}}
\end{figure*}

\begin{figure*}
    \plotone{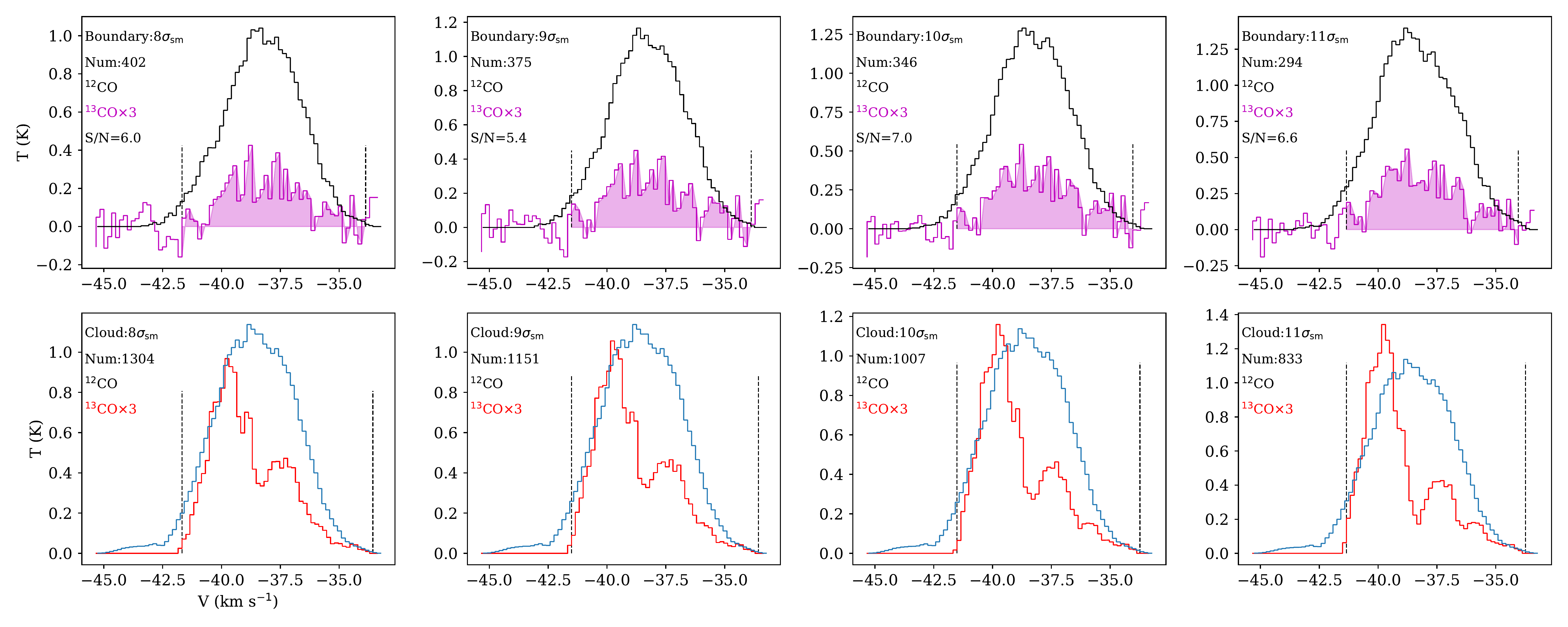}
    \caption{Same as Figure \ref{fig:fedgeline_clip}, but for the $^{13}$CO structures identified by the 
    moment mask at the cutoff level of 8$\sigma_{\rm sm}$, 9$\sigma_{\rm sm}$, 10$\sigma_{\rm sm}$ and 11$\sigma_{\rm sm}$. 
    The noise $\sigma_{\rm sm}$ is calculated from the smoothed data. Its value is 0.05 K. \label{fig:fedgeline_sm}}
\end{figure*}

\begin{figure*}
\plotone{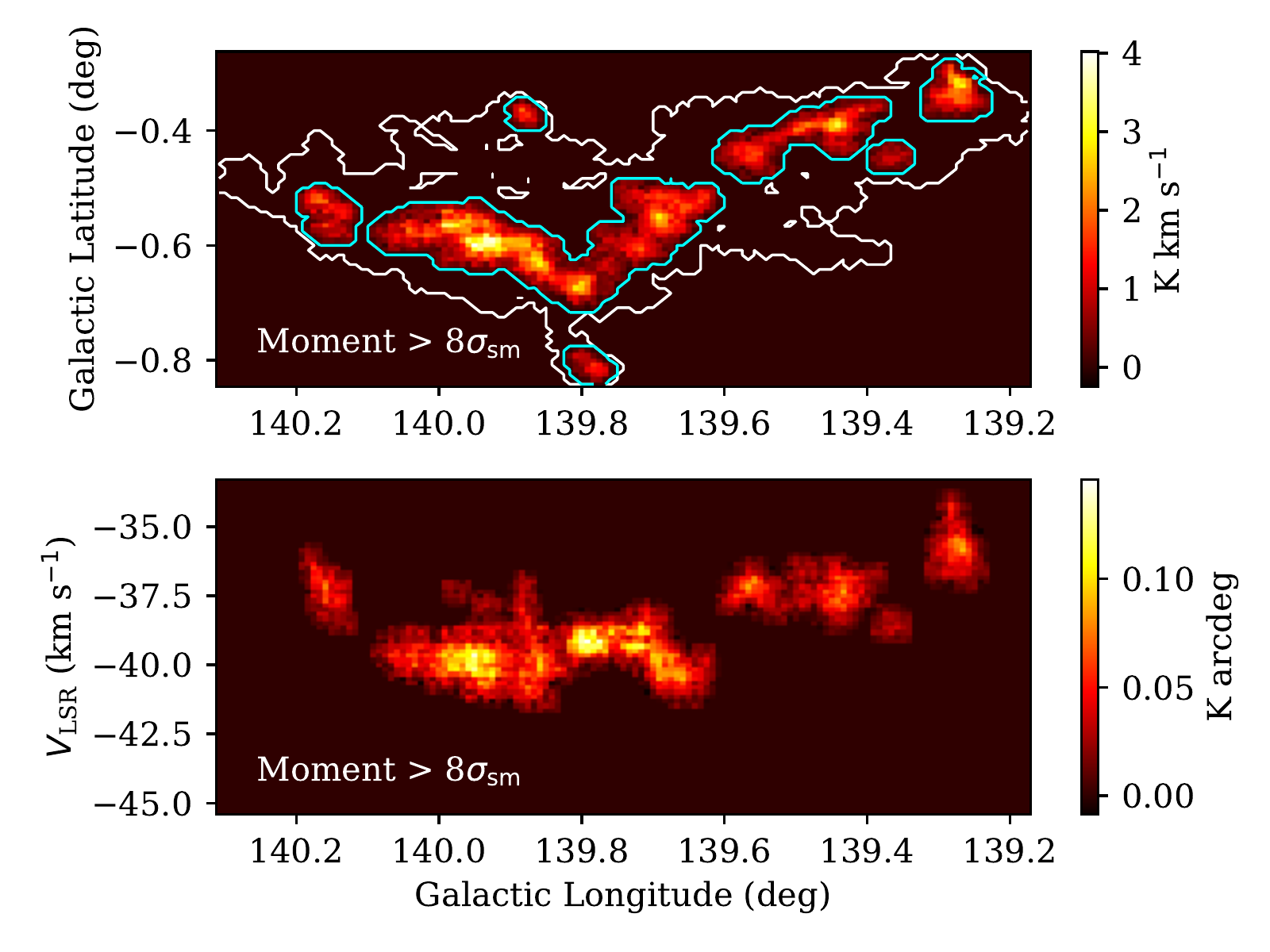}
\caption{Same as Figure \ref{fig:fclip3}, but using the moment mask at the cutoff level of 8$\sigma_{\rm sm}$ (0.4 K). \label{fig:fmom8}}
\end{figure*}

\begin{figure*}
\plotone{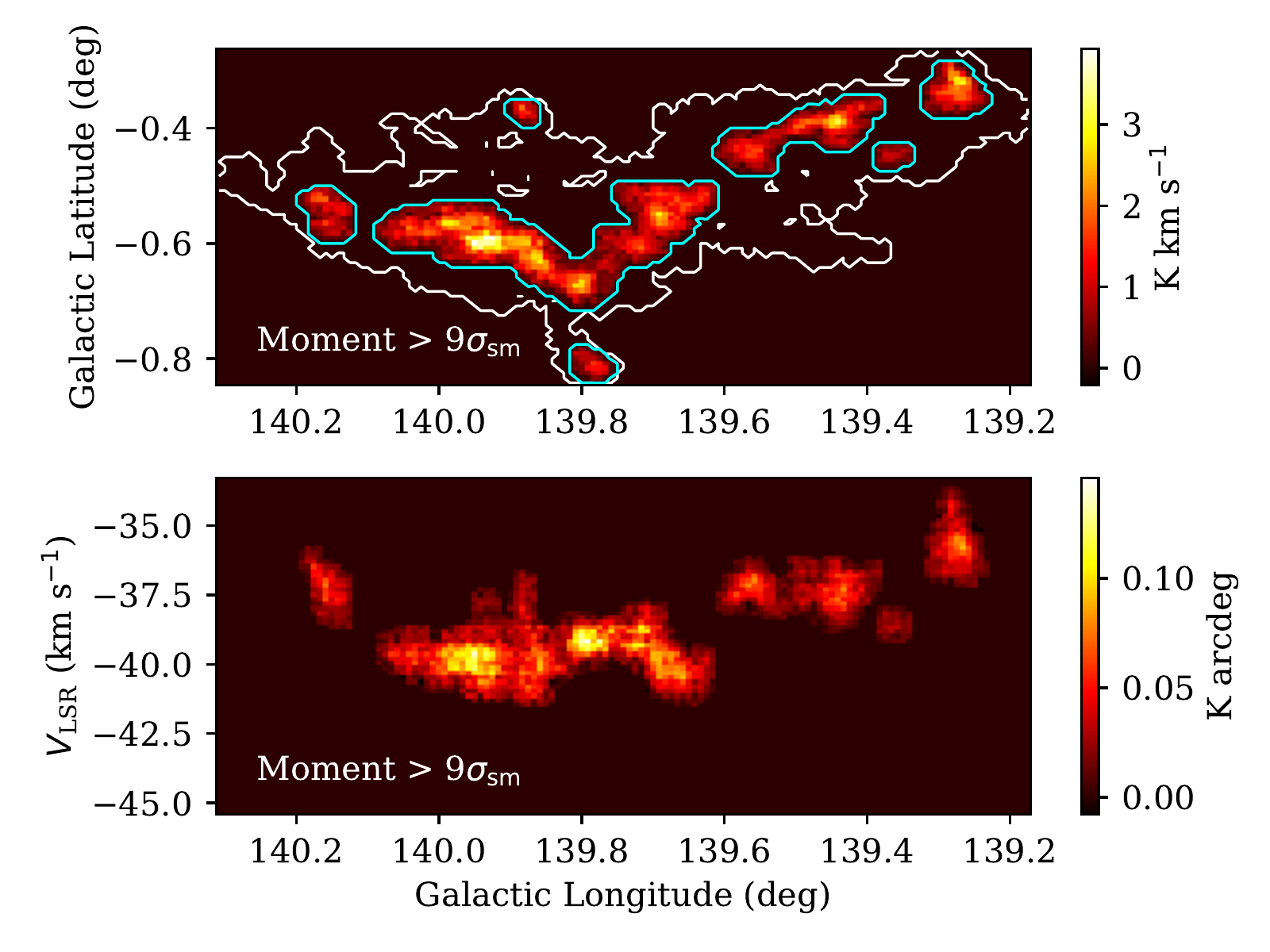}
\caption{Same as Figure \ref{fig:fclip3}, but using the moment mask at the cutoff level of 9$\sigma_{\rm sm}$ (0.45 K). \label{fig:fmom9}}
\end{figure*}

\section{Random examples of molecular clouds with $^{13}$CO structures}

\begin{figure*}
\plotone{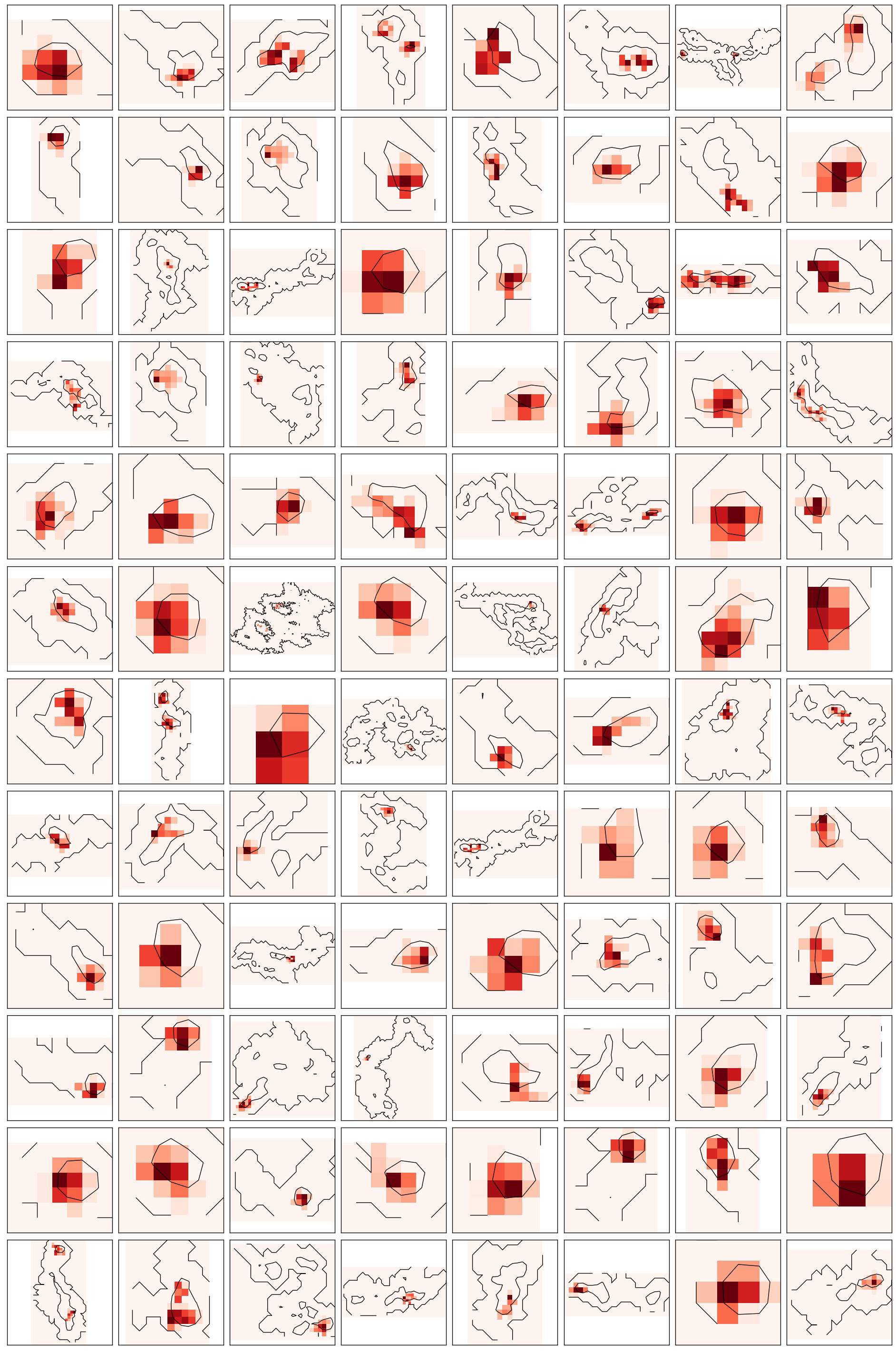}
\caption{$^{13}$CO structures identified by the DBSCAN, but not by the moment mask. The colormaps represent 
the distributions of the velocity-integrated intensities of $^{13}$CO line emission. The black contours 
indicate the boundaries and the level of 50$\%$ at the $^{12}$CO line velocity-integrated intensities. \label{fig:fdbscan_diff1}}
\end{figure*}

\begin{figure*}
\plotone{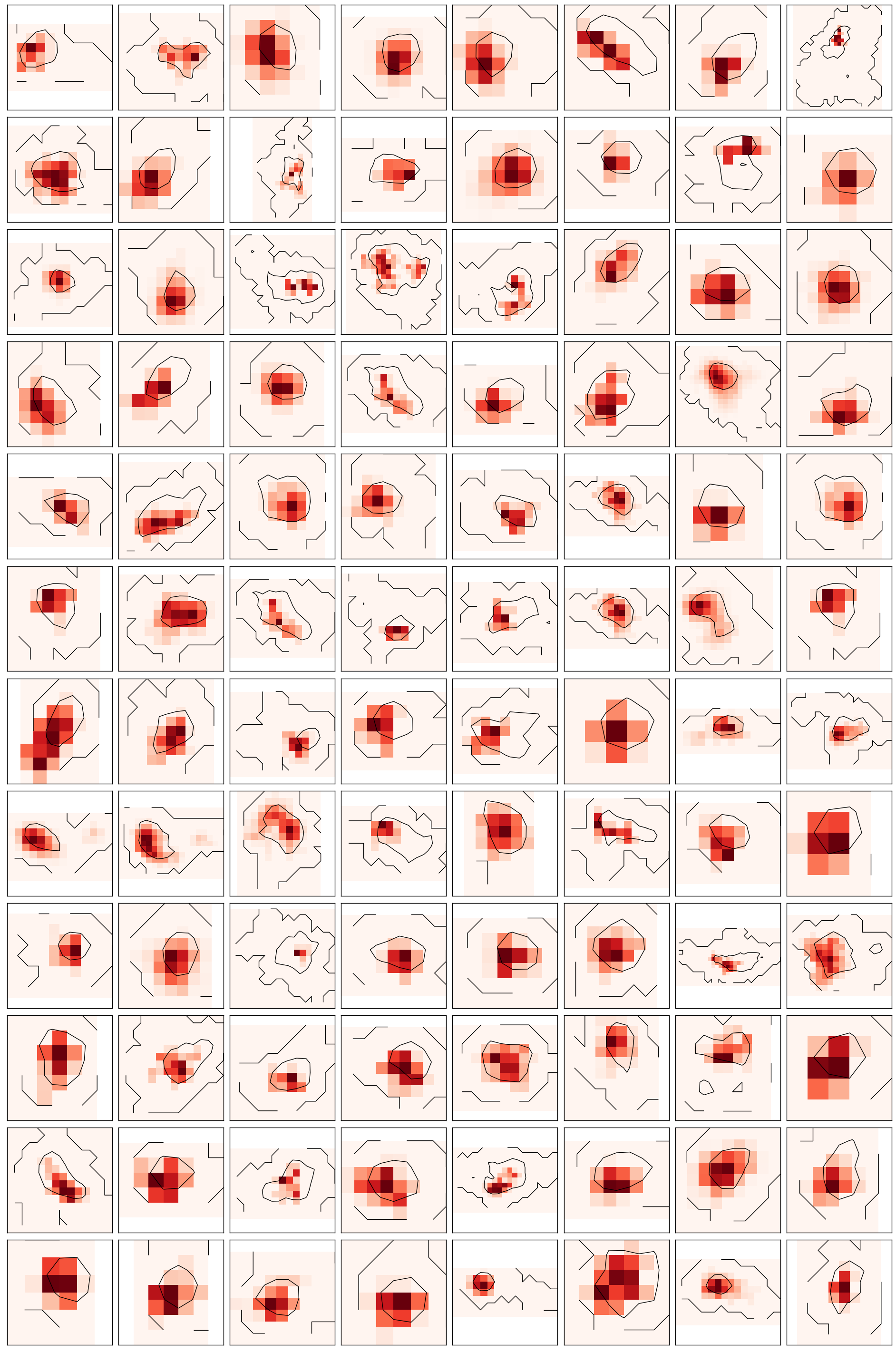}
\caption{$^{13}$CO structures identified by the DBSCAN. The colormaps represent 
the distributions of the velocity-integrated intensities of $^{13}$CO line emission. The black contours 
indicate the boundaries and the level of 50$\%$ at the $^{12}$CO line velocity-integrated intensities. \label{fig:fdbscan_clump}}
\end{figure*}

\begin{figure*}
\plotone{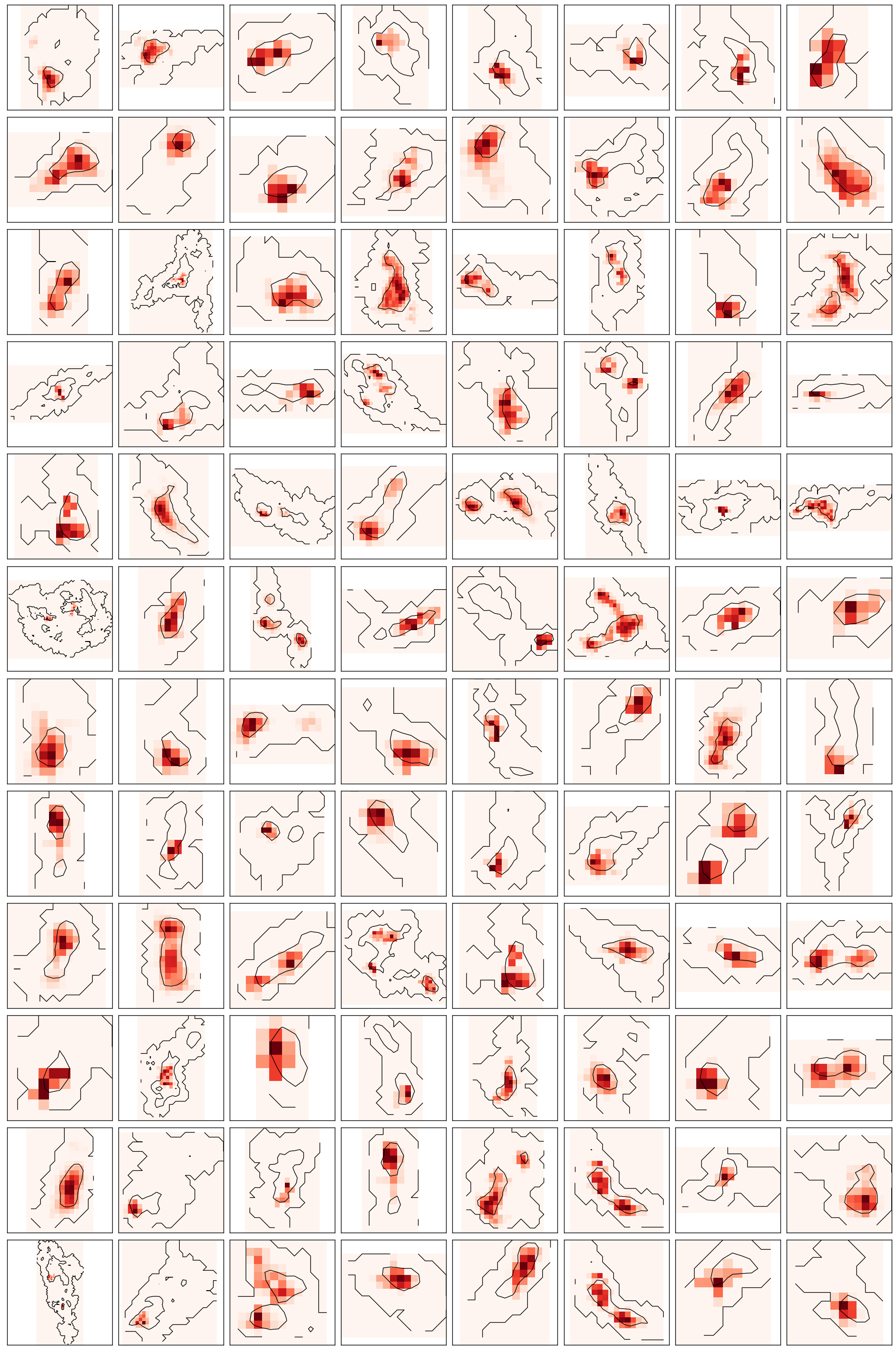}
\caption{Same as above. \label{fig:fdbscan_es}}
\end{figure*}

\begin{figure*}
\plotone{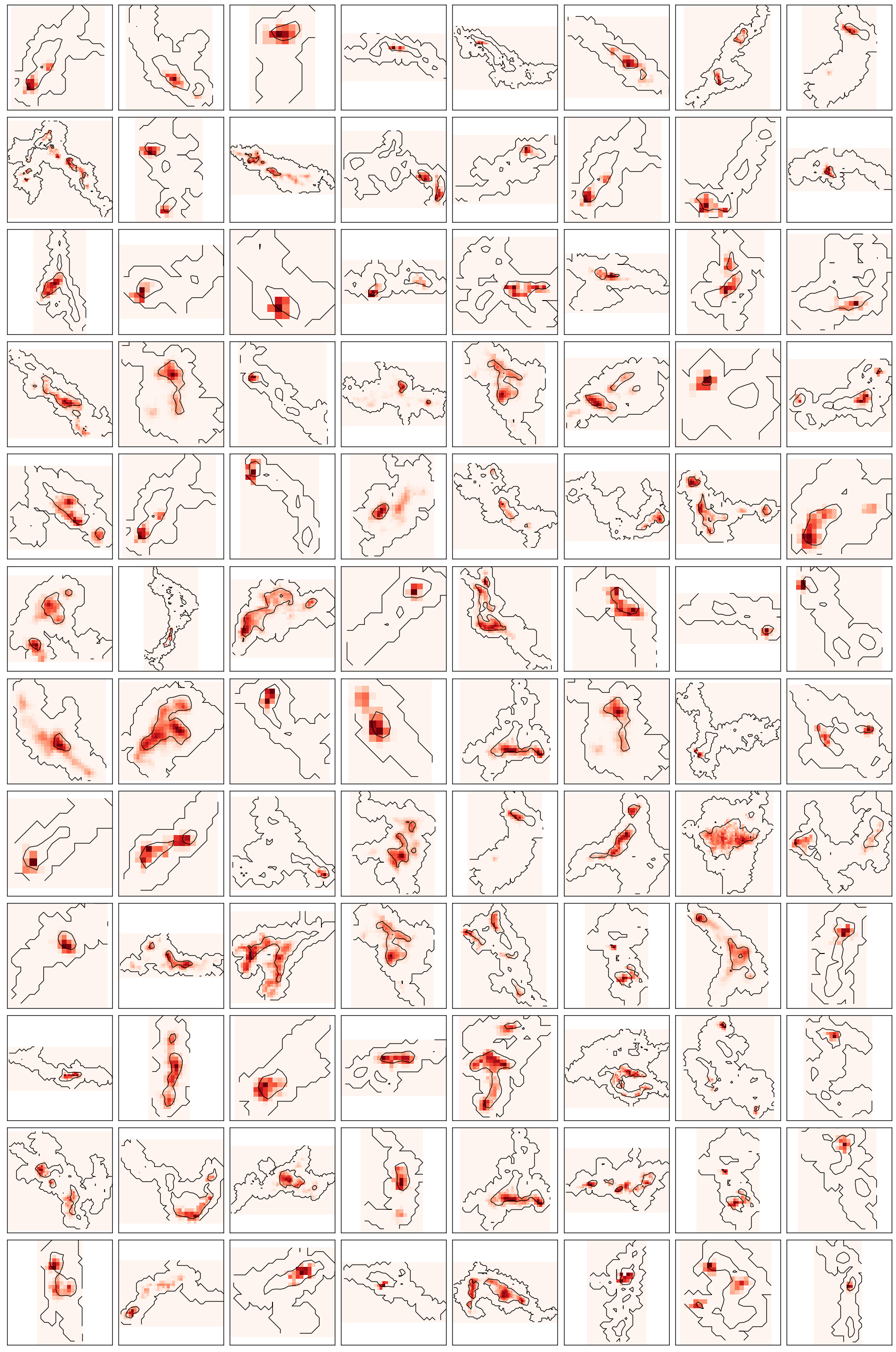}
\caption{Same as above. \label{fig:fdbscan_fil}}
\end{figure*}

\begin{figure*}
\plotone{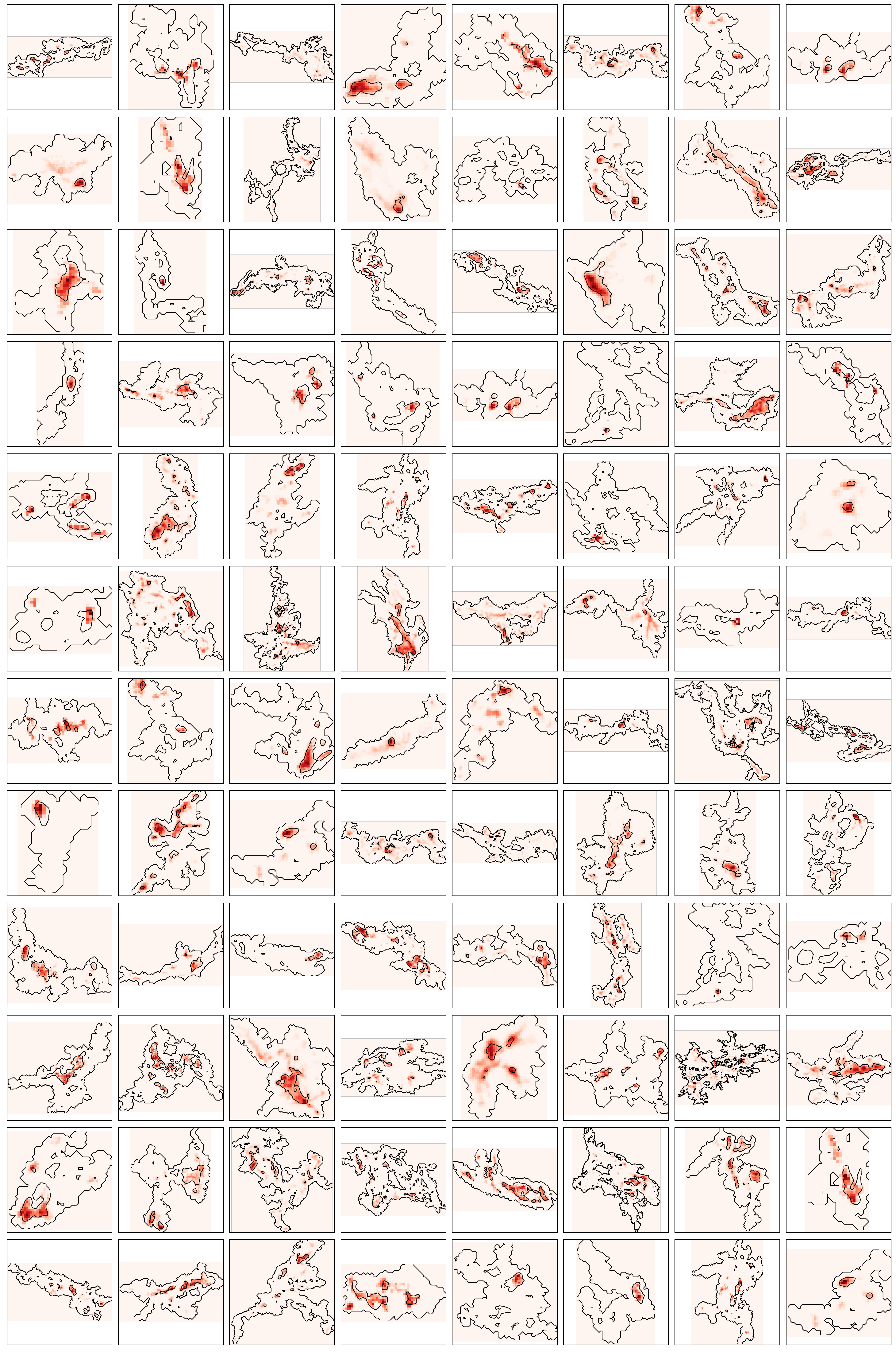}
\caption{Same as above. \label{fig:fdbscan_net}}
\end{figure*}

\clearpage

\bibliography{MC_13co.bib}

\begin{thebibliography}{}
\expandafter\ifx\csname natexlab\endcsname\relax\def\natexlab#1{#1}\fi
\providecommand{\url}[1]{\href{#1}{#1}}
\providecommand{\dodoi}[1]{doi:~\href{http://doi.org/#1}{\nolinkurl{#1}}}
\providecommand{\doeprint}[1]{\href{http://ascl.net/#1}{\nolinkurl{http://ascl.net/#1}}}
\providecommand{\doarXiv}[1]{\href{https://arxiv.org/abs/#1}{\nolinkurl{https://arxiv.org/abs/#1}}}

\bibitem[{{Abe} {et~al.}(2021){Abe}, {Inoue}, {Inutsuka}, \&
  {Matsumoto}}]{Abe2021}
{Abe}, D., {Inoue}, T., {Inutsuka}, S.-i., \& {Matsumoto}, T. 2021, \apj, 916,
  83, \dodoi{10.3847/1538-4357/ac07a1}

\bibitem[{{Andr{\'e}} {et~al.}(2014){Andr{\'e}}, {Di Francesco},
  {Ward-Thompson}, {Inutsuka}, {Pudritz}, \& {Pineda}}]{Andre2014}
{Andr{\'e}}, P., {Di Francesco}, J., {Ward-Thompson}, D., {et~al.} 2014, in
  Protostars and Planets VI, ed. H.~{Beuther}, R.~S. {Klessen}, C.~P.
  {Dullemond}, \& T.~{Henning}, 27,
  \dodoi{10.2458/azu\_uapress\_9780816531240-ch002}

\bibitem[{{Andr{\'e}} {et~al.}(2010){Andr{\'e}}, {Men'shchikov}, {Bontemps},
  {K{\"o}nyves}, {Motte}, {Schneider}, {Didelon}, {Minier}, {Saraceno},
  {Ward-Thompson}, {di Francesco}, {White}, {Molinari}, {Testi}, {Abergel},
  {Griffin}, {Henning}, {Royer}, {Mer{\'\i}n}, {Vavrek}, {Attard},
  {Arzoumanian}, {Wilson}, {Ade}, {Aussel}, {Baluteau}, {Benedettini},
  {Bernard}, {Blommaert}, {Cambr{\'e}sy}, {Cox}, {di Giorgio}, {Hargrave},
  {Hennemann}, {Huang}, {Kirk}, {Krause}, {Launhardt}, {Leeks}, {Le Pennec},
  {Li}, {Martin}, {Maury}, {Olofsson}, {Omont}, {Peretto}, {Pezzuto}, {Prusti},
  {Roussel}, {Russeil}, {Sauvage}, {Sibthorpe}, {Sicilia-Aguilar}, {Spinoglio},
  {Waelkens}, {Woodcraft}, \& {Zavagno}}]{Andre2010}
{Andr{\'e}}, P., {Men'shchikov}, A., {Bontemps}, S., {et~al.} 2010, \aap, 518,
  L102, \dodoi{10.1051/0004-6361/201014666}

\bibitem[{{Andr{\'e}} {et~al.}(2016){Andr{\'e}}, {Rev{\'e}ret}, {K{\"o}nyves},
  {Arzoumanian}, {Tig{\'e}}, {Gallais}, {Roussel}, {Le Pennec}, {Rodriguez},
  {Doumayrou}, {Dubreuil}, {Lortholary}, {Martignac}, {Talvard}, {Delisle},
  {Visticot}, {Dumaye}, {De Breuck}, {Shimajiri}, {Motte}, {Bontemps},
  {Hennemann}, {Zavagno}, {Russeil}, {Schneider}, {Palmeirim}, {Peretto},
  {Hill}, {Minier}, {Roy}, \& {Rygl}}]{Andre2016}
{Andr{\'e}}, P., {Rev{\'e}ret}, V., {K{\"o}nyves}, V., {et~al.} 2016, \aap,
  592, A54, \dodoi{10.1051/0004-6361/201628378}

\bibitem[{{Arzoumanian} {et~al.}(2013){Arzoumanian}, {Andr{\'e}}, {Peretto}, \&
  {K{\"o}nyves}}]{Arzoumanian2013}
{Arzoumanian}, D., {Andr{\'e}}, P., {Peretto}, N., \& {K{\"o}nyves}, V. 2013,
  \aap, 553, A119, \dodoi{10.1051/0004-6361/201220822}

\bibitem[{{Arzoumanian} {et~al.}(2018){Arzoumanian}, {Shimajiri}, {Inutsuka},
  {Inoue}, \& {Tachihara}}]{Arzoumanian2018}
{Arzoumanian}, D., {Shimajiri}, Y., {Inutsuka}, S.-i., {Inoue}, T., \&
  {Tachihara}, K. 2018, \pasj, 70, 96, \dodoi{10.1093/pasj/psy095}

\bibitem[{{Arzoumanian} {et~al.}(2022){Arzoumanian}, {Russeil}, {Zavagno},
  {Chen}, {Andr{\'e}}, {Inutsuka}, {Misugi}, {S{\'a}nchez-Monge}, {Schilke},
  {Men'shchikov}, \& {Kohno}}]{Arzoumanian2022}
{Arzoumanian}, D., {Russeil}, D., {Zavagno}, A., {et~al.} 2022, arXiv e-prints,
  arXiv:2201.04267.
\newblock \doarXiv{2201.04267}

\bibitem[{{Astropy Collaboration} {et~al.}(2013){Astropy Collaboration},
  {Robitaille}, {Tollerud}, {Greenfield}, {Droettboom}, {Bray}, {Aldcroft},
  {Davis}, {Ginsburg}, {Price-Whelan}, {Kerzendorf}, {Conley}, {Crighton},
  {Barbary}, {Muna}, {Ferguson}, {Grollier}, {Parikh}, {Nair}, {Unther},
  {Deil}, {Woillez}, {Conseil}, {Kramer}, {Turner}, {Singer}, {Fox}, {Weaver},
  {Zabalza}, {Edwards}, {Azalee Bostroem}, {Burke}, {Casey}, {Crawford},
  {Dencheva}, {Ely}, {Jenness}, {Labrie}, {Lim}, {Pierfederici}, {Pontzen},
  {Ptak}, {Refsdal}, {Servillat}, \& {Streicher}}]{astropy2013}
{Astropy Collaboration}, {Robitaille}, T.~P., {Tollerud}, E.~J., {et~al.} 2013,
  \aap, 558, A33, \dodoi{10.1051/0004-6361/201322068}

\bibitem[{{Astropy Collaboration} {et~al.}(2018){Astropy Collaboration},
  {Price-Whelan}, {Sip{\H{o}}cz}, {G{\"u}nther}, {Lim}, {Crawford}, {Conseil},
  {Shupe}, {Craig}, {Dencheva}, {Ginsburg}, {VanderPlas}, {Bradley},
  {P{\'e}rez-Su{\'a}rez}, {de Val-Borro}, {Aldcroft}, {Cruz}, {Robitaille},
  {Tollerud}, {Ardelean}, {Babej}, {Bach}, {Bachetti}, {Bakanov}, {Bamford},
  {Barentsen}, {Barmby}, {Baumbach}, {Berry}, {Biscani}, {Boquien}, {Bostroem},
  {Bouma}, {Brammer}, {Bray}, {Breytenbach}, {Buddelmeijer}, {Burke},
  {Calderone}, {Cano Rodr{\'\i}guez}, {Cara}, {Cardoso}, {Cheedella}, {Copin},
  {Corrales}, {Crichton}, {D'Avella}, {Deil}, {Depagne}, {Dietrich}, {Donath},
  {Droettboom}, {Earl}, {Erben}, {Fabbro}, {Ferreira}, {Finethy}, {Fox},
  {Garrison}, {Gibbons}, {Goldstein}, {Gommers}, {Greco}, {Greenfield},
  {Groener}, {Grollier}, {Hagen}, {Hirst}, {Homeier}, {Horton}, {Hosseinzadeh},
  {Hu}, {Hunkeler}, {Ivezi{\'c}}, {Jain}, {Jenness}, {Kanarek}, {Kendrew},
  {Kern}, {Kerzendorf}, {Khvalko}, {King}, {Kirkby}, {Kulkarni}, {Kumar},
  {Lee}, {Lenz}, {Littlefair}, {Ma}, {Macleod}, {Mastropietro}, {McCully},
  {Montagnac}, {Morris}, {Mueller}, {Mumford}, {Muna}, {Murphy}, {Nelson},
  {Nguyen}, {Ninan}, {N{\"o}the}, {Ogaz}, {Oh}, {Parejko}, {Parley}, {Pascual},
  {Patil}, {Patil}, {Plunkett}, {Prochaska}, {Rastogi}, {Reddy Janga},
  {Sabater}, {Sakurikar}, {Seifert}, {Sherbert}, {Sherwood-Taylor}, {Shih},
  {Sick}, {Silbiger}, {Singanamalla}, {Singer}, {Sladen}, {Sooley},
  {Sornarajah}, {Streicher}, {Teuben}, {Thomas}, {Tremblay}, {Turner},
  {Terr{\'o}n}, {van Kerkwijk}, {de la Vega}, {Watkins}, {Weaver}, {Whitmore},
  {Woillez}, {Zabalza}, \& {Astropy Contributors}}]{astropy2018}
{Astropy Collaboration}, {Price-Whelan}, A.~M., {Sip{\H{o}}cz}, B.~M., {et~al.}
  2018, \aj, 156, 123, \dodoi{10.3847/1538-3881/aabc4f}

\bibitem[{{Ballesteros-Paredes}
  {et~al.}(1999{\natexlab{a}}){Ballesteros-Paredes}, {V{\'a}zquez-Semadeni}, \&
  {Scalo}}]{Ballesteros1999}
{Ballesteros-Paredes}, J., {V{\'a}zquez-Semadeni}, E., \& {Scalo}, J.
  1999{\natexlab{a}}, \apj, 515, 286, \dodoi{10.1086/307007}

\bibitem[{{Ballesteros-Paredes}
  {et~al.}(1999{\natexlab{b}}){Ballesteros-Paredes}, {V{\'a}zquez-Semadeni}, \&
  {Scalo}}]{Parades1999}
---. 1999{\natexlab{b}}, \apj, 515, 286, \dodoi{10.1086/307007}

\bibitem[{{Beuther} {et~al.}(2020){Beuther}, {Wang}, {Soler}, {Linz},
  {Henshaw}, {Vazquez-Semadeni}, {Gomez}, {Ragan}, {Henning}, {Glover}, {Lee},
  \& {G{\"u}sten}}]{Beuther2020}
{Beuther}, H., {Wang}, Y., {Soler}, J., {et~al.} 2020, \aap, 638, A44,
  \dodoi{10.1051/0004-6361/202037950}

\bibitem[{{Bolatto} {et~al.}(2013){Bolatto}, {Wolfire}, \&
  {Leroy}}]{Bolatto2013}
{Bolatto}, A.~D., {Wolfire}, M., \& {Leroy}, A.~K. 2013, \araa, 51, 207,
  \dodoi{10.1146/annurev-astro-082812-140944}

\bibitem[{{Cormier} {et~al.}(2018){Cormier}, {Bigiel}, {Jim{\'e}nez-Donaire},
  {Leroy}, {Gallagher}, {Usero}, {Sandstrom}, {Bolatto}, {Hughes}, {Kramer},
  {Krumholz}, {Meier}, {Murphy}, {Pety}, {Rosolowsky}, {Schinnerer}, {Schruba},
  {Sliwa}, \& {Walter}}]{Cormier2018}
{Cormier}, D., {Bigiel}, F., {Jim{\'e}nez-Donaire}, M.~J., {et~al.} 2018,
  \mnras, 475, 3909, \dodoi{10.1093/mnras/sty059}

\bibitem[{{Dame}(2011)}]{Dame2011}
{Dame}, T.~M. 2011, arXiv e-prints, arXiv:1101.1499.
\newblock \doarXiv{1101.1499}

\bibitem[{{Dobbs} \& {Baba}(2014)}]{Dobbs2014}
{Dobbs}, C., \& {Baba}, J. 2014, \pasa, 31, e035, \dodoi{10.1017/pasa.2014.31}

\bibitem[{Ester {et~al.}(1996)Ester, Kriegel, Sander, \& Xu}]{Ester1996}
Ester, M., Kriegel, H.-P., Sander, J., \& Xu, X. 1996, in Proceedings of the
  Second International Conference on Knowledge Discovery and Data Mining,
  KDD'96 (AAAI Press), 226--231.
\newblock \url{http://dl.acm.org/citation.cfm?id=3001460.3001507}

\bibitem[{{Field} \& {Saslaw}(1965)}]{Field1965}
{Field}, G.~B., \& {Saslaw}, W.~C. 1965, \apj, 142, 568, \dodoi{10.1086/148318}

\bibitem[{{Gallagher} {et~al.}(2018){Gallagher}, {Leroy}, {Bigiel}, {Cormier},
  {Jim{\'e}nez-Donaire}, {Ostriker}, {Usero}, {Bolatto}, {Garc{\'\i}a-Burillo},
  {Hughes}, {Kepley}, {Krumholz}, {Meidt}, {Meier}, {Murphy}, {Pety},
  {Rosolowsky}, {Schinnerer}, {Schruba}, \& {Walter}}]{Gallagher2018}
{Gallagher}, M.~J., {Leroy}, A.~K., {Bigiel}, F., {et~al.} 2018, \apj, 858, 90,
  \dodoi{10.3847/1538-4357/aabad8}

\bibitem[{{Goldreich} \& {Lynden-Bell}(1965)}]{Goldreich1965}
{Goldreich}, P., \& {Lynden-Bell}, D. 1965, \mnras, 130, 97,
  \dodoi{10.1093/mnras/130.2.97}

\bibitem[{{Gong} {et~al.}(2021){Gong}, {Belloche}, {Du}, {Menten}, {Henkel},
  {Li}, {Wyrowski}, \& {Mao}}]{Gong2021}
{Gong}, Y., {Belloche}, A., {Du}, F.~J., {et~al.} 2021, \aap, 646, A170,
  \dodoi{10.1051/0004-6361/202039465}

\bibitem[{{Gratier} {et~al.}(2021){Gratier}, {Pety}, {Bron}, {Roueff},
  {Orkisz}, {Gerin}, {de Souza Magalhaes}, {Gaudel}, {Vono}, {Bardeau},
  {Chanussot}, {Chainais}, {Goicoechea}, {Guzm{\'a}n}, {Hughes}, {Kainulainen},
  {Languignon}, {Le Bourlot}, {Le Petit}, {Levrier}, {Liszt}, {Peretto},
  {Roueff}, \& {Sievers}}]{Gratier2021}
{Gratier}, P., {Pety}, J., {Bron}, E., {et~al.} 2021, \aap, 645, A27,
  \dodoi{10.1051/0004-6361/202037871}

\bibitem[{{Hacar} \& {Tafalla}(2011)}]{Hacar2011}
{Hacar}, A., \& {Tafalla}, M. 2011, \aap, 533, A34,
  \dodoi{10.1051/0004-6361/201117039}

\bibitem[{{Heitsch} {et~al.}(2006){Heitsch}, {Slyz}, {Devriendt}, {Hartmann},
  \& {Burkert}}]{Heitsch2006}
{Heitsch}, F., {Slyz}, A.~D., {Devriendt}, J. E.~G., {Hartmann}, L.~W., \&
  {Burkert}, A. 2006, \apj, 648, 1052, \dodoi{10.1086/505931}

\bibitem[{{Henshaw} {et~al.}(2016){Henshaw}, {Caselli}, {Fontani},
  {Jim{\'e}nez-Serra}, {Tan}, {Longmore}, {Pineda}, {Parker}, \&
  {Barnes}}]{Henshaw2016}
{Henshaw}, J.~D., {Caselli}, P., {Fontani}, F., {et~al.} 2016, \mnras, 463,
  146, \dodoi{10.1093/mnras/stw1794}

\bibitem[{{Heyer} \& {Dame}(2015)}]{Heyer2015}
{Heyer}, M., \& {Dame}, T.~M. 2015, \araa, 53, 583,
  \dodoi{10.1146/annurev-astro-082214-122324}

\bibitem[{Hunter(2007)}]{Hunter2007}
Hunter, J.~D. 2007, Computing in Science \& Engineering, 9, 90,
  \dodoi{10.1109/MCSE.2007.55}

\bibitem[{{Inoue} \& {Fukui}(2013)}]{Inoue2013}
{Inoue}, T., \& {Fukui}, Y. 2013, \apjl, 774, L31,
  \dodoi{10.1088/2041-8205/774/2/L31}

\bibitem[{{Inoue} {et~al.}(2018){Inoue}, {Hennebelle}, {Fukui}, {Matsumoto},
  {Iwasaki}, \& {Inutsuka}}]{Inoue2018}
{Inoue}, T., {Hennebelle}, P., {Fukui}, Y., {et~al.} 2018, \pasj, 70, S53,
  \dodoi{10.1093/pasj/psx089}

\bibitem[{{Kainulainen} {et~al.}(2017){Kainulainen}, {Stutz}, {Stanke},
  {Abreu-Vicente}, {Beuther}, {Henning}, {Johnston}, \&
  {Megeath}}]{Kainulainen2017}
{Kainulainen}, J., {Stutz}, A.~M., {Stanke}, T., {et~al.} 2017, \aap, 600,
  A141, \dodoi{10.1051/0004-6361/201628481}

\bibitem[{{Koyama} \& {Inutsuka}(2002)}]{Koyama2002}
{Koyama}, H., \& {Inutsuka}, S.-i. 2002, \apjl, 564, L97,
  \dodoi{10.1086/338978}

\bibitem[{{Kwan} \& {Valdes}(1983)}]{Kwan1983}
{Kwan}, J., \& {Valdes}, F. 1983, \apj, 271, 604, \dodoi{10.1086/161227}

\bibitem[{{Lin} \& {Shu}(1964)}]{Lin1964}
{Lin}, C.~C., \& {Shu}, F.~H. 1964, \apj, 140, 646, \dodoi{10.1086/147955}

\bibitem[{{Lin} {et~al.}(2019){Lin}, {Csengeri}, {Wyrowski}, {Urquhart},
  {Schuller}, {Weiss}, \& {Menten}}]{Lin2019}
{Lin}, Y., {Csengeri}, T., {Wyrowski}, F., {et~al.} 2019, \aap, 631, A72,
  \dodoi{10.1051/0004-6361/201935410}

\bibitem[{{Lu} {et~al.}(2018){Lu}, {Zhang}, {Liu}, {Sanhueza}, {Tatematsu},
  {Feng}, {Smith}, {Myers}, {Sridharan}, \& {Gu}}]{Lu2018}
{Lu}, X., {Zhang}, Q., {Liu}, H.~B., {et~al.} 2018, \apj, 855, 9,
  \dodoi{10.3847/1538-4357/aaad11}

\bibitem[{{Matsumoto} {et~al.}(2015){Matsumoto}, {Dobashi}, \&
  {Shimoikura}}]{Matsumoto2015}
{Matsumoto}, T., {Dobashi}, K., \& {Shimoikura}, T. 2015, \apj, 801, 77,
  \dodoi{10.1088/0004-637X/801/2/77}

\bibitem[{{M{\'e}ndez-Hern{\'a}ndez} {et~al.}(2020){M{\'e}ndez-Hern{\'a}ndez},
  {Ibar}, {Knudsen}, {Cassata}, {Aravena}, {Micha{\l}owski}, {Zhang},
  {Lara-L{\'o}pez}, {Ivison}, {van der Werf}, {Villanueva}, {Herrera-Camus}, \&
  {Hughes}}]{Mendez-Hernandez2020}
{M{\'e}ndez-Hern{\'a}ndez}, H., {Ibar}, E., {Knudsen}, K.~K., {et~al.} 2020,
  \mnras, 497, 2771, \dodoi{10.1093/mnras/staa1964}

\bibitem[{{Molinari} {et~al.}(2010){Molinari}, {Swinyard}, {Bally}, {Barlow},
  {Bernard}, {Martin}, {Moore}, {Noriega-Crespo}, {Plume}, {Testi}, {Zavagno},
  {Abergel}, {Ali}, {Anderson}, {Andr{\'e}}, {Baluteau}, {Battersby},
  {Beltr{\'a}n}, {Benedettini}, {Billot}, {Blommaert}, {Bontemps}, {Boulanger},
  {Brand}, {Brunt}, {Burton}, {Calzoletti}, {Carey}, {Caselli}, {Cesaroni},
  {Cernicharo}, {Chakrabarti}, {Chrysostomou}, {Cohen}, {Compiegne}, {de
  Bernardis}, {de Gasperis}, {di Giorgio}, {Elia}, {Faustini}, {Flagey},
  {Fukui}, {Fuller}, {Ganga}, {Garcia-Lario}, {Glenn}, {Goldsmith}, {Griffin},
  {Hoare}, {Huang}, {Ikhenaode}, {Joblin}, {Joncas}, {Juvela}, {Kirk},
  {Lagache}, {Li}, {Lim}, {Lord}, {Marengo}, {Marshall}, {Masi}, {Massi},
  {Matsuura}, {Minier}, {Miville-Desch{\^e}nes}, {Montier}, {Morgan}, {Motte},
  {Mottram}, {M{\"u}ller}, {Natoli}, {Neves}, {Olmi}, {Paladini}, {Paradis},
  {Parsons}, {Peretto}, {Pestalozzi}, {Pezzuto}, {Piacentini}, {Piazzo},
  {Polychroni}, {Pomar{\`e}s}, {Popescu}, {Reach}, {Ristorcelli}, {Robitaille},
  {Robitaille}, {Rod{\'o}n}, {Roy}, {Royer}, {Russeil}, {Saraceno}, {Sauvage},
  {Schilke}, {Schisano}, {Schneider}, {Schuller}, {Schulz}, {Sibthorpe},
  {Smith}, {Smith}, {Spinoglio}, {Stamatellos}, {Strafella}, {Stringfellow},
  {Sturm}, {Taylor}, {Thompson}, {Traficante}, {Tuffs}, {Umana}, {Valenziano},
  {Vavrek}, {Veneziani}, {Viti}, {Waelkens}, {Ward-Thompson}, {White},
  {Wilcock}, {Wyrowski}, {Yorke}, \& {Zhang}}]{Molinari2010}
{Molinari}, S., {Swinyard}, B., {Bally}, J., {et~al.} 2010, \aap, 518, L100,
  \dodoi{10.1051/0004-6361/201014659}

\bibitem[{{Neralwar} {et~al.}(2022{\natexlab{a}}){Neralwar}, {Colombo},
  {Duarte-Cabral}, {Urquhart}, {Mattern}, {Wyrowski}, {Menten}, {Barnes},
  {Sanchez-Monge}, {Beuther}, {Rigby}, {Mazumdar}, {Eden}, {Csengeri}, {Dobbs},
  {Veena}, {Neupane}, {Henning}, {Schuller}, {Leurini}, {Wienen}, {Yang},
  {Ragan}, {Medina}, \& {Nguyen-Luong}}]{Neralwar2022a}
{Neralwar}, K.~R., {Colombo}, D., {Duarte-Cabral}, A., {et~al.}
  2022{\natexlab{a}}, arXiv e-prints, arXiv:2203.02504.
\newblock \doarXiv{2203.02504}

\bibitem[{{Neralwar} {et~al.}(2022{\natexlab{b}}){Neralwar}, {Colombo},
  {Duarte-Cabral}, {Urquhart}, {Mattern}, {Wyrowski}, {Menten}, {Barnes},
  {Sanchez-Monge}, {Rigby}, {Mazumdar}, {Eden}, {Csengeri}, {Dobbs}, {Veena},
  {Neupane}, {Henning}, {Schuller}, {Leurini}, {Wienen}, {Yang}, {Ragan},
  {Medina}, \& {Nguyen-Luong}}]{Neralwar2022b}
---. 2022{\natexlab{b}}, arXiv e-prints, arXiv:2205.02253.
\newblock \doarXiv{2205.02253}

\bibitem[{{Oort}(1954)}]{Oort1954}
{Oort}, J.~H. 1954, \bain, 12, 177

\bibitem[{{Padoan} \& {Nordlund}(1999)}]{Padoan1999}
{Padoan}, P., \& {Nordlund}, {\r{A}}. 1999, \apj, 526, 279,
  \dodoi{10.1086/307956}

\bibitem[{{Passot} {et~al.}(1995){Passot}, {Vazquez-Semadeni}, \&
  {Pouquet}}]{Passot1995}
{Passot}, T., {Vazquez-Semadeni}, E., \& {Pouquet}, A. 1995, \apj, 455, 536,
  \dodoi{10.1086/176603}

\bibitem[{{Peretto} {et~al.}(2022){Peretto}, {Adam}, {Ade}, {Ajeddig},
  {Andr{\'e}}, {Artis}, {Aussel}, {Bacmann}, {Beelen}, {Beno{\^\i}t}, {Berta},
  {Bing}, {Bourrion}, {Calvo}, {Catalano}, {De Petris}, {D{\'e}sert}, {Doyle},
  {Driessen}, {Gomez}, {Goupy}, {K{\'e}ruzor{\'e}}, {Kramer}, {Ladjelate},
  {Lagache}, {Leclercq}, {Lestrade}, {Mac{\'\i}as-P{\'e}rez}, {Maury},
  {Mauskopf}, {Mayet}, {Monfardini}, {Mu{\~n}oz-Echeverr{\'\i}a}, {Perotto},
  {Pisano}, {Ponthieu}, {Rev{\'e}ret}, {Rigby}, {Ristorcelli}, {Ritacco},
  {Romero}, {Roussel}, {Ruppin}, {Schuster}, {Shu}, {Sievers}, {Tucker}, \&
  {Zylka}}]{peretto2022}
{Peretto}, N., {Adam}, R., {Ade}, P., {et~al.} 2022, in European Physical
  Journal Web of Conferences, Vol. 257, European Physical Journal Web of
  Conferences, 00037, \dodoi{10.1051/epjconf/202225700037}

\bibitem[{{Pudritz} \& {Kevlahan}(2013)}]{Pudritz2013}
{Pudritz}, R.~E., \& {Kevlahan}, N.~K.~R. 2013, Philosophical Transactions of
  the Royal Society of London Series A, 371, 20120248,
  \dodoi{10.1098/rsta.2012.0248}

\bibitem[{{Reid} {et~al.}(2016){Reid}, {Dame}, {Menten}, \&
  {Brunthaler}}]{Reid2016}
{Reid}, M.~J., {Dame}, T.~M., {Menten}, K.~M., \& {Brunthaler}, A. 2016, \apj,
  823, 77, \dodoi{10.3847/0004-637X/823/2/77}

\bibitem[{{Roberts}(1969)}]{Roberts1969}
{Roberts}, W.~W. 1969, \apj, 158, 123, \dodoi{10.1086/150177}

\bibitem[{{Roman-Duval} {et~al.}(2016){Roman-Duval}, {Heyer}, {Brunt}, {Clark},
  {Klessen}, \& {Shetty}}]{Roman-Duval2016}
{Roman-Duval}, J., {Heyer}, M., {Brunt}, C.~M., {et~al.} 2016, \apj, 818, 144,
  \dodoi{10.3847/0004-637X/818/2/144}

\bibitem[{{Su} {et~al.}(2019){Su}, {Yang}, {Zhang}, {Gong}, {Wang}, {Zhou},
  {Wang}, {Chen}, {Sun}, {Chen}, {Xu}, \& {Jiang}}]{Su2019}
{Su}, Y., {Yang}, J., {Zhang}, S., {et~al.} 2019, \apjs, 240, 9,
  \dodoi{10.3847/1538-4365/aaf1c8}

\bibitem[{{Tasker} \& {Tan}(2009)}]{Tasker2009}
{Tasker}, E.~J., \& {Tan}, J.~C. 2009, \apj, 700, 358,
  \dodoi{10.1088/0004-637X/700/1/358}

\bibitem[{{Tokuda} {et~al.}(2019){Tokuda}, {Fukui}, {Harada}, {Saigo},
  {Tachihara}, {Tsuge}, {Inoue}, {Torii}, {Nishimura}, {Zahorecz}, {Nayak},
  {Meixner}, {Minamidani}, {Kawamura}, {Mizuno}, {Indebetouw}, {Sewi{\l}o},
  {Madden}, {Galametz}, {Lebouteiller}, {Chen}, \& {Onishi}}]{Tokuda2019}
{Tokuda}, K., {Fukui}, Y., {Harada}, R., {et~al.} 2019, \apj, 886, 15,
  \dodoi{10.3847/1538-4357/ab48ff}

\bibitem[{{Tomisaka}(1984)}]{Tomisaka1984}
{Tomisaka}, K. 1984, \pasj, 36, 457

\bibitem[{{Torii} {et~al.}(2019){Torii}, {Fujita}, {Nishimura}, {Tokuda},
  {Kohno}, {Tachihara}, {Inutsuka}, {Matsuo}, {Kuriki}, {Tsuda}, {Minamidani},
  {Umemoto}, {Kuno}, \& {Miyamoto}}]{Torii2019}
{Torii}, K., {Fujita}, S., {Nishimura}, A., {et~al.} 2019, \pasj, 71, S2,
  \dodoi{10.1093/pasj/psz033}

\bibitem[{{Vazquez-Semadeni} {et~al.}(1995){Vazquez-Semadeni}, {Passot}, \&
  {Pouquet}}]{Vazquez1995}
{Vazquez-Semadeni}, E., {Passot}, T., \& {Pouquet}, A. 1995, \apj, 441, 702,
  \dodoi{10.1086/175393}

\bibitem[{{V{\'a}zquez-Semadeni} {et~al.}(2006){V{\'a}zquez-Semadeni}, {Ryu},
  {Passot}, {Gonz{\'a}lez}, \& {Gazol}}]{Semadeni2006}
{V{\'a}zquez-Semadeni}, E., {Ryu}, D., {Passot}, T., {Gonz{\'a}lez}, R.~F., \&
  {Gazol}, A. 2006, \apj, 643, 245, \dodoi{10.1086/502710}

\bibitem[{{Wilson} {et~al.}(1970){Wilson}, {Jefferts}, \&
  {Penzias}}]{Wilson1970}
{Wilson}, R.~W., {Jefferts}, K.~B., \& {Penzias}, A.~A. 1970, \apjl, 161, L43,
  \dodoi{10.1086/180567}

\bibitem[{{Yan} {et~al.}(2020){Yan}, {Yang}, {Su}, {Sun}, \& {Wang}}]{Yan2020}
{Yan}, Q.-Z., {Yang}, J., {Su}, Y., {Sun}, Y., \& {Wang}, C. 2020, \apj, 898,
  80, \dodoi{10.3847/1538-4357/ab9f9c}

\bibitem[{{Yan} {et~al.}(2021){Yan}, {Yang}, {Sun}, {Su}, {Xu}, {Wang}, {Zhou},
  \& {Wang}}]{Yan2021}
{Yan}, Q.-Z., {Yang}, J., {Sun}, Y., {et~al.} 2021, \aap, 645, A129,
  \dodoi{10.1051/0004-6361/202039768}

\bibitem[{Yuan {et~al.}(2022)Yuan, Yang, Du, \& Su}]{Yuan2022}
Yuan, L., Yang, J., Du, F., \& Su, Y. 2022, {The 12CO and 13CO lines data of
  18,190 molecular clouds from the MWISP CO survey}, V1,  Science Data Bank,
  \dodoi{10.57760/sciencedb.j00001.00427}

\bibitem[{{Yuan} {et~al.}(2019){Yuan}, {Zhu}, {Liu}, {Yuan}, {Wu}, {Kim},
  {Wang}, {Zhou}, {Tatematsu}, \& {Kuno}}]{Yuan2019}
{Yuan}, L., {Zhu}, M., {Liu}, T., {et~al.} 2019, \mnras, 487, 1315,
  \dodoi{10.1093/mnras/stz1266}

\bibitem[{{Yuan} {et~al.}(2020){Yuan}, {Li}, {Zhu}, {Liu}, {Wang}, {Liu},
  {Kim}, {Tatematsu}, {Yuan}, \& {Wu}}]{Yuan2020}
{Yuan}, L., {Li}, G.-X., {Zhu}, M., {et~al.} 2020, \aap, 637, A67,
  \dodoi{10.1051/0004-6361/201936625}

\bibitem[{{Yuan} {et~al.}(2021){Yuan}, {Yang}, {Du}, {Liu}, {Zhang}, {Lin},
  {Sun}, {Yan}, {Ma}, {Su}, {Sun}, \& {Zhou}}]{Yuan2021}
{Yuan}, L., {Yang}, J., {Du}, F., {et~al.} 2021, \apjs, 257, 51,
  \dodoi{10.3847/1538-4365/ac242a}

\end{thebibliography}
\bibliographystyle{aasjournal}

%% This command is needed to show the entire author+affiliation list when
%% the collaboration and author truncation commands are used.  It has to
%% go at the end of the manuscript.
%\allauthors

%% Include this line if you are using the \added, \replaced, \deleted
%% commands to see a summary list of all changes at the end of the article.
%\listofchanges

\end{document}